\documentclass[11pt,a4paper]{article}
\usepackage{jcappub}
\usepackage{multirow}
\usepackage{graphicx}
\usepackage{caption}
\usepackage{subcaption}
\usepackage{color}
\usepackage{amsmath,amsfonts,amsthm}

\usepackage[normalem]{ulem}

\definecolor{myred}{RGB}{102,0,0}
\definecolor{Thomas}{RGB}{128,0,128}
\definecolor{blueArturo}{RGB}{95,158,160}

\newcommand{\change}[1]{{\color{black} #1}}

\newcommand{\beq}{\begin{equation}}
\newcommand{\eeq}{\end{equation}}
\newcommand{\bsubeq}{\begin{subequations}}
\newcommand{\esubeq}{\end{subequations}}
\newcommand{\ben}{\begin{eqnarray}}
\newcommand{\een}{\end{eqnarray}}
\newcommand{\bal}{\begin{align}}
\newcommand{\eal}{\end{align}}
\newcommand{\balt}{\begin{alignat}}
\newcommand{\ealt}{\end{alignat}}
\newcommand{\bi}{\begin{itemize}}
\newcommand{\ei}{\end{itemize}}
\newcommand{\nn}{\nonumber}

\newcommand{\ie}{\textit{i.e.}}
\newcommand{\eg}{\textit{e.g.}}
\newcommand{\etc}{\textit{etc.}}

\newcommand{\citeeq}[1]{Eq.~(\ref{#1})}

\newcommand{\citeeqs}[1]{Eqs.~(\ref{#1})}
\newcommand{\citeeqtoeq}[2]{Eqs.~(\ref{#1}-\ref{#2})}

\newcommand{\citeeqp}[1]{Eq.~\ref{#1}}
\newcommand{\citesec}[1]{Sect.~\ref{#1}}
\newcommand{\citesecs}[1]{Sects.~\ref{#1}}
\newcommand{\citeapp}[1]{App.~\ref{#1}}

\newcommand{\citetab}[1]{Table~\ref{#1}}
\newcommand{\citefig}[1]{Fig.~\ref{#1}}
\newcommand{\citefigs}[1]{Figs.~\ref{#1}}

\newcommand{\fsol}{{\ifmmode f_{\odot} \else $f_{\odot}$\fi}}

\newcommand{\sqgev}{{\ifmmode {\rm GeV}^2 \else ${\rm GeV}^2$\fi}}

\newcommand{\sr}{{\ifmmode {\rm sr} \else ${\rm sr}$\fi}}
\newcommand{\invsr}{{\ifmmode {\rm sr}^{-1} \else ${\rm sr}^{-1}$\fi}}

\newcommand{\scnd}{{\ifmmode {\rm s} \else ${\rm s}$\fi}}
\newcommand{\invscnd}{{\ifmmode {\rm s}^{-1} \else ${\rm s}^{-1}$\fi}}

\newcommand{\kpc}{{\ifmmode {\rm kpc} \else ${\rm kpc}$\fi}}
\newcommand{\invkpc}{{\ifmmode {\rm kpc}^{-1} \else ${\rm kpc}^{-1}$\fi}}

\newcommand{\sqkpc}{{\ifmmode {\rm kpc}^{2} \else ${\rm kpc}^{2}$\fi}}
\newcommand{\invsqkpc}{{\ifmmode {\rm kpc}^{-2} \else ${\rm kpc}^{-2}$\fi}}

\newcommand{\cm}{{\ifmmode {\rm cm} \else ${\rm cm}$\fi}}
\newcommand{\invcm}{{\ifmmode {\rm cm}^{-1} \else ${\rm cm}^{-1}$\fi}}

\newcommand{\sqcm}{{\ifmmode {\rm cm}^2 \else ${\rm cm}^2$\fi}}
\newcommand{\invsqcm}{{\ifmmode {\rm cm}^{-2} \else ${\rm cm}^{-2}$\fi}}

\newcommand{\meter}{{\ifmmode {\rm m} \else ${\rm m}$\fi}}
\newcommand{\invmeter}{{\ifmmode {\rm m}^{-1} \else ${\rm m}^{-1}$\fi}}

\newcommand{\sqmeter}{{\ifmmode {\rm m}^2 \else ${\rm m}^2$\fi}}
\newcommand{\invsqmeter}{{\ifmmode {\rm m}^{-2} \else ${\rm m}^{-2}$\fi}}

\newcommand{\lcdm}{{\ifmmode \Lambda{\rm CDM} \else $\Lambda{\rm CDM}$\fi}}

\newcommand{\Rvirh}{{\ifmmode R_{\rm vir}^{\rm h} \else 
    $R_{\rm vir}^{\rm h}$\fi}}

\newcommand{\vesc}{{\ifmmode v_{\rm esc} \else $v_{\rm esc}$\fi}}

\graphicspath{{./Figs/}{./}}

\subheader{LUPM:20-024, IFT-UAM/CSIC-20-64}

\title{Predicting the dark matter velocity distribution in galactic structures: tests against hydrodynamic cosmological simulations}
\author[a,b]{Thomas Lacroix,}
\author[c,d]{Arturo N\'u{\~{n}}ez-Casti{\~{n}eyra},}
\author[e,b]{Martin Stref,}
\author[b]{Julien Lavalle}
\author[c]{and Emmanuel Nezri}

\affiliation[a]{Instituto de F\'isica Te\'orica UAM/CSIC, Universidad Aut\'onoma de Madrid, 28049 Madrid, Spain}
\affiliation[b]{Laboratoire Univers et Particules de Montpellier (LUPM), Universit\'e de Montpellier \& CNRS, Place Eug\`ene Bataillon, 34095 Montpellier Cedex 05, France}
\affiliation[c]{Aix Marseille Univ, CNRS, CNES, Laboratoire d'Astrophysique de Marseille (LAM),\\
38 rue F. Joliot-Curie, 13388 Marseille Cedex 13, France}
\affiliation[d]{Centre de Physique des Particules de Marseille (CPPM), 163 av. de Luminy, 13288 Marseille Cedex 09, France}
\affiliation[e]{Univ. Grenoble Alpes, USMB, CNRS, LAPTh, F-74000 Annecy, France}

\emailAdd{thomas.lacroix@uam.es}
\emailAdd{arturo.nunez@lam.fr}
\emailAdd{martin.stref@lapth.cnrs.fr}
\emailAdd{lavalle@in2p3.fr}
\emailAdd{emmanuel.nezri@lam.fr}

\abstract{
  Reducing theoretical uncertainties in Galactic dark matter (DM) searches is an important challenge
  as several experiments are now delving into the parameter space relevant to popular (particle or
  not) candidates. Since many DM signal predictions rely on the knowledge of the DM velocity
  distribution---direct searches, capture by stars, $p$-wave-suppressed or Sommerfeld-enhanced
  annihilation rate, microlensing of primordial black holes, \etc---it is necessary to assess the
  accuracy of our current theoretical handle. Beyond Maxwellian approximations or ad-hoc
  extrapolations of fits on cosmological simulations, approaches have been proposed to
  self-consistently derive the DM phase-space distribution only from the detailed mass content of
  the Galaxy and some symmetry assumptions (\eg~the Eddington inversion and its anisotropic
  extensions). Although theoretically sound, these methods are still based on simplifying
  assumptions and their relevance to real galaxies can be questioned. In this paper, we use
  zoomed-in cosmological simulations to quantify the associated uncertainties. Assuming isotropy,
  we predict the speed distribution and its moments from the DM and baryonic content measured in
  simulations, and compare them with the true ones. \change{Taking as input galactic mass models
    fitted on full simulation data, we reach a predictivity down to $\sim 10$\% for some
    velocity-related observables, significantly better than some Maxwellian models.} This
  moderate theoretical error is particularly encouraging at a time when stellar surveys like the
  \textit{Gaia} mission should allow us to improve constraints on Galactic mass models.}

\arxivnumber{2005.03955}

\begin{document}
\maketitle

\section{Introduction}
\label{sec:intro}
The phase-space distribution function (PSDF) of dark matter (DM) in galactic halos is very often a central ingredient of predictions related to DM searches (\eg~\cite{PressEtAl1985,DrukierEtAl1986,Griest1991,SpergelEtAl2000,HisanoEtAl2005,EssigEtAl2013a,MassariEtAl2015,ZhaoEtAl2016,LiuEtAl2016,BoudaudEtAl2019}), and as such potentially a major source of astrophysical uncertainties. A realistic characterization of this ubiquitous quantity is therefore essential in order to determine the properties of a DM candidate in the event of a detection, or to turn null results into constraints. This is particularly relevant to DM scenarios in which several signatures are velocity-dependent, but may have different spatial origins (consider for instance a particle DM scenario that would lead to both a non-negligible scattering cross section off of nuclei, possibly enabling direct detection or stellar capture, and a Sommerfeld-enhanced annihilation rate). Further assessing and reducing the associated systematic uncertainties is also critical in order to reach meaningful conclusions on one's DM models of interest.

There are several approaches to this problem---from now on we concentrate on the case of cold and collisionless DM on galactic scales. A very common one, mostly dedicated to proof-of-principle estimates, relies on radical simplifications like assuming a Maxwell-Boltzmann distribution function (DF), calibrated on more or less educated guesses for the velocity dispersion often based on the virial theorem (\eg~\cite{PressEtAl1985,DrukierEtAl1986,Griest1991,SpergelEtAl2000,HisanoEtAl2005}). However, even if this might provide a reasonable description of the DM PSDF at some {\em specific} locations in DM structures \cite{BozorgniaEtAl2016}, the aforementioned calibration can hardly be made but in a rather ad hoc way. Moreover, it has been known for a long time that the Maxwell-Boltzmann DF provides a poor {\em general} description of gravitational systems (\eg~\cite{King1966,BinneyTremaine2008}), which has also been confirmed from simulations (\eg~\cite{KazantzidisEtAl2004,SanchisEtAl2004,WojtakEtAl2008}). Another approach consists in extrapolating results from ``Milky Way-like'' cosmological simulations (\eg~\cite{HelmiEtAl2002,LavalleEtAl2008a,VogelsbergerEtAl2009,LingEtAl2010,MaoEtAl2013,PillepichEtAl2014a,BozorgniaEtAl2016,BozorgniaEtAl2019,NunezCastineyraEtAl2019,CallinghamEtAl2020}).

\change{There is no doubt about the essential role of highly resolved simulations to more deeply understand the complex sequence of structure formation, and to try to single out generic properties of DM. This has very often helped deepen or even develop our understanding in this field. Here, however, an issue is whether averaging velocity DFs from many different simulations, even with similar total masses, can really lead to a realistic description of one particular {\em real} object like the Milky Way (MW), with a specific history and specific distributions of its intimate components. Is it supported by physical principles? How to quantify theoretical errors in the absence of established and/or understood physical correlations? How far from averaged properties of simulated galaxies can a real single galaxy lie, especially when baryonic processes are described on rather empirical bases in simulations? Empirical methods are particularly useful when theory is limited by complexity, but are in the meantime more predictive when they rely on or comply with physical guidance. There have been examples in which semi-empirical fits provided deep insight into complex non-linear physical processes, like parametric fits of DM density profiles \cite{Einasto1965,Hernquist1990,Zhao1996,NavarroEtAl1996a,MerrittEtAl2006} or other generic properties found in simulations \cite{NavarroEtAl1997,EkeEtAl2001,BullockEtAl2001b}. Nevertheless, they were justified a priori or a posteriori by some scale invariance arguments (\eg~\cite{Bertschinger1985a,MoEtAl2010}), and very often strengthened or guided by physical principles independently. It is much more delicate to ``rescale'' or even to use a velocity DF averaged from simulations, because we know from physical principles that it should respond to the gravitational influence of all ingredients in the specific object under consideration. An alternative averaging procedure might resort to theoretically sound ways of rescaling each individual DF {\em before} the averaging, as was done in another context in Ref.~\cite{CallinghamEtAl2020}. Below, we focus instead on approaches based upon physical principles.}

A PSDF model should at least be made consistent with existing kinematic constraints on the galactic system under scrutiny. This is a minimal requirement that becomes increasingly important for Galactic DM searches with the advent of the \textit{Gaia} space mission \cite{GaiaCollab2016,GaiaCollab2018}, which is now characterizing the components of the Milky Way with an unprecedented level of detail, and is helping to refine Galactic mass models \cite{CautunEtAl2020}. A strong limitation of the approaches described above is precisely that they do not account for these constraints. Fortunately, methods of PSDF predictions have been developed which can be made consistent with existing kinematic observations. They are usually grounded on solutions to the steady-state Boltzmann equation, and can vary in the level of simplifying assumptions and implemented complexity (see a non-exhaustive list of examples in Refs.~\cite{BinneyTremaine2008,Eddington1916,Ollongren1962,King1966,Osipkov1979,KentEtAl1982,Merritt1985,Cuddeford1991,UllioEtAl2001a,WidrowEtAl2005,Wojtak2008,McMillanEtAl2008,CatenaEtAl2012,Strigari2013,FerrerEtAl2013,FornasaEtAl2014,Hunter2014,WilliamsEtAl2015,CuddefordLouis1995,LavalleEtAl2015,PostiEtAl2015,BinneyEtAl2015,SandersEtAl2016,CerdenoEtAl2016,LacroixEtAl2018,Petac2019axisymmetric}). An important advantage of these methods, in addition to their theoretical self-consistency, is that they are by construction physically consistent with the observational constraints they are based on (or trained from)---therefore, the predicted PSDF for DM does respond to the gravitational information given in input. However, quantifying their level of predictivity for the PSDF itself remains an important step to clear in order to assess the corresponding theoretical uncertainties. This is precisely what we aim at doing here on both qualitative and quantitative bases, focusing on Milky Way-like galaxies in the broad sense. Such a study can be performed by comparing the predictions of the DM PSDF inferred from a global mass model (including the profiles of both the baryons and the DM) fitted on a high-resolution zoomed-in cosmological simulation, with the true PSDF measured in the same simulation. Note that here, contrary to the extrapolations mentioned above, we do not really care about how much ``Milky Way-like'' our virtual test galaxies are, since the primary physical effect that matters at first order is the dynamical correlation of all ingredients through gravity. For the prediction part, we will actually assume the maximally symmetric case, \ie~dynamical equilibrium, isotropy, and spherical symmetry for the DM component. In that case, the PSDF depends only on energy by virtue of the Jeans theorem \cite{Ollongren1962}, and one can use the well-known Eddington inversion to predict the PSDF from the DM profile and the total gravitational potential of the system (see \eg~Refs.~\cite{Eddington1916,BinneyTremaine2008} for an introduction, and Ref.~\cite{LacroixEtAl2018} for a detailed review in the context of DM searches). This should be generalized to more complex prediction methods, like those based on action-angle coordinates \cite{McMillanEtAl2008,PostiEtAl2015,BinneyEtAl2015,SandersEtAl2016,ColeEtAl2017,Binney2019}, but we will see in the sequel that maximally symmetric approaches can already reach a precision of order 10-20\% on the moments and inverse moments of the DM speed or relative speed at different positions in simulated spiral galaxies.

Finally, it is worth mentioning yet another complementary approach, based on data-driven modeling, which consists in inferring the DM velocity DF from observed stars. The idea is to identify those stellar populations (typically old stars) which are more likely to trace the DM velocity DF, and use them to infer the latter \cite{Herzog-ArbeitmanEtAl2018,Herzog-ArbeitmanEtAl2018a,NecibEtAl2019}. The systematic uncertainties associated with this approach are of different nature, since it also appeals to star formation and evolution, but they can also be examined using cosmological simulations \cite{BozorgniaEtAl2019}. An advantage of this kind of data-driven reconstruction methods, even though they require some modeling, is that they can be used to infer non-equilibrium features due to possible past mergers of dwarf galaxies directly from observations \cite{NecibEtAl2019a,OHareEtAl2019,EvansEtAl2019,BozorgniaEtAl2019}. The approach under study here does not allow us to account for such details, but we will see that it still performs reasonably well even in the presence of unrelaxed components (simulations do contain such features).

The paper is structured as follows. In \citesec{sec:theory}, we recall the main ingredients of the Eddington inversion method. We further introduce two benchmark Maxwellian models widely used in the literature, which will serve as references when discussing the predictivity of the Eddington model---we emphasize that the goal of this paper is to compare {\em predictions} with one another, not fits. Then, in \citesec{sec:sims}, we describe the suite of cosmological simulations used in our study, as well as the main properties of the synthetic halos. In \citesec{sec:comp} we confront the predictions of the Eddington method with various quantities associated with the velocity distribution of DM particles extracted from simulations. We further make relative comparisons with the {\em predictions} obtained from our reference Maxwellian models. We summarize our results and conclude in \citesec{sec:concl}, to which we refer the expert reader for a quick summary. Some technical details, as well as comparisons of theoretical predictions and simulation outputs for other halos of interest in our set of simulations, can be found in the Appendices.

\section{Eddington approach to the DM phase space and theoretical validity range}
\label{sec:theory}
\subsection{Generalities}
\label{ssec:gen}
In this section, we give a brief overview of the Eddington inversion method \cite{Eddington1916}---see Ref.~\cite{LacroixEtAl2018} for an extensive review also discussing theoretical issues in the context of DM searches. The underlying idea of the Eddington approach (and its extensions) is to predict, from first principles, the equilibrium PSDF of a gravitating system, $f(\vec{r},\vec{v})$. For a system of identical particles, the latter is the solution to a coupled system of equations, comprising the steady-state collisionless Boltzmann equation, 
\ben
\vec{v} \cdot \dfrac{\partial f}{\partial \vec{r}} - \dfrac{\partial \Phi}{\partial \vec{r}} \cdot \dfrac{\partial f}{\partial \vec{v}} = 0\,, 
\een
and the Poisson equation,
\ben
\label{eq:poisson}
\Delta \Phi = 4 \pi G \rho\,, 
\een
where $\Phi$ is the gravitational potential and
\ben
\label{eq:rho}
\rho(\vec{r}) = \int \! \mathrm{d}^{3}\vec{v}\, f(\vec{r},\vec{v}) 
\een
is the mass density.\footnote{According to our definition of the mass density, the PSDF is normalized to the total mass of the system---this holds throughout the paper unless specified otherwise.}

A clue as to the quantities on which the PSDF should depend is provided by the Jeans theorem, which states that any steady-state solution of the collisionless Boltzmann equation can be expressed as a function of isolating integrals of motion \cite{Jeans1915,Ollongren1962,BinneyTremaine2008}. Consequently, for spherically symmetric systems, the PSDF can be written $f(\vec{r},\vec{v})\equiv f(\mathcal{E},L)$, where $L=|\vec{r}\times\vec{v}|$ is the modulus of the angular momentum per unit mass, and
\ben
\label{eq:energy}
{\cal E} = \Psi(r) - \dfrac{v^{2}}{2}
\een
is the relative energy per unit mass. Here, $r$ stands for the distance to the center of the system, and we have introduced the relative gravitational potential
\ben
\label{eq:rel_pot}
\Psi(r) = \Phi(R_{\rm max}) - \Phi(r)\,,
\een
where $\Phi(r)$ is the solution to the Poisson equation (bound to vanish at infinity), and $R_{\rm max}$ is a radius chosen to represent the spatial extension of the system. The definition of this boundary is made more explicit in \citesec{ssec:rmax}. It should be noted that $\Psi$ is positive-definite everywhere, except at $R_{\rm max}$ where it vanishes. This means that the maximal value of the relative energy is attained for a particle at rest at the center of the object, where the relative potential is maximum: ${\cal E}_{\rm max}=\Psi_{\rm max}=\Psi(r=0)$. On the other hand, ${\cal E}\to 0$ whenever the speed approaches the escape speed $ v\to v_{\rm e}(r)\equiv\sqrt{2 \psi(r)}$. Otherwise, ${\cal E}$ globally decreases for particles getting farther and farther away from the center of the system.

In practice, for a realistic galactic object comprising DM and baryons, the relative potential can be written as
\ben
\Psi(r) =\Psi_{\rm D}(r) + \Psi_{\rm B}(r)\,,
\label{eq:pot_sum}
\een
where $\Psi_{\rm D}$ and $\Psi_{\rm B}$ are the relative potentials induced by DM and baryons, respectively. The relative potential of the whole system (or its components) is directly related to the underlying mass distribution via the Poisson equation, and can be expressed as
\ben
\Psi(r) = \int_{r}^{R_{\rm max}} \!  \mathrm{d}r' \, \dfrac{G m(r')}{r'^{2}} \,,
\label{eq:relative_potential}
\een
where the mass enclosed in a sphere of radius $r$ is
\ben
m(r) = 4 \pi \int_{0}^{r} \! {\rm d}r'\, \rho(r') r'^{2} \,.
\label{eq:mass}
\een
Since we restrict our study to the framework of the Eddington inversion, which is only valid for spherically symmetric and isotropic systems, we account for non-spherical components---such as stellar and gaseous disks---by considering the associated mass within a radius $r$
\ben
\label{eq:mass_axisymm}
m(r) = \int_{|\vec{x}|\le r}{\rm d}^{3}\vec{x} \,\rho(\vec{x})\,,
\een
which allows us to compute a spherically symmetric approximation of the gravitational potential.\footnote{In spiral galaxies, the baryonic matter distribution is mostly confined in a disk, and therefore strongly departs from spherical symmetry. However, the associated potential is closer to spherical than the matter distribution is, and thus ``sphericizing'' the baryonic potential remains a good approximation as for the precision we want to reach in the PSDF prediction \cite{BinneyTremaine2008,LacroixEtAl2018}.} The comparisons we make in this study between predictions obtained with the Eddington inversion and simulation outputs actually serve as a test of the very validity of this approach.
\subsection{Maximal symmetry: the Eddington equation}
\label{ssec:edd}
In the maximally symmetric case of an isotropic and spherically symmetric system, the angular momentum plays no role and the PSDF can be written as a function of the energy only: $f(\vec{r},\vec{v}) \equiv f(\mathcal{E})$---see \citeeq{eq:energy}. By performing an Abel inversion of \citeeq{eq:rho}, one can derive the unique steady-state solution of the collisionless Boltzmann equation corresponding to a given density-potential pair \cite{Eddington1916},
\ben
\label{eq:Eddington_formula}
f(\mathcal{E}) = \dfrac{1}{\sqrt{8}\pi^{2}} \left[ \dfrac{1}{\sqrt{\mathcal{E}}} \left( \dfrac{\mathrm{d}\rho}{\mathrm{d}\Psi} \right)_{\Psi=0} + \int_{0}^{\mathcal{E}} \! \dfrac{\mathrm{d}\Psi}{\sqrt{\mathcal{E} - \Psi}} \,  \dfrac{\mathrm{d}^{2}\rho}{\mathrm{d}\Psi^{2}}     \right] \,.   
\een
It should be noted that here $\rho$ refers to the DM density, while $\Psi$ is the total relative potential of the system, comprising all of its additional baryonic components if present. In the following we confront the Eddington prediction with both DM-only and DM+baryons simulations. The Eddington formula and its implications are discussed extensively in Ref.~\cite{LacroixEtAl2018}. Below, we provide a brief summary of the main features. 

In \citeeq{eq:Eddington_formula}, the Eddington PSDF features a term $\propto 1/\sqrt{\mathcal{E}}$ which only vanishes for an infinite (\ie~isolated) system, and otherwise diverges in the limit ${\cal E} \rightarrow 0$ for a non-isolated halo. This divergence of the PSDF translates into a divergence of the speed distribution for $v \rightarrow v_{\rm e}$, \ie~at the escape speed at any location in the halo. We describe the procedure we use to define the radial boundary of a simulated halo in \citesec{ssec:rmax}. This discussion is particularly relevant in the context of DM-related observables that depend on the high-speed tail of the speed distribution such as direct searches for (sub-)GeV DM particles. In principle, in order for the PSDF model to be fully consistent with the input density used in the Eddington inversion, while at the same time accounting for the radial boundary of the system, it is a priori not correct to simply drop the diverging term defining $f(\mathcal{E})$, as detailed in Ref.~\cite{LacroixEtAl2018}, where some prescriptions to solve this issue are presented. More specifically, one can either very slightly modify the input DM profile so that the diverging term is analytically zero, or directly regularize the PSDF itself via a method similar to the King method \cite{King1966,WidrowEtAl2005,DrakosEtAl2017}. \change{In the Milky Way, related systematic
  errors in the velocity moments at the solar position (relevant to \eg~direct DM particle
  searches) may exceed the percent level only when calculated above a rather high threshold speed,
  $v_{\rm min}\sim 450$~km/s (for a fixed mass model)---see Ref.~\cite{LacroixEtAl2018}.}

However, in the present work, we want to compare the global predictivity of the Eddington inversion with respect to synthetic halos, focusing in particular on the radial profiles of the moments of the speed and relative speed distributions, but not specifically on observables sensitive to the high-speed tail of the distribution. For these quantities, unless one is interested in the region close to the boundary of the system, this theoretical inconsistency of the Eddington formalism actually plays a negligible part and it is in practice sufficient to drop the diverging term in the Eddington PSDF. This is the simplifying approach we use in the following.
\subsection{Extensions of the Eddington formalism}
In this work, as a first step toward a better characterization of theoretical uncertainties, we test the simplest Eddington predictions against cosmological simulations, and therefore stick to the maximally symmetric assumptions. This is actually rather well motivated at first order, considering the evolution of the anisotropy with radius throughout the galactic halos considered here, as illustrated in \citefig{fig:anisotropy-beta-param}. This plot shows that the halos of interest are close to isotropic in the inner $\sim 10\, \rm kpc$ and become on average radially anisotropic further out. Some extensions of the Eddington formalism are discussed in Refs.~\cite{Osipkov1979,Merritt1985,Cuddeford1991,CuddefordLouis1995,BinneyTremaine2008,Wojtak2008,Bozorgnia2013} for spherically symmetric anisotropic systems, and the theoretical issues they raise especially in the context of DM searches are emphasized in Ref.~\cite{LacroixEtAl2018}. It is also possible to extend the Eddington inversion to axisymmetric systems, as discussed in Refs.~\cite{HunterQian1993,Petac2019axisymmetric}, or to implement more general theoretical setups based on action-angle coordinates \cite{McMillanEtAl2008,BinneyEtAl2015}. The comparison of PSDF predictions derived in the framework of these more complex models with simulations goes beyond the scope of this paper and will be the object of upcoming studies.
\subsection{An aside on the Jeans equation}
\label{ssec:Jeans}
Information on the PSDF of a system can be gained without solving the full collisionless Boltzmann equation. In particular, taking the second velocity moment of the steady-state Boltzmann equation leads to the Jeans equation \cite{Jeans1915,Ollongren1962}, which reads for a spherical system \cite{BinneyTremaine2008}
\ben
\frac{1}{\rho}\frac{\mathrm{d}(\rho\left\langle v_{r}^2\right>)}{\mathrm{d}r}+2\,\frac{\beta}{r}\left<v_{r}^2\right\rangle = \frac{\mathrm{d}\Psi}{\mathrm{d}r}\,.
\een
The $\beta$ parameter measures the velocity anisotropy of the system (see \citesec{ssec:beta}). If the system is isotropic, $\beta=0$ and the velocity dispersion is
\ben
\left\langle v^2 \right\rangle(r)=3\,\left<v_{r}^2\right> &=& -\frac{3}{\rho}\int_{r}^{R_{\rm max}}{\rm d}r'\,\rho(r')\frac{{\rm d}\Psi}{{\rm d}r}(r') \nn \\
&=& \frac{3\,G}{\rho}\int_{r}^{R_{\rm max}}{\rm d}r'\,\frac{\rho(r')\,m(r')}{r'^2}\,.
\label{eq:velocity_dispersion_jeans}
\een
By construction, the same velocity dispersion can also be obtained from the Eddington PSDF
\ben
\left\langle v^2  \right \rangle(r)= \frac{1}{\rho(r)}\int {\rm d}^{3}\vec{v}\,v^2\,f({\cal E})\,,
\een
which is rigorously equivalent to \citeeq{eq:velocity_dispersion_jeans}.

\subsection{Self-consistency range for DM-baryons configurations}
\label{ssec:sc_range}
For multi-component systems, such as a realistic galaxy comprising DM and baryons, some density-potential pairs may lead to ill-defined Eddington PSDFs. This is fundamentally connected to the simplifying assumptions used to derive the Eddington solution, which in some cases lacks for the additional degrees of freedom necessary to describe a given galactic mass model in phase space. Here again, we refer the reader to Ref.~\cite{LacroixEtAl2018} for a more detailed discussion.

The most basic requirement that a PSDF must satisfy is for it to be positive-definite everywhere, i.e.~$f(\vec{r},\vec{v}) \geqslant 0$ for any phase-space point $(\vec{r},\vec{v})$, and to vanish on the phase-space boundaries. Although this condition is in general satisfied for DM-only systems, the Eddington inversion procedure may not always result in a positive-definite DF for multi-component systems, such as galactic halos featuring a baryonic component. This particularly affects DM halos with cored profiles, whose Eddington PSDFs are well-behaved in the absence of baryons but get spoiled when baryons come to contribute significantly to the central potential. This already provides a qualitative criterion to ensure that the Eddington inversion yields a physically reasonable result (see Ref.~\cite{LacroixEtAl2018} for more quantitative criteria to select parameters of the DM profile).

However, even a positive-definite PSDF can be a dynamically unstable solution to the collisionless Boltzmann equation for some DM-baryon configurations. Nevertheless, a sufficient condition for an ergodic PSDF $f(\mathcal{E})$ to be stable against all perturbations exists\footnote{It should be noted that for anisotropic systems, only criteria against radial perturbations have been derived \cite{DoremusEtAl1973}.} (see Refs.~\cite{BinneyTremaine2008,LacroixEtAl2018} and references therein),
\ben
\dfrac{\mathrm{d}f}{\mathrm{d}\mathcal{E}} > 0 \Leftrightarrow \dfrac{\mathrm{d}^{2}\rho}{\mathrm{d}\Psi^{2}} > 0\,.
\een
Satisfying this condition ensures that the Eddington inversion can be applied to the considered density-potential pairs without leading to unstable solutions of the collisionless Boltzmann equation. In practice, to circumvent the problem of unstable PSDFs, one can still try to vary the mass model parameters (in particular those of the DM halo profile) within their uncertainty range until the global model configuration satisfies both the positivity and stability criteria. This trick can only work if the central values of parameters are not too far from the stability region in the first place. We will actually resort to such tiny adjustments in some of the synthetic halos we consider in the following to avoid dynamically unstable configurations.
\subsection{Quantities of interest and observables related to DM searches}
\label{ssec:observables}
The velocity distribution (assuming spherical symmetry) is simply defined from the Eddington PSDF in the following way:
\ben
f_{\vec{v}}(r,\vec{v}) &\equiv& \frac{f({\cal E})}{\rho (r)}\,,
\een
such that it is normalized to 1 over velocity space, and thus represents a probability distribution function. The speed distribution, \ie~the distribution of the modulus of the velocity, $v \equiv |\vec{v}|$, is further defined by\footnote{In the following, for simplicity, and when the position in the galaxy under study is explicit, $f_{v}(r,v)$ is sometimes abbreviated as $f(v)$.}
\ben
f_{v}(r,v) &\equiv& v^2 \int {\rm d}\Omega_v \, f_{\vec{v}}(r,\vec{v})\,.
\een
For an isotropic system, integration over solid angle in velocity space is trivial, and the speed distribution boils down to the following expression: 
\ben
f_{v}(r,v) = \dfrac{4 \pi v^{2}}{\rho(r)} f \left( {\cal E} = \Psi(r) - \dfrac{v^{2}}{2} \right)\,.
\een
Although the velocity distribution encapsulates all the directional information on a given system, observables in the context of DM searches are in fact sensitive to some specific scalar moments of the distribution. We thus recall the expression of the moment of order $n$ of the velocity distribution:
\ben
\left\langle v^{n}  \right\rangle(r) = \int \! \mathrm{d}^{3}\vec{v} \, v^{n}\, f_{\vec{v}}(r,\vec{v}) \,.
\een
The $n=-1$ moment is relevant for the capture of DM by astrophysical objects (\eg~\cite{PressEtAl1985,Gould1987,BouquetEtAl1989}), while the $n=1$ moment appears in the computation of the rate of microlensing of stars by primordial black holes (\eg~\cite{Griest1991,Green2017}). Indirect searches involving pair annihilation of DM particles \cite{LavalleEtAl2012,BringmannEtAl2012c,Strigari2013} are instead sensitive to moments of the \textit{relative} velocity distribution:
\ben
\left\langle v_{\rm rel}^n \right\rangle(r) = \int\,\mathrm{d}^3\vec{v}_{1}\int\,\mathrm{d}^3\vec{v}_2\,|\vec{v}_1-\vec{v}_2|^n\,f_{\vec{v}}(r,\vec{v}_1)\,f_{\vec{v}}(r,\vec{v}_2)\,.
\een
The $n=2$ moment is relevant for $p$-wave annihilating DM (\eg~\cite{BoddyEtAl2018,BoudaudEtAl2019}), while the $n=-1$ and $n=-2$ moments are relevant for Sommerfeld-enhanced annihilation processes (\eg~\cite{HisanoEtAl2005,BoddyEtAl2017}). Technical details of the computation of moments of the relative speed distribution can be found in Ref.~\cite{LacroixEtAl2018}.

\subsection{Maxwell-Boltzmann approximation: reference models}
\label{ssec:MB}

We start by emphasizing that the goal of the paper is to assess the precision of models able to predict the PSDF of DM in galactic structures, by comparing some predictions with measurements directly made in cosmological simulations. If we restrict ourselves to the velocity DF and factorize out the mass density profile of DM, we can also design velocity DF models based on the Maxwell-Boltzmann (MB) approximation. Here we summarize a couple of ways to design such models, in order to get references to compare with the Eddington method. By no means will we discuss fits of generalized Maxwellian or Tsallis functions to velocity DFs in cosmological simulations, which usually give a decent match to the data when peaks and widths are inferred from the data themselves (even better a match for integrated quantities like the mean speed and associated standard deviation) \cite{VogelsbergerEtAl2009,LingEtAl2010,MaoEtAl2013,PillepichEtAl2014a,BozorgniaEtAl2016,NunezCastineyraEtAl2019}. However, in real life, those peaks and widths can actually not be measured for the DM component (otherwise predictions would be useless), so a fit on a simulated object can hardly be used as a prediction valid for any other real object with different properties. Therefore, we shall instead discuss potential ways of {\em predicting} the velocity DF from the Maxwellian approximation, rather than Gaussian fits which are not the purpose of this work.

In the DM literature, in particular the one related to direct searches (\eg~\cite{DrukierEtAl1986,LewinEtAl1996,JungmanEtAl1996,FreeseEtAl2013}), it is common to approximate the local velocity distribution by a MB DF. The latter is usually further truncated at the escape velocity. Considering a MB distribution is formally equivalent to assuming an isothermal DM density profile \cite{BinneyTremaine2008}, an oversimplification which, unlike the Eddington inversion, cannot account for the global properties of the dark halo in general. Indeed, the MB DF is not a generic solution to the Boltzmann equation for realistic galactic gravitational potentials. However, in this study we still use it as a point of comparison with the Eddington solution. This may also characterize the extent to which both our predictions and the simulation data depart from Maxwellianity.

More specifically, we focus on the following parameterization for the MB velocity DF:
\ben
f_{\vec{v}}^{\rm MB}(r,\vec{v}) = \frac{1}{K(r)}\left(e^{-v^2/v_{0}^2}-e^{-v_{\rm e}(r)^2/v_{0}^2}\right)\,,
\label{eq:maxwell_boltzmann}
\een
where $K(r)$ ensures the normalization to unity after integration over velocity space, $v_{\rm e}(r)$ is the position-dependent escape velocity and $v_{0}$ the peak speed, not specified for the moment and which may also exhibit a radial dependence. The second term on the right-hand side ensures a smooth truncation at $v_{\rm e}$.

We can now define two benchmark MB DF models simply by considering two different cases for the peak speed:
\ben
v_{0} =
\begin{cases}
  v_{\rm circ}(r) = \sqrt{\frac{G \, m(r)}{r}} \; &\text{(isothermal model)}\\
  \sqrt{\frac{2}{3}}\, \sigma_{v}(r) & \text{(Jeans model)}
\end{cases}\,,
\label{eq:MB_models}
\een
where in the isothermal approximation the peak velocity is consistently set to the circular speed, and where
\ben
\sigma_{v}(r) = \sqrt{\left\langle v^2 \right\rangle(r)}
\een
is the velocity dispersion computed by solving the Jeans equation given in \citeeq{eq:velocity_dispersion_jeans}. The Jeans model is supposed to contain more information on the system than the isothermal model, even though the resulting MB DF is still not a solution to the Boltzmann equation. In any case, the MB approximation (both in the isothermal or Jeans cases discussed above) is generally known not to lead to dynamically stable phase-space configurations even when enforced to a dark halo with a realistic mass profile (\eg~\cite{KazantzidisEtAl2004}).

These models define two reference MB models widely used in the literature to make predictions for DM searches. We stress again here that we do not refer to ``MB model'' as a mere Gaussian fit, since we are interested in the ability of a model to {\em predict} velocity DFs {\em from galactic mass models}, and by no means in {\em fitting} velocity DFs in simulations, which has no predicting power. We can already anticipate that the isothermal model is likely to behave badly away from isothermal regimes, in particular in the central regions of galactic structures, where the circular speed is known to provide a very poor estimate of the velocity dispersion. On the other hand, the Jeans model should better capture the phase-space information, at least partially. Indeed, while not a proper solution to the Boltzmann solution, it is here tuned to contain some relevant physical input inferred from the mass model, by including both the escape speed and a theoretically more consistent velocity dispersion.

\section{Main relevant features of the test virtual galaxies}
\label{sec:sims}
In this section, we present the main characteristics of the set of cosmological numerical simulations that we use in this work.
\subsection{General description of the cosmological simulations used in this work}
\label{ssec:descr}

\begin{figure}[t!]
\begin{center}
\includegraphics[width=0.32\textwidth]{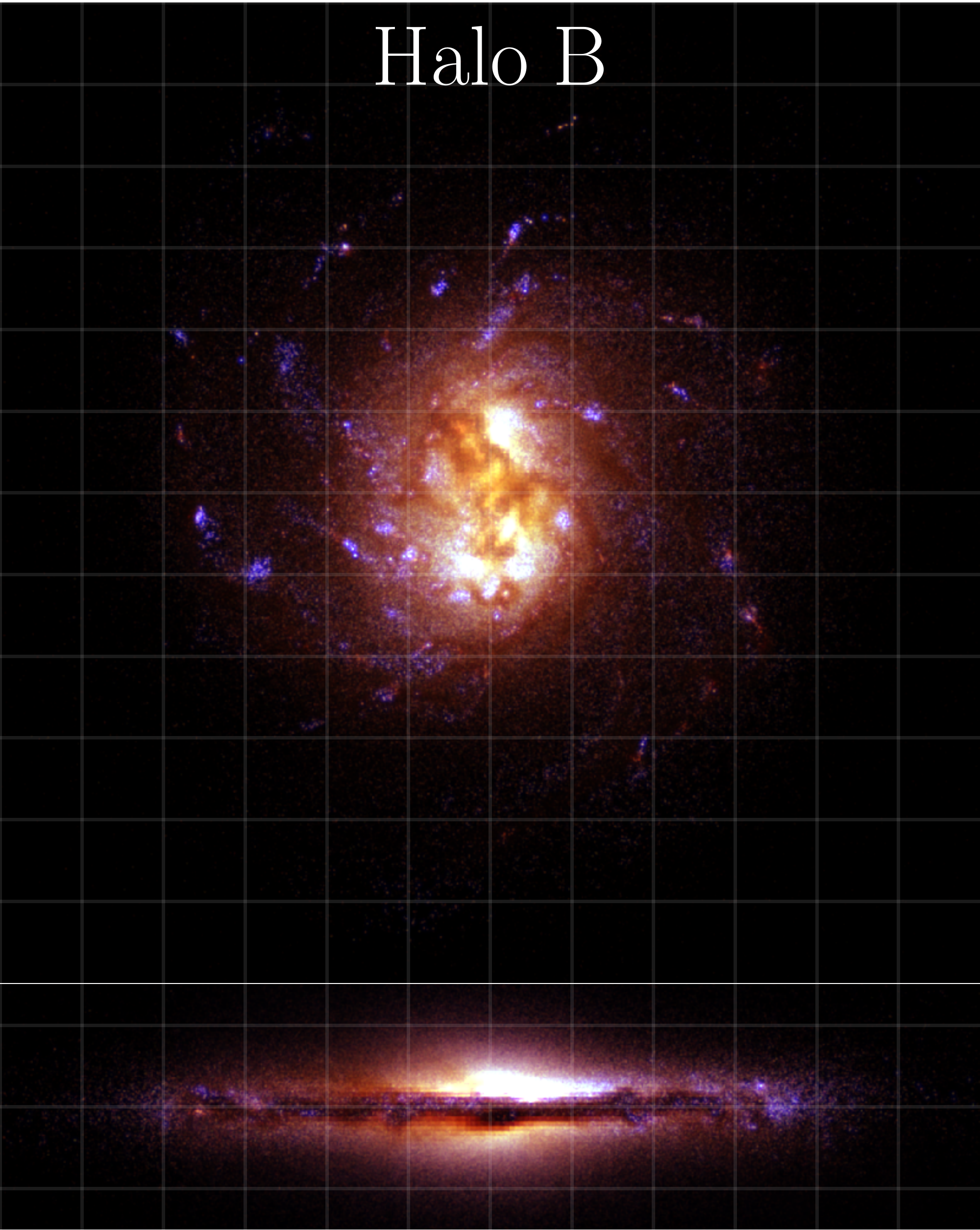}
\includegraphics[width=0.32\textwidth]{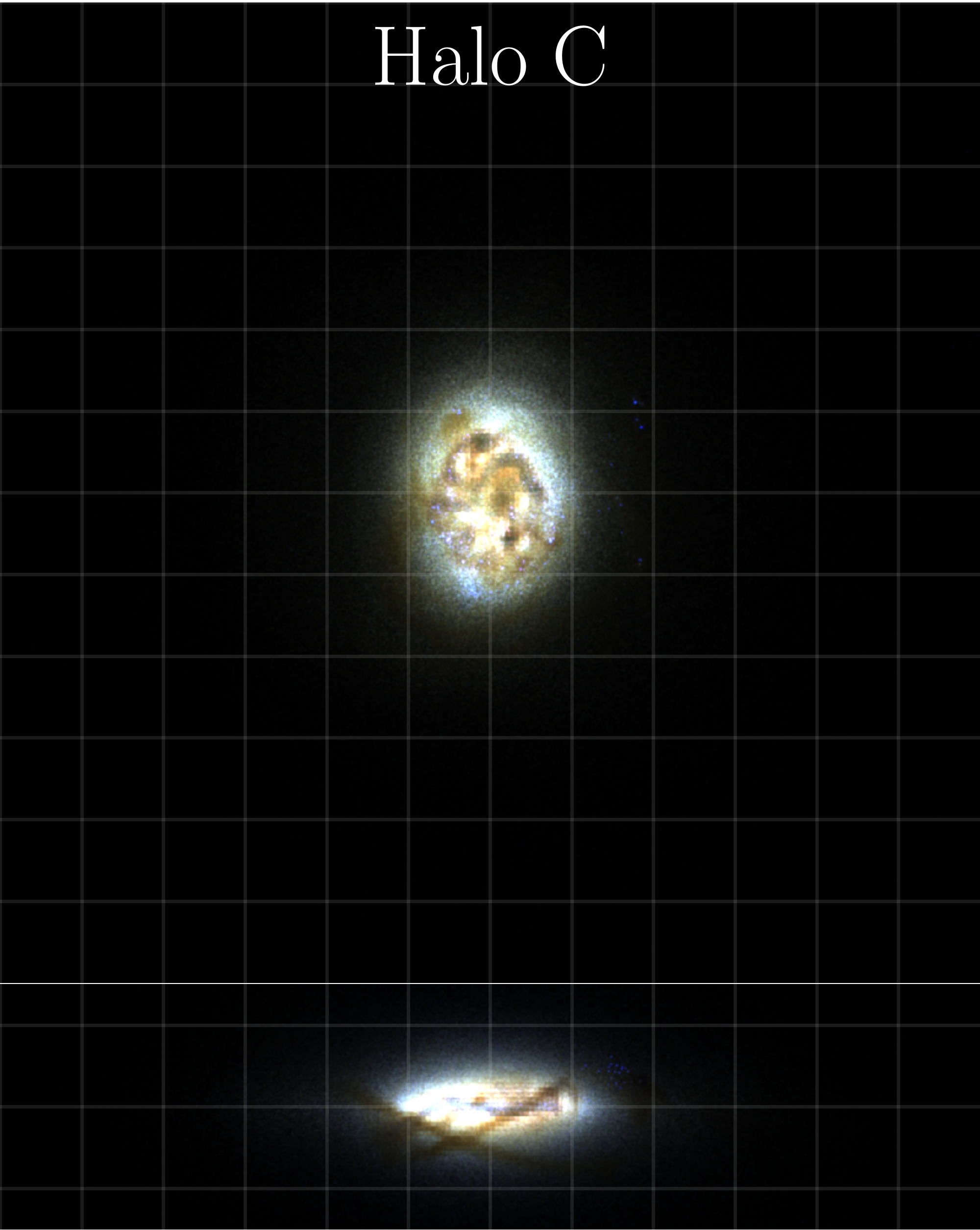}
\includegraphics[width=0.32\textwidth]{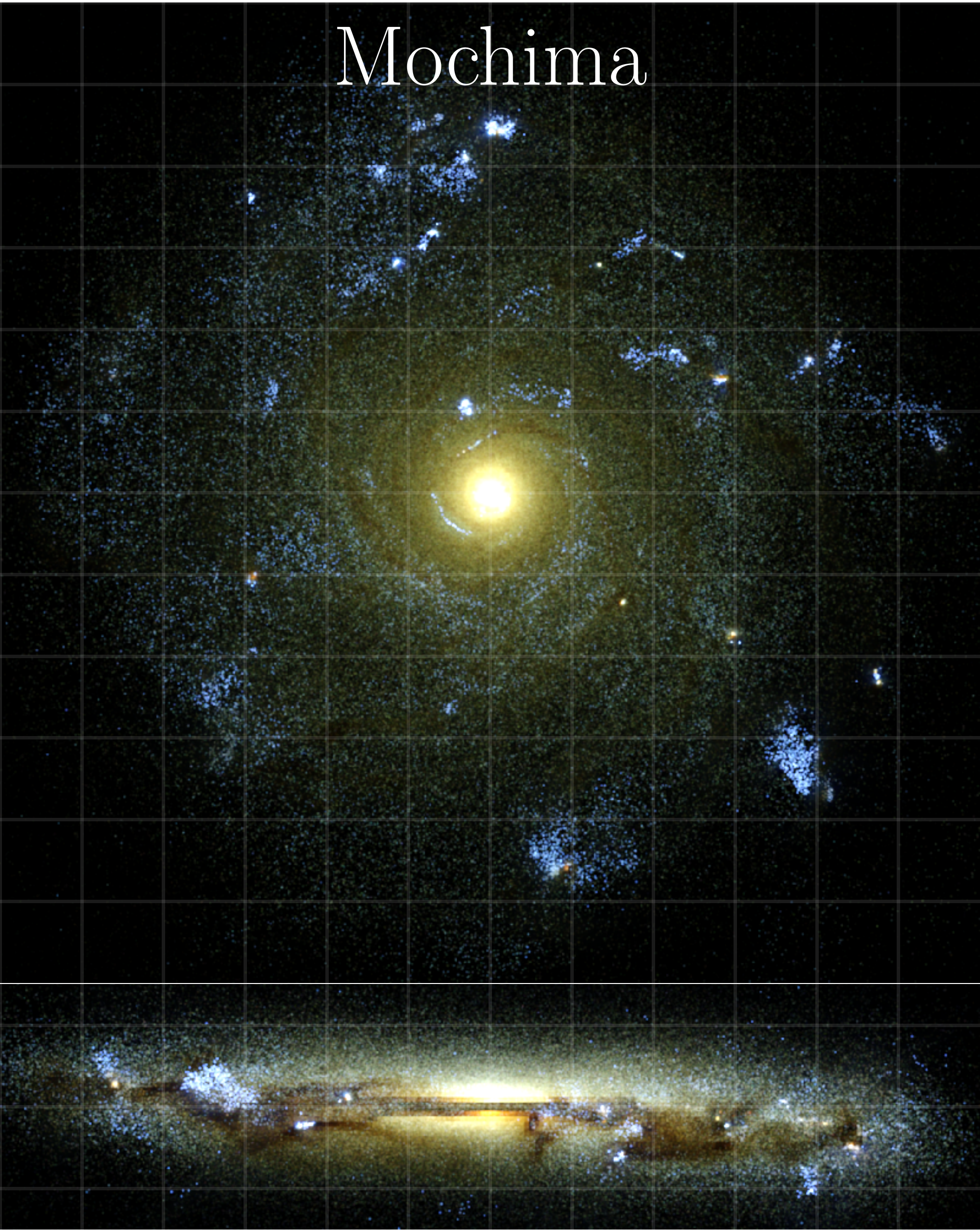}
\caption{\small Present-day true color simulated images in the SDSS (Sloane Digital Sky Survey) bands including dust absorption. Face-on (top) and edge-on (bottom) views are shown for Halo~B (left panel), Halo~C (middle panel), and Mochima (right panel). These images are rendered with the SKIRT code \cite{SKIRT2015A} using the gas and stars in the simulations, and assuming a dust-to-gas ratio of 0.01 \cite{Sandstrom2013}, see Ref.~\cite{Trayford2017} for details. The grid is made of 5~kpc $\times$ 5~kpc cells.}
\label{fig:pictures-galaxies}
\end{center}
\end{figure}

\begin{table}[t!]
\begin{tabular}{|c|l|c|c|c|c|c|c|}
\hline
\multicolumn{2}{|c|}{\multirow{2}{*}{Run}} & $M_{200}$  & $R_{200}$ & $M_{\star}$ & $m_{\mathrm{DM}}$  & $m_{\star}$ & $l_{\rm sm}$ \\
\multicolumn{2}{|c|}{}                     & $(10^{10}\; \mathrm{M}_{\odot})$ & (kpc)     & $(10^{10}\; \mathrm{M}_{\odot})$ & (M$_{\odot}$)      & (M$_{\odot}$)      & (pc)                    \\
\hline
\multirow{2}{*}{Halo B}       & DMO        & 49.60                            & 163.5     &                         & $2.757\times 10^5$ &                    & \multirow{2}{*}{151.67} \\ \cline{2-7}
& Hydro      & 50.06                            & 177.54    & 7.96                    & $2.308\times 10^5$ & $2.87\times 10^4$  &                         \\
\hline
\multirow{2}{*}{Halo C}       & DMO        & 62.48                            & 176.36    &                         & $2.757\times 10^5$ &                    & \multirow{2}{*}{151.67} \\ \cline{2-7}
& Hydro      & 55.02                            & 182.23    & 9.56                    & $2.308\times 10^5$ & $2.87\times 10^4$  &                         \\
\hline
\multirow{2}{*}{Mochima}      & DMO        & 91.31                            & 205.66    &                         & $2.279\times 10^5$ &                    & \multirow{2}{*}{35.13}  \\ \cline{2-7}
& Hydro      & 99.80                            & 199.80    & 3.128                   & $1.947\times 10^5$ & $1.567\times 10^4$ &                         \\
\hline
\end{tabular}
\caption{\label{tab:halo-global-features} Main features of the simulations at $z=0$. The columns correspond to the virial radius $R_{200}$, the associated mass $M_{200}$, the total stellar mass $M_{\star}$, the DM and stellar particle masses $m_{\rm DM}$ and $m_{\star}$ respectively, and the size $l_{\rm sm}$ of the smallest cell in the simulation (usually dubbed {\sf hsml} in numerical routines).}
\end{table}

In this paper we use three spiral-type galaxy simulations coming from two different studies, but performed with the same prescriptions and code. These high-resolution simulations were run with the Eulerian code RAMSES \cite{Teyssier2002} from cosmological initial conditions generated with the MUSIC package \cite{HahnEtAl2011}. Galaxies are evolved down to $z = 0$ focusing the computing power on a carefully decontaminated Lagrangian region of the galactic neighborhood as in typical zoom-in simulations \cite{OnorbeEtAl2014}. The first two simulated galaxies, published in Ref.~\cite{MollitorEtAl2015}, are referred to as Halo~B and Halo~C, and the third galaxy, which corresponds to the control run presented in Ref.~\cite{ArturoSimu}, is dubbed Mochima.

For the baryonic physics, both sets of simulations use conventional recipes for gas cooling, ultraviolet background and self-shielding. Star formation (SF) follows a standard Schmidt law with adapted density threshold and efficiency. The supernova (SN) feedback prescription, known as delayed cooling, relies on a non-thermal energy injection 10~Myr after the birth of the star particles, generated with a Chabrier initial mass function (IMF), and with an energy of $10^{51}$~erg per SN event. A detailed description of this method as implemented in the RAMSES code is presented in Ref.~\cite{TeyssierEtAl2013}.

The first two galaxies, Halo~B and Halo~C, are located at the centers of 20~Mpc cosmological boxes and reach a resolution of $l_{\rm sm}\sim 150$~pc. A detailed analysis of these simulations can be found in Ref.~\cite{MollitorEtAl2015}. One of the main features of the DM profiles of the associated galactic halos is a compression around the scale radius of the disk and a flattening in the more central regions due to the violent SN feedback scheme that was adopted.

The third galaxy, Mochima, is the object of a dedicated study where different subgrid physics prescriptions are considered for the SF and SN feedback. A detailed description of the resulting simulations can be found in Ref.~\cite{ArturoSimu}. This galaxy lies at the center of a 36~Mpc box with a spatial resolution of $l_{\rm sm}\sim 35$~pc. Here, we only consider the control run, labeled KSlaw-DCool in the reference paper, since it uses the same baryonic prescriptions than Halo~B and Halo~C, with the related free parameters adjusted according to the new resolution following Ref.~\cite{DuboisEtAl2015}---see Ref.~\cite{ArturoSimu} for details. Notably, the DM profile of this galaxy has a steeper inner slope than in Halo~B and Halo~C, presumably due to a weaker delayed cooling SN feedback that results from the different choices in the free parameters of the model.

Regardless of their ``MW-like'' features, we use the synthetic halos produced by these simulations as dynamically consistent frameworks governed by gravity to study the PSDF of the DM and to test the Eddington inversion in a numerical galaxy environment.

For all three objects we consider, we study both DM only (DMO) runs, and runs accounting for both DM and baryonic matter (hydrodynamical runs, or hydro runs for short). Note that the DMO runs might, to some extent, reflect the dynamics relevant to dwarf spheroidal galaxies, which, although much lighter than our galaxies, are DM-dominated objects (keeping in mind that baryons can still play a role in the formation and evolution of DM-dominated dwarf galaxies, \eg~\cite{WadepuhlEtAl2011,ReadEtAl2016,WetzelEtAl2016,ZhuEtAl2016,MaccioEtAl2017,FringsEtAl2017}). Face-on and edge-on pictures of the resulting numerical galaxies are shown in \citefig{fig:pictures-galaxies}. The virial mass $M_{200}$ and radius $R_{200}$, as well as the total stellar mass $M_{\star}$ and the mass and spatial resolutions for the different runs are summarized in \citetab{tab:halo-global-features}.\footnote{$M_{200} \equiv m(\leqslant R_{200})$ and $R_{200}$ is defined by $\frac{M_{200}}{4/3\pi R^3_{200}}=200 \rho_{\rm c}$, where $\rho_{\rm c}$ is the critical cosmological density---the reduced Hubble parameter is fixed to $h=0.7$ for Halo~B and Halo~C, and to $h=0.67$ for Mochima. We define $M_{\star}$ as the total stellar mass inside $0.1\, R_{200}$.}

\subsection{Level of equilibrium}
\label{ssec:eq}

The Eddington inversion method assumes dynamical equilibrium. However, it is well known that simulated galaxies in a cosmological context, like real galaxies, do exhibit some local departures from equilibrium (recent minor mergers, ongoing disruption of subhalos, \etc). It is therefore interesting to quantify the level of equilibrium in the simulated halos, to check whether it correlates with the precision of the Eddington inversion for the PSDF prediction.

\begin{figure}[t!]
\begin{center}
\includegraphics[width=.45\textwidth]{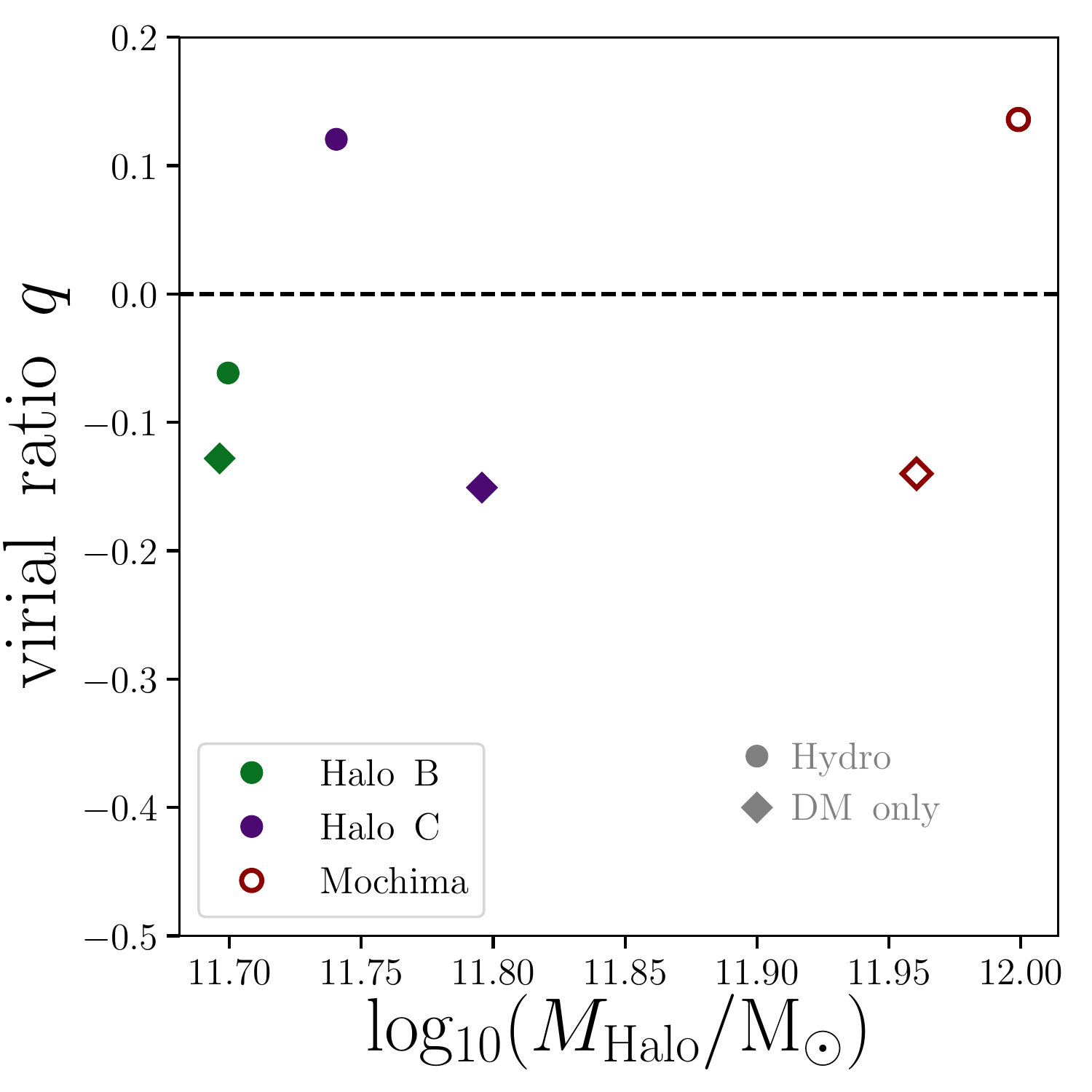}
\includegraphics[width=.45\textwidth]{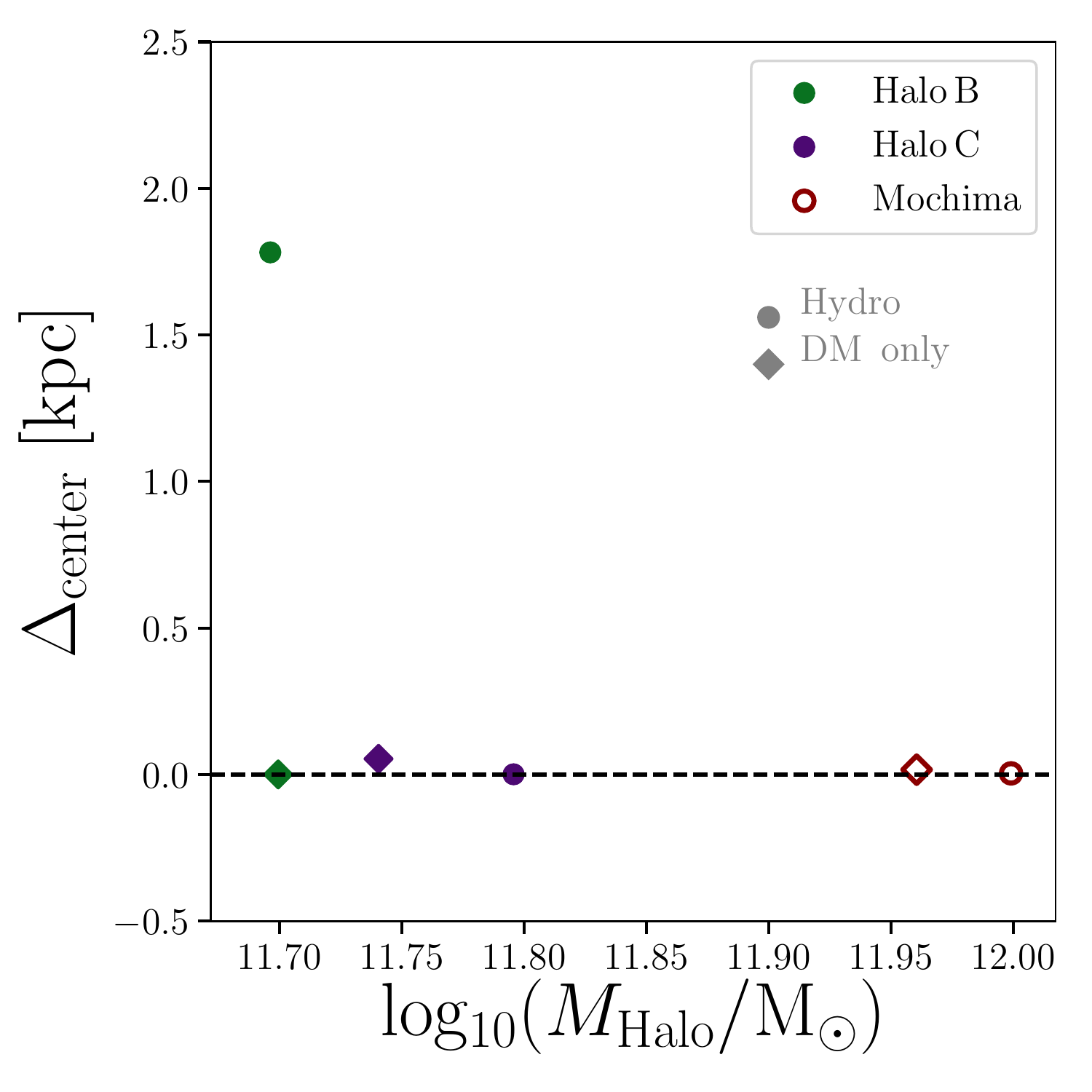}
\caption{\small Virialization parameter $q$ (left panel) and center offset $\Delta_{\mathrm{center}}$ (right panel), as a function of halo mass, for DMO runs (diamonds) and hydro runs (circles). In both panels, green filled, dark blue filled and red open symbols correspond to the Halo~B, Halo~C, and Mochima simulations, respectively.}
\label{fig:virialization}
\end{center}
\end{figure}

In cosmological simulations, there is no definitive test to assess the level of virialization of a halo. We follow the work presented in Ref.~\cite{ZjupaEtAl2017} (see also Ref.~\cite{CallinghamEtAl2020}), where a virial parameter is defined from the virial ratio as:
\ben
\label{eq:q_param}
q = \frac{2\,E_{\rm kin}}{E_{\rm pot}} + 1\,,
\een
where the total kinetic and potential energies follow the definitions of Ref.~\cite{ShapiroEtAl2004}. More specifically, the kinetic energy for the DM or star particles is constructed from their individual velocities, whereas for gas cells it can be inferred from thermal energy, such that the total kinetic energy is given by
\ben
E_{\rm kin} = \sum_{i,{\rm particles}} \frac{1}{2}\,m_i\, v_i^2 +\sum_{j,{\rm cells}} \frac{3}{2} k_{\rm B} T_j\,,
\een
where $k_{\rm B}$ is the Boltzmann constant. The values $m_i$ and $v_i$ correspond to the mass and speed of the $i$-th particle and $T_j$ is the temperature of the gas inside the $j$-th cell. For the negative-definite potential energy, spherical symmetry is assumed, which leads to the expression
\ben
E_{\rm pot} = -\sum_i \frac{G m(<r_i)m_i}{r_i}\,,
\een 
where $m(<r_i)$ is the mass contained inside the radius $r_i$ of the $i$-th particle of mass $m_i$.

As a consequence of the virial theorem, a fully relaxed system should have its virial ratio parameter $q$, defined in \citeeq{eq:q_param}, close to 0. A negative (positive) value of $q$ corresponds to a kinetic-energy (potential-energy, respectively) dominated halo. The $q$ parameters for the three halos, DMO and hydro runs, are shown in the left panel of \citefig{fig:virialization}. We observe that once baryons are included in the simulation, $q$ increases to values closer to 0. For Halo~C and Mochima it even becomes slightly positive, indicating that the system has turned to potential-energy domination. However, this is not the case for Halo~B, for which even though the halo is less dominated by the kinetic energy than in the DMO run, $q$ still lies below -0.1. Overall, the values of the virial ratio $q$ we find seem to indicate a departure from equilibrium that does not exceed $\sim 10$-20\% in terms of energy redistribution. This is fully consistent with the results of Ref.~\cite{ZjupaEtAl2017}. We can therefore consider our galactic objects to be reasonably close to equilibrium, which further legitimizes the use of the Eddington model.

Another potential indicator of virialization is the offset $\Delta_{\mathrm{center}}$, if any, between positions that could be considered the center of a halo, \ie~(i) the one of highest mass density (in stars and/or DM), (ii) the one of the minimum of the gravitational potential well, or (iii) the center of mass found by the recursive procedure presented in Ref.~\cite{Schaller2015} (namely a ``shrinking sphere'' down to a sample of 200 particles, whose final center is defined as the halo center). Note that in this work, the halo center is set from the latter method. As an example, we define the offset as 
\ben
\Delta_{\rm center}=|\vec{c}_{\rm com}-\vec{c}_{\rm pot}|,
\een
where $\vec{c}_{\rm com}$ is the position vector of the center found by the recursive center-of-mass method, and $\vec{c}_{\rm pot}$ the position vector of the minimum of the gravitational potential in the simulated galaxy. For virialized structures, $\Delta_{\rm center}\to 0$. Values of the offsets $\Delta_{\rm center}$ obtained in this work are shown in the right panel of \citefig{fig:virialization} as a function of the halo mass. Only Halo~B exhibits a significant offset between the two centers, of almost 2~kpc.\footnote{The offset for Halo B is however much smaller than the upper limit of $0.07 R_{200}$ defined in Ref.~\cite{NetoEtAl2007} as one of the criteria that a halo must satisfy in order to be relaxed.} This can be attributed to a recent merger big enough to have displaced the center of mass from the potential center. Finding that the halo with the biggest center offset is also the one featuring the most negative virial parameter $q$, indicating kinetic-energy domination, makes perfect sense. The other synthetic halos feature offsets that fall within the resolution limit of the simulation.

Note that in addition to potentially measuring departures from equilibrium, the offset parameter $\Delta_{\rm center}$ also provides a useful diagnosis tool as to whether imposing spherical symmetry when fitting components in simulations may lead to additional systematic errors. Indeed, if for instance an offset exists between the baryonic and DM distributions, an actual cuspy profile may be fitted as a cored profile depending on where the center of the object is defined. This is actually what happens for Halo~B, for which the DM profile is best-fitted with a core when our nominal definition of the galactic center is used, while it would have been found more cuspy by using another definition.  The geometry of our synthetic halos is discussed in the next paragraph.

\subsection{Sphericity and triaxiality}
\label{ssec:shape}

The Eddington inversion assumes spherical symmetry. Therefore, it is an important check to determine how close our synthetic halos are to spherical, since departure from spherical symmetry could help understand systematic biases between predictions deriving from the Eddington inversion and actual PSDFs. A way to characterize the shape of a halo is to build the sphericity $S$ and triaxiality $T$ parameters from the mass distribution tensor ${\cal M}$ as follows \cite{ColeEtAl1996}:
\ben
\label{eq:mass_tensor}
{\cal M}_{ij} = \sum_k m_k r_{k,i} r_{k,j}\,.
\een
Then, denoting the square root of the eigenvalues of ${\cal M}$ as $a$, $b$, and $c$ (where $a>b>c$), the sphericity, the elongation and the triaxiality of the halo can be defined as $S=c/a$, $e = b/a$ and $T = (a^2 - b^2)/ (a^2 - c^2)$, respectively. The evolution of $S$ and $T$ with radius is shown in \citefigs{fig:sphericity} and \ref{fig:triaxiality}, for DMO runs (left panels) and hydrodynamical runs (right panels). A purely spherical halo has $S=e=1$, with $T$ undefined. Otherwise, low values of $T$ ($T \ll 1$) correspond to oblate halos and $T\to1$ to prolate halos, as defined in Ref.~\cite{BryanEtAl2013}.
\begin{figure}[t!]
\begin{center}
\includegraphics[width=.45\textwidth]{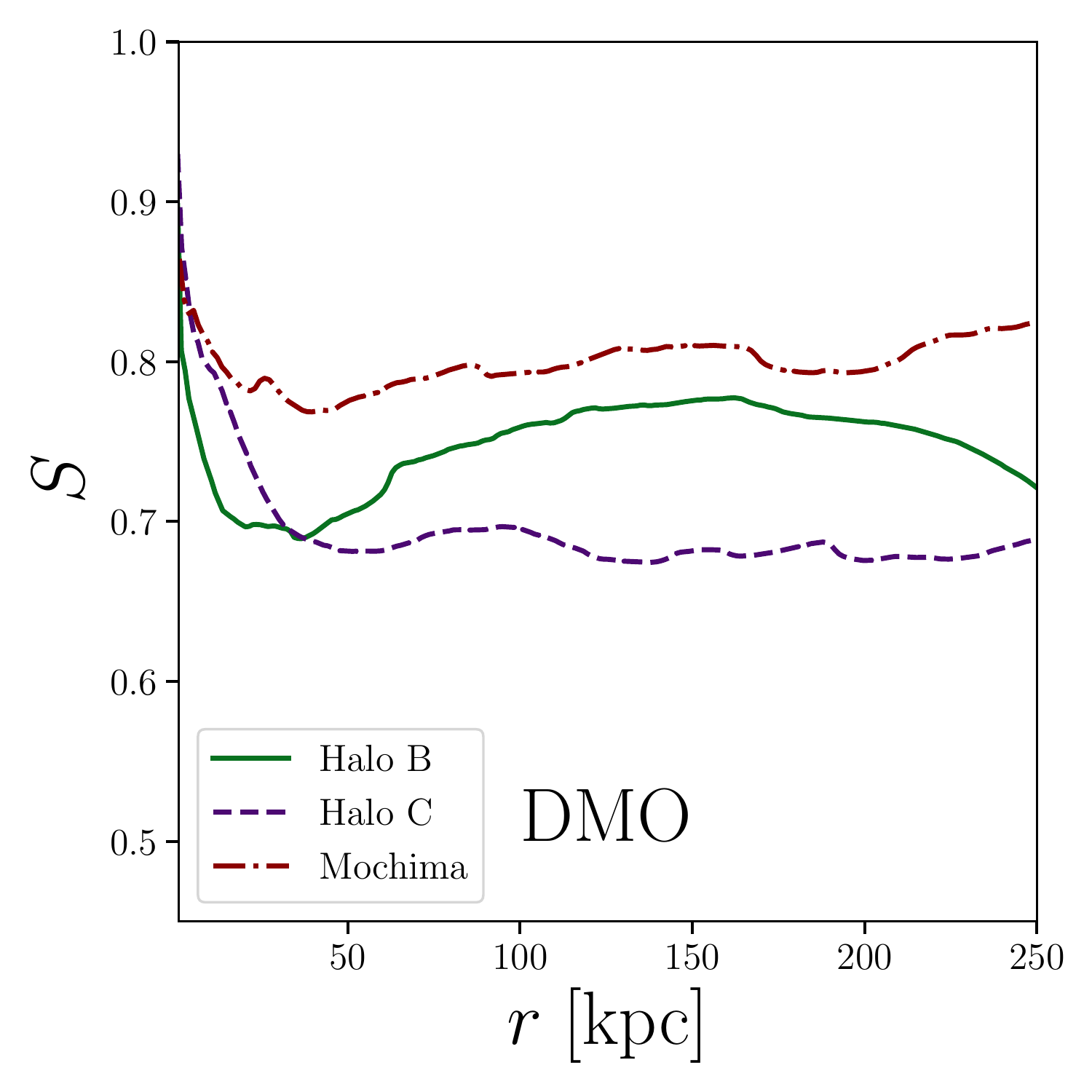} \hfill 
\includegraphics[width=.45\textwidth]{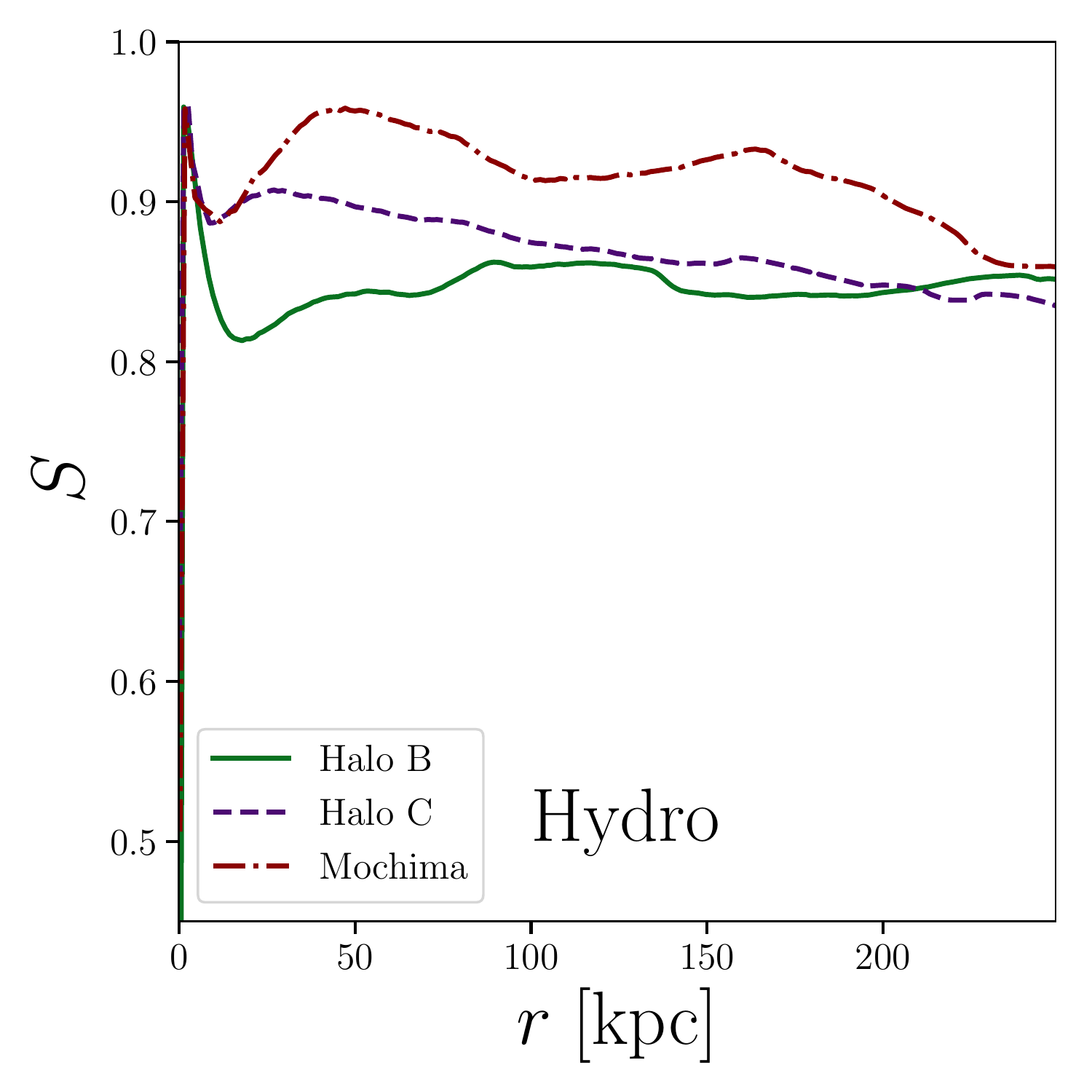}
\caption{\small Sphericity parameter $S$ for runs with DMO (left panel) and hydro (right panel), for the Halo~B (green solid), Halo~C (blue dashed), and Mochima (red dot-dashed) simulations.}
\label{fig:sphericity}
\end{center}
\end{figure}

\begin{figure}[h!]
\begin{center}
\includegraphics[width=.45\textwidth]{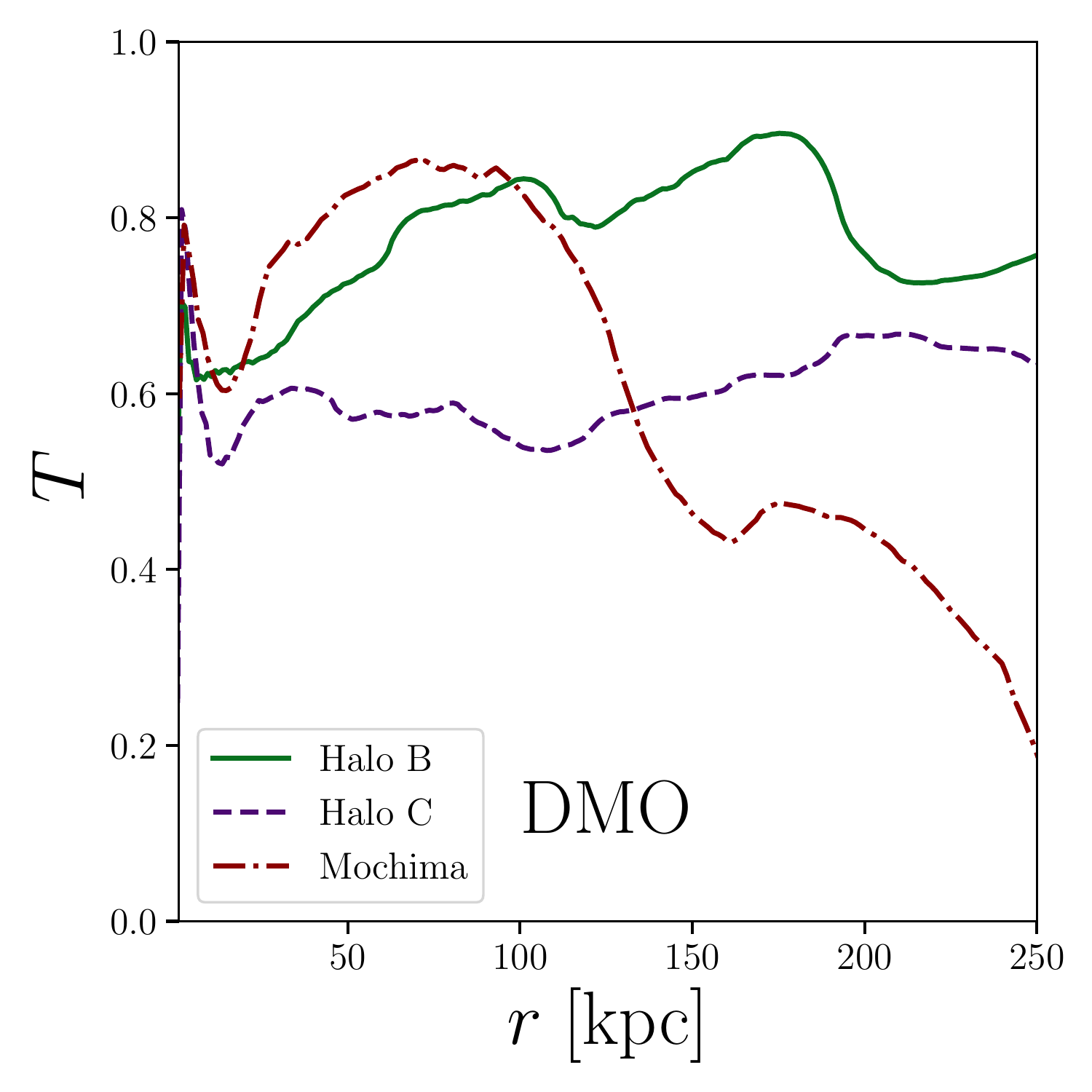} \hfill 
\includegraphics[width=.45\textwidth]{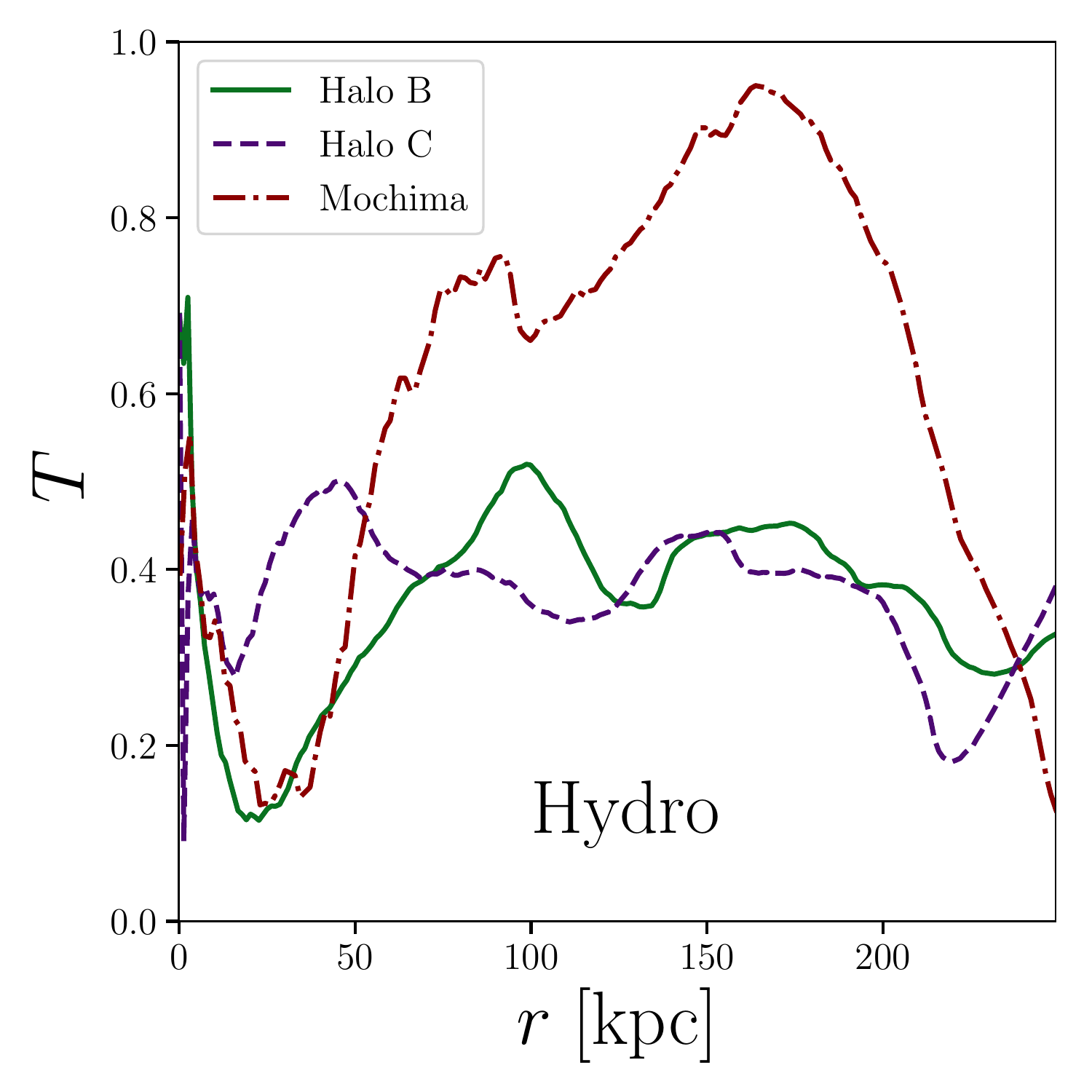}
\caption{\small Triaxiality parameter $T$ for runs with DMO (left panel) and hydro (right panel), for the Halo~B (green solid), Halo~C (blue dashed), and Mochima (red dot-dashed) simulations.}
\label{fig:triaxiality}
\end{center}
\end{figure}

We obtain values of the sphericity closer to 1 for the hydro runs, with values close to 0.9 for all halos in our sample. This indicates that halos are made more spherical by baryonic cooling processes, in agreement with other studies \cite{ChuaEtAl2019,KatzEtAl1991,KatzEtAl1993,Dubinski1994,AbadiEtAl2010,BryanEtAl2013}. This can further be confirmed by inspecting the morphology of the gravitational potential. In \citefig{fig:potential-maps-haloB}, we show maps of the potential projected in three different planes, for DMO (top panels) and hydrodynamical (bottom panels) runs, for our Halo~B. The $x$-$y$ plane is defined as the one comprising the baryonic disk. A quick comparison of the top with the bottom panels shows that the potential is indeed more spherical in the presence of baryons. Regardless, even for DMO runs, the synthetic halos of interest in this work can be considered as reasonably close to spherical (with $S \approx 0.7$-0.8), which further justifies the spherical symmetry assumption the Eddington inversion method relies upon.

\begin{figure}[t!]
\begin{center}
\includegraphics[width=\textwidth]{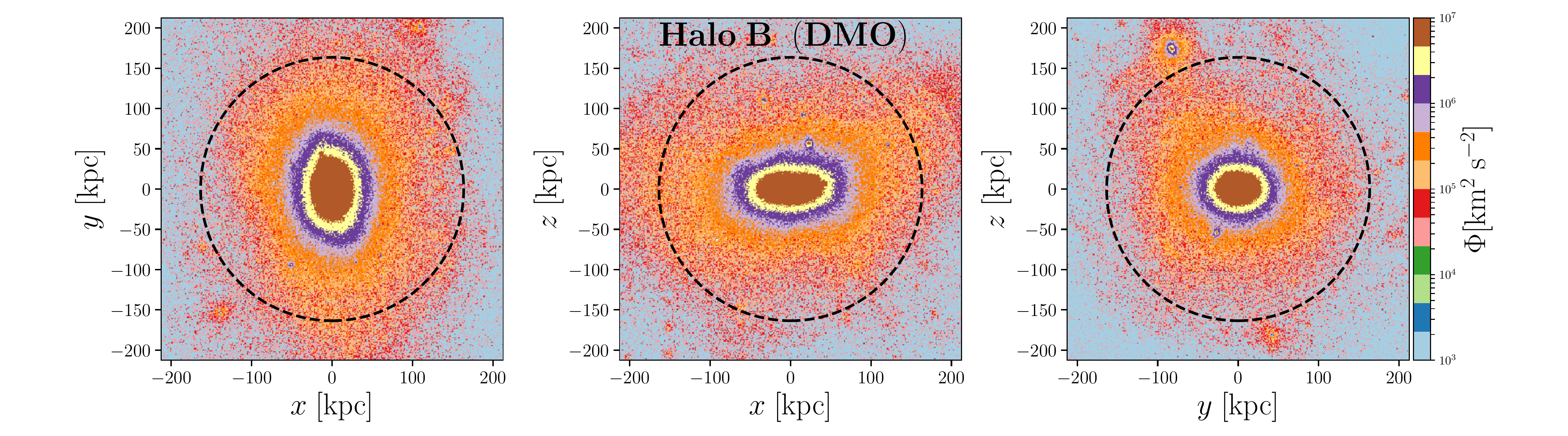}
\includegraphics[width=\textwidth]{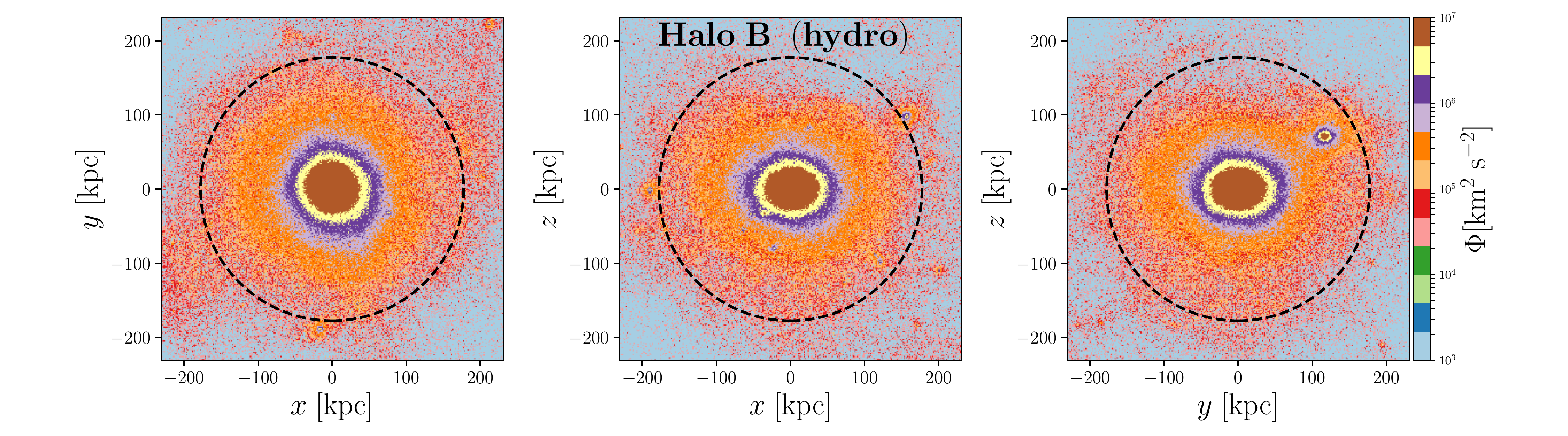}
\caption{\small Projected maps of the gravitational potential for Halo B. From left to right: $(x, y)$, $(x, z)$, and $(y, z)$ planes. Top panels: DMO run. Bottom panels: hydrodynamical run.}
\label{fig:potential-maps-haloB}
\end{center}
\end{figure}

\subsection{Level of anisotropy}
\label{ssec:beta}
A further interesting check prior to using a model that assumes isotropy is precisely to quantify the level of anisotropy in our simulated galaxies. A system with an anisotropic velocity tensor can be characterized in terms of an anisotropy parameter \cite{Binney1980,BinneyTremaine2008}:
\ben
\label{eq:beta}
\beta(r) = 1 - \frac{\sigma_{\theta}^2 + \sigma_{\phi}^2}{2\sigma_{r}^2},
\een
where $\sigma_{r}$, $\sigma_{\theta}$, and $\sigma_{\phi}$ are the velocity dispersion components in spherical coordinates. In each simulation we estimate the velocity dispersions in logarithmic bins in $r$ from the resolution limit to twice the virial radius $R_{200}$ of the halo. The resulting $\beta$ parameter as a function of $r$ is shown in \citefig{fig:anisotropy-beta-param}.

\begin{figure}[t!]
  \centering
  \includegraphics[width=0.49\textwidth]{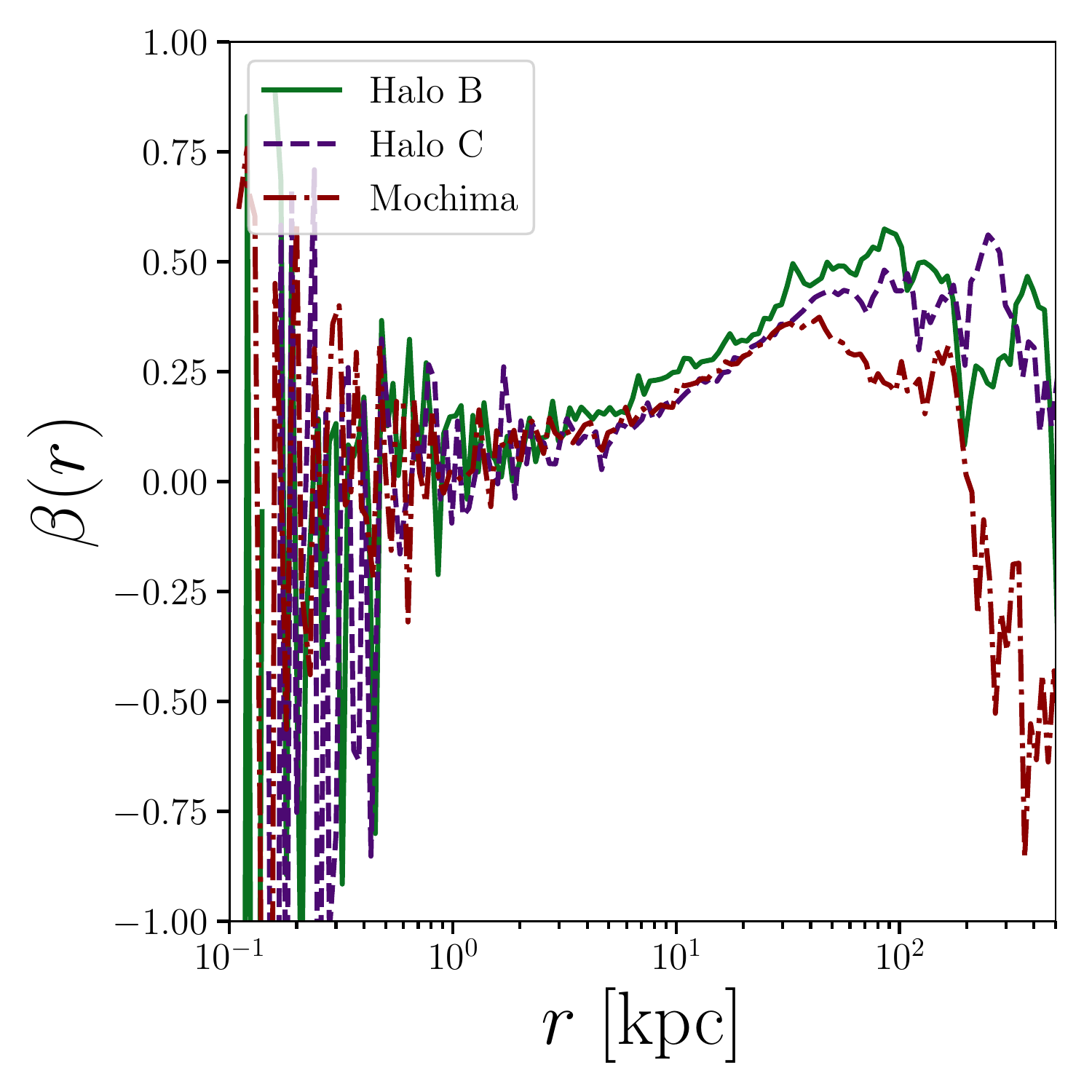}
  \includegraphics[width=0.49\textwidth]{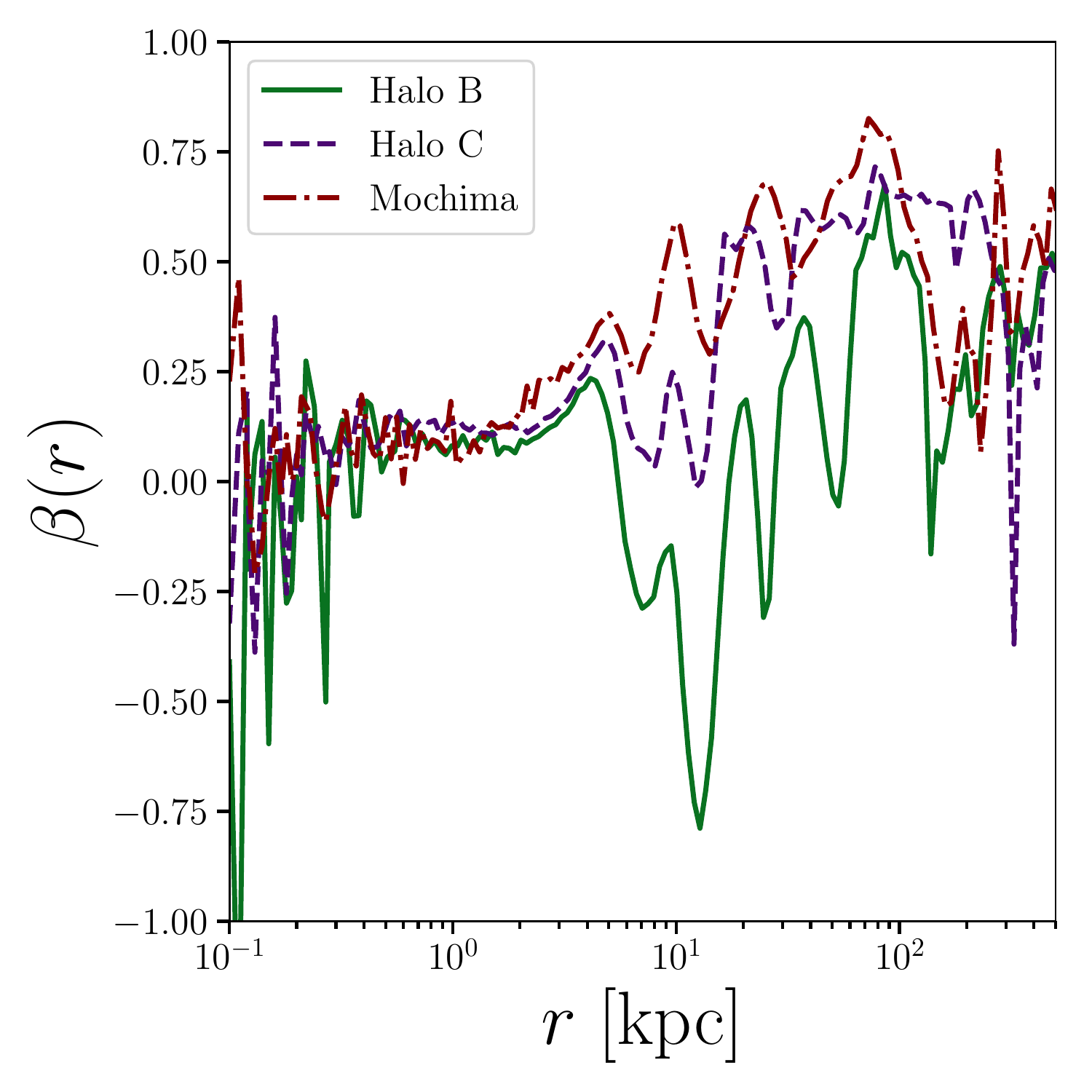}
  \caption{Anisotropy parameter $\beta(r)$ as a function of galactocentric radius, for DMO runs (left panel) and hydro runs (right panel) of the Halo~B (green solid), Halo~C (blue dashed), and Mochima (red dot-dashed) simulations.}
  \label{fig:anisotropy-beta-param}
\end{figure}  

For all halos and runs, the anisotropy is, while noisy, on average close to 0 below $\sim 10$~kpc and slowly increases to reach positive values of the order of 0.5 around $\sim 100$~kpc. This means that on average over our sample of simulated halos, the velocity tensor tends to be more radially anisotropic in the outer regions, consistently with Refs.~\cite{HansenEtAl2006,WojtakEtAl2008,WojtakEtAl2009,LudlowEtAl2011,SparreEtAl2012,ChuaEtAl2019} for instance. However, for some individual objects, there can be very large deviations from positive anisotropy in the outskirts; moreover, there is significant scatter from one halo to another, especially in the hydro case. 

In the context of DM searches, at first order we are mostly interested in regions below or slightly beyond the equivalent of the solar circle for MW-like galaxies. In these regions, as a matter of fact, assuming isotropy seems to be reasonable to try to derive a representative PSDF. We further recall that in a real system, estimating the actual anisotropy of the DM component would by itself be extremely difficult. It might still be possible to infer the DM anisotropy from the overall mass model (\eg~\cite{HunterEtAl1993,Petac2019axisymmetric}), but we leave a dedicated comparison of more complex anisotropic equilibrium models with simulation data for future work.

\subsection{Fitting the virtual galaxy components: deriving mass models}
\label{ssec:fits}
\begin{figure}[t!]
\centering
\begin{subfigure}[t]{0.5\textwidth}
  \centering
  \includegraphics[width=1.\textwidth]{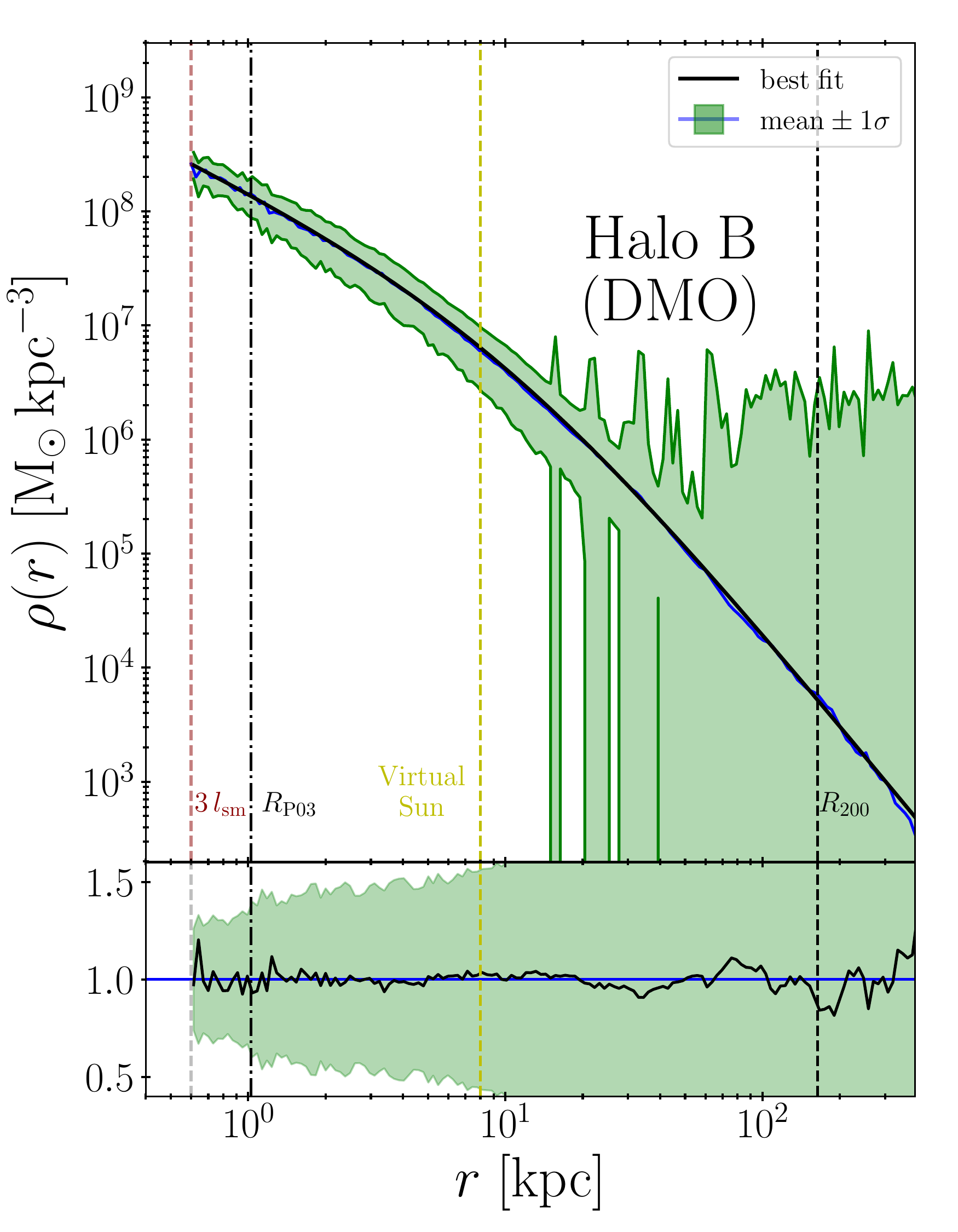}
\end{subfigure}%
\begin{subfigure}[t]{0.5\textwidth}
  \centering
  \includegraphics[width=1.\textwidth]{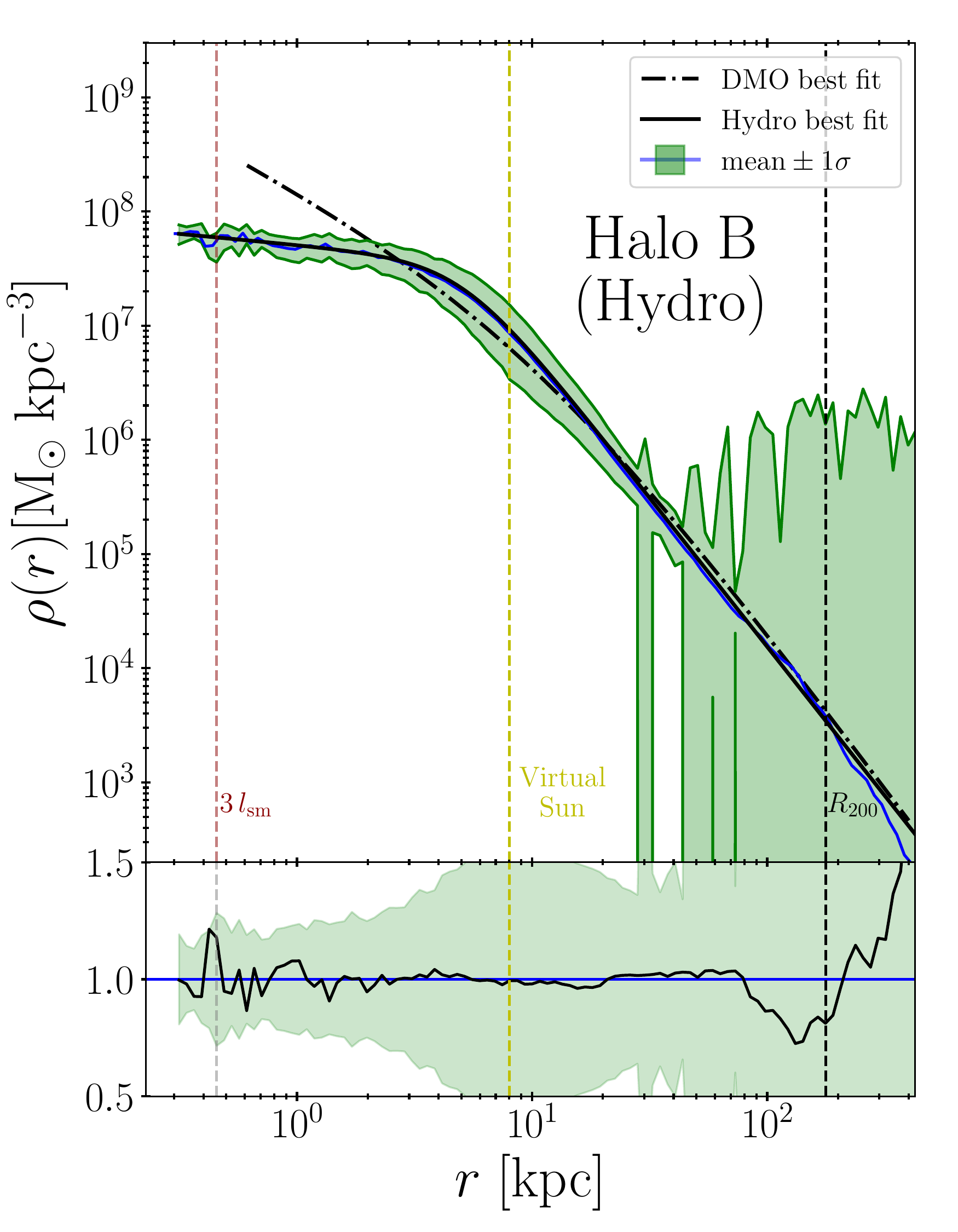}
\end{subfigure}
\caption{\small DM density profiles (blue solid lines) for Halo~B in the DMO run (left panel) and in the hydro run (right panel), with associated 1-$\sigma$ statistical error (green shaded areas). See text for details. The best-fit profiles are obtained using the $\chi^2$ based on the density, \citeeq{eq:chi_rho}. As an illustration we show the resolution limit of the hydro simulation ($3\times l_{\rm sm}$) and the DMO conservative reliable region ($R_{\mathrm{P}03}$) from Ref.~\cite{PowerEtAl2003}. We also report the distance of the Sun to the Galactic center in the real MW as a reference, reminding the reader about the risks of making too detailed comparisons with the MW.}
\label{fig:profilesHALOB}
\end{figure}

In order to use the Eddington inversion in the same way it can be used in the MW, we need to resort to the density profiles of all components \cite{LacroixEtAl2018}. The DM mass density profile is needed for the inversion itself, while the baryonic one is necessary to evaluate the full gravitational potential, see \citeeqtoeq{eq:rho}{eq:Eddington_formula}. Therefore, the starting point of our simulation data analysis is to study the DM density and total gravitational potential for a given simulated object. To place ourselves in a situation similar to what would be derived for the MW
\footnote{\change{We have to be clear about what we mean here. There are basically two independent
    steps in deriving the velocity DFs from observational data in the context of the Eddington
    inversion (this would change in \eg~the action-angle procedure): (i) infer the mass
    model itself from kinematic data (\ie~star or gas surveys); and (ii) translate this information
    in terms of PSDFs. In both steps, systematic uncertainties (on top of statistical ones) can bias
    the results. Here, we assume that step (i) is as perfect as possible, since we infer the mass
    distributions of baryons and DM directly from the full simulation data---in contrast, analyses
    of \eg~the \textit{Gaia} data would lead to much larger (statistical and systematic) uncertainties on the
    mass model because of the incompleteness of the data (and surveys do not collect any direct
    information on DM). In this paper, we only focus on the sources of biases or systematic errors
    affecting step (ii). Implicitly, we assume (or hope) that current or future stellar surveys will
    help getting us closer and closer to the actual mass distributions of all Galactic components,
    so that we only pay attention to the remaining theoretical errors. Improving on step (i) is a
    full and active research field that we do not address at all in this paper.}}
(\eg,~using kinematic data), we do not use the raw density profiles directly extracted from the simulation but we perform fits based on generic \change{parametric functions (see the profiles used for baryons and DM in \citeapp{app:mass_models})}. Using the raw simulation data instead would not lead to significant changes except in the very centers of simulated galaxies below the resolution limit, where the density fluctuations can be large and nonphysical, and in the outskirts where big subhalos can be encountered. In any case, these fluctuations should only moderately affect the smooth DM distribution in the outer halo, and the smearing of the density profiles induced by our fitting procedure is not likely to induce extra non-physical dynamical features.

For DM (star, respectively) particles, we use two different methods to estimate the density. The first one computes the average density $\rho^{\rm shell}$ by accumulating the mass inside a spherical (cylindrical) shell and by dividing it by the volume of the shell (blue solid curves in \citefig{fig:profilesHALOB}), for an array of shells. The second one relies on averaging the local density values obtained from the definition of Ref.~\cite{Dehnen2002} for particles in those spherical (cylindrical) shells. We checked that both methods converge to the same results---the agreement naturally degrades below the resolution limit. The 1-$\sigma$ statistical uncertainty is then determined from the square root of the canonical variance estimate computed from the second method (green shaded areas in \citefig{fig:profilesHALOB}). For the gas cells, we use the density information that we average in cylindrical shells obtaining a mean and the corresponding 1-$\sigma$ statistical error.

The best-fit parameters of the density profile of component $j$ (DM, stars or gas) are finally determined by minimizing the following $\chi^2$:
\ben
\label{eq:chi_rho}
\chi^2_{\rho_{j}} = \sum_{i} \left\{ \frac{\log_{10}(\rho_{j}(r_i)) - \log_{10}(\rho_{j}^{{\rm shell}\, i})}{\sigma_{\log,i}}\right\}^2\,,
\een
where $\rho_{j}^{{\rm shell}\, i}$ is the mean density of shell $i$ at radius $r_i$. Shells are defined as a logarithmic array from 100~pc to 30~kpc with 100 bins. Parameter $\sigma_{\log,i}$ is the standard deviation of $\log_{10}(\rho_{j}^{{\rm shell}\, i})$, also evaluated from the data as described above.

For completeness we also use a second approach which consists in defining a $\chi^2$ based on the mass per shell:
\ben
\label{eq:chi_m}
\chi^2_{j,m_{\rm shell}} = \sum \left\{\log_{10}\left(m_{j}(r_{i})\right) - \log_{10}\left(m^{{\rm shell}\,i}_{j}\right)  \right\}^2\,,
\een
where $m^{{\rm shell}\,i}_{j}$ is the mass of shell $i$, and $m_{j}(r)$ is the integrated mass enclosed in a sphere of radius $r$ obtained from the density profile---with parameters to be adjusted. It should be noted that the mass is known exactly in the simulation, hence the absence of standard deviation in the formula of \citeeq{eq:chi_m}.

We find that both minimization procedures of \citeeqs{eq:chi_rho} and \eqref{eq:chi_m} give the same results. For definiteness, in subsequent calculations, we use the method relying on the density, defined in \citeeq{eq:chi_rho}. The corresponding best-fit DM profiles for Halo~B are illustrated in the left and right panels of \citefig{fig:profilesHALOB}, for the DMO and hydro runs, respectively. We define the resolution limit for the hydro run and for the DMO one as $3\times l_{\rm sm}$ and
$R_{\rm P03}$ \cite{PowerEtAl2003}, respectively, where the latter is calculated from the implicit equation
\ben
R_{\rm P03} = 0.6 \times \frac{8\,\ln N(R_{\rm P03}) \,v_{\rm circ}(R_{\rm P03})}{N(R_{\rm P03}) \,v_{\rm circ}(R_{200})}\,R_{200}\,,
\een
where $N(r)$ is the number of particles within a sphere of radius $r$, and $v_{\rm circ}(r)$ is the circular velocity at radius $r$ computed from the enclosed mass. The radius $R_{\rm P03}$ corresponds to the radius at which the 2-body relaxation time is 0.6 times the circular orbit period at $R_{200}$; this was found as a reliable resolution estimator in Ref.~\cite{PowerEtAl2003}. We stress that both resolution limits, $3\times l_{\rm sm}$ for hydro runs and $R_{\rm P03}$ for DMO runs, should not be considered as strict resolution limits, but rather indicative ones. The functional forms, based on Ref.~\cite{McMillan2017}, used for the fits of the density profiles of the various components, are described in \citeapp{app:mass_models}, and the best-fit values for all parameters and all halos of interest in this study are summarized in \citetab{tab:bf_parameters} of \citeapp{app:mass_models_params}.

In \citefig{fig:profilesHALOB}, we also report the distance of the Sun to the Galactic center, $\sim 8$~kpc, simply as a very naive indicator of what the local environment would look like to the virtual observer. However, we strongly warn the reader about the temptation of reaching any quantitative conclusion by inspecting physical quantities at this particular distance, since all of our simulated galaxies are by no means {\em the} MW itself.

\subsection{Assessing the spatial boundary of the virtual galaxy}
\label{ssec:rmax}

\begin{figure}[t!]
  \centering
  \includegraphics[width=0.49\textwidth]{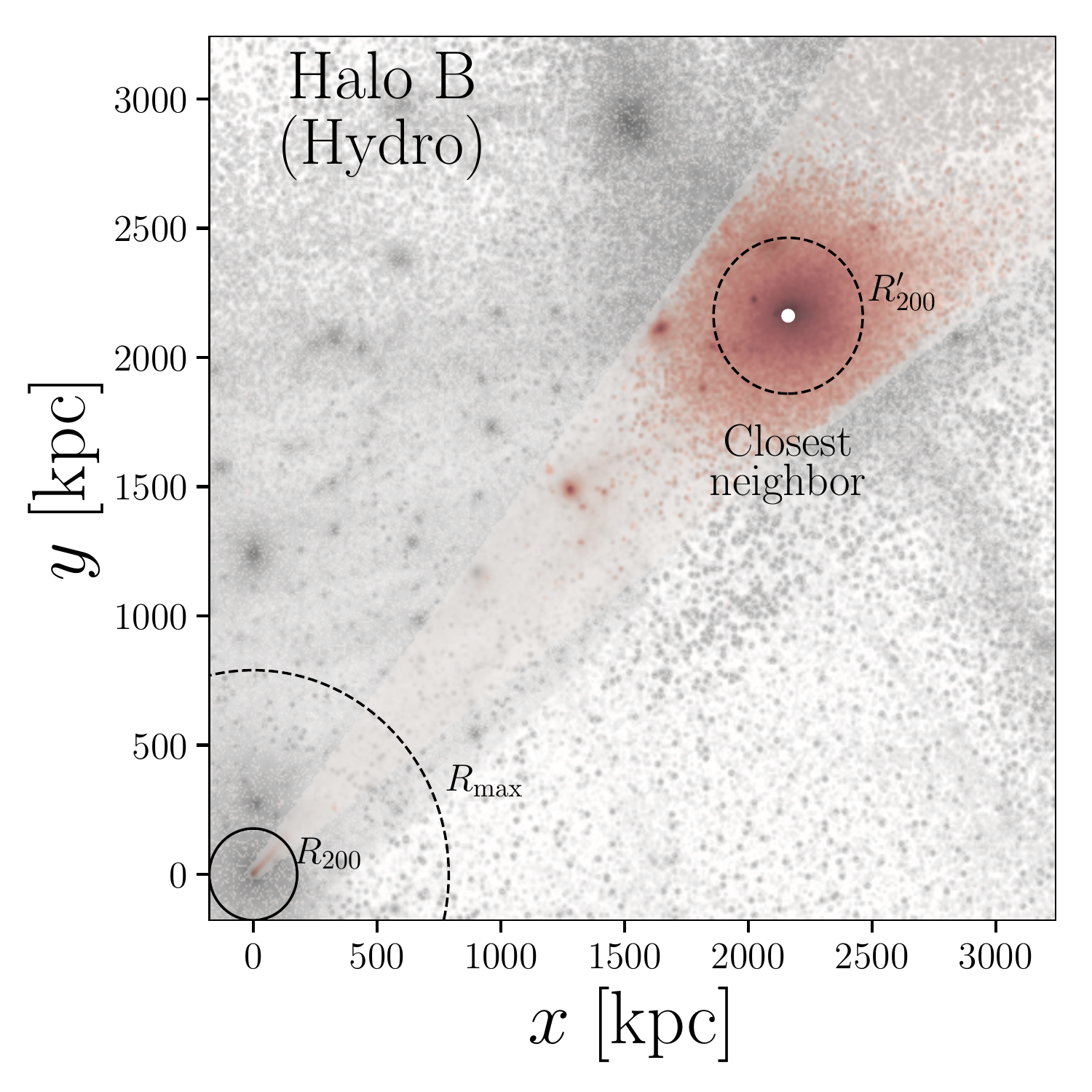}
  \hfill
  \includegraphics[width=0.49\textwidth]{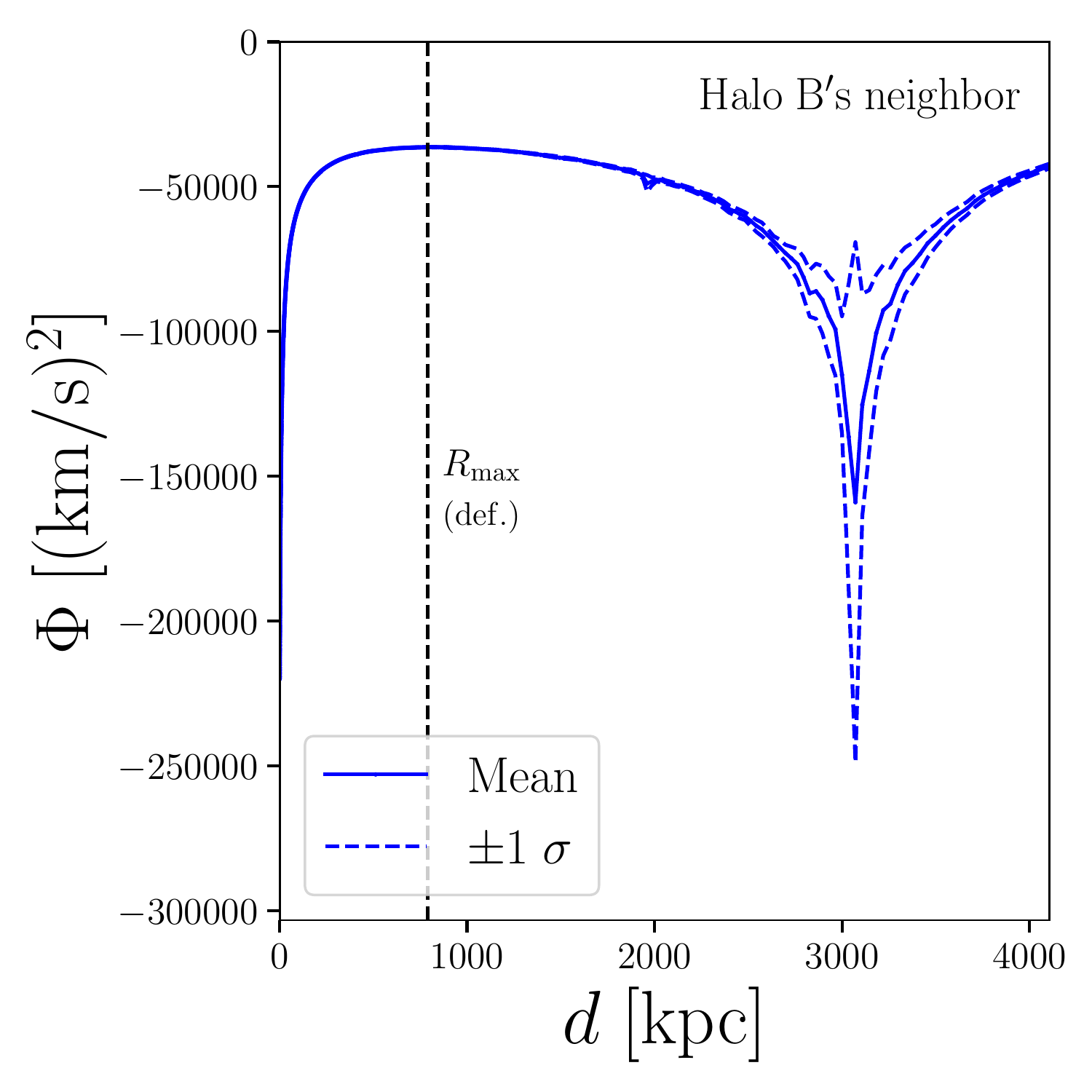}
  \caption{\small Determination of the radial extension $R_{\mathrm{max}}$ for the Halo~B simulation, for the hydro run. \textbf{Left panel:} Definition of the cone around the imaginary line joining the halo of interest (in the lower left corner) to the next massive neighbor of virial radius $R_{200}'$. The virial radius $R_{200}$ and the boundary radius $R_{\rm max}$ of the halo of interest are identified by solid and dashed circles, respectively. \textbf{Right panel:} The mean gravitational potential computed in spherical shells inside the cone along the imaginary line between the center of the halo and that of its next massive neighbor, as a function of distance.}
\label{fig:rmax-HALOB} 
\end{figure}

In order to estimate the extension of a bound structure and define the escape speed $v_{\rm e}$ in a self-consistent way, we compute the radius $R_{\rm max}$ at which the gravitational potential reaches a maximum and then starts falling down again due to the gravitational influence of the next biggest neighbor. More specifically, we define a cone around the imaginary line connecting the center of the halo of interest with the one of its next massive neighbor, as shown in the left panel of \citefig{fig:rmax-HALOB}. We then compute the mean gravitational potential in spherical shells inside the cone selection, as shown in the right panel, from which we extract the maximum. The boundary radius allows us to set the 0 of the relative potential $\Psi$ given in \citeeq{eq:rel_pot} by fixing $\Phi(R_{\rm max})$, where the gravitational potential $\Phi$ is calculated by solving the Poisson equation \citeeq{eq:poisson} with vanishing boundary conditions at infinity (isolated system). Note that the relative potential $\Psi$ is the one entering the Eddington inversion formula given in \citeeq{eq:Eddington_formula}. For reference, we also show in \citefig{fig:rmax-HALOB} the usual virial radius $R_{200}$ of the system, which is smaller than $R_{\rm max}$ here.

This boundary radius $R_{\rm max}$ allows us to define the escape velocity as $v_{\rm e}(r)=\sqrt{2\,\psi(r)}$. \change{However, it should still be noted that in real objects, the escape velocity, or similarly the spatial boundary, cannot be defined from the maximum of the potential. Furthermore, it is not easy to determine from observational data (see for instance Ref.~\cite{PifflEtAl2014a} in light of Ref.~\cite{LavalleEtAl2015}).} Without access to as detailed information as in a simulation, it can still be estimated from the mass $M_{200}^{\rm neighbor}$ and distance $D$ of the closest neighboring galaxy, by finding the point where the potential from the halo of interest is equal to that of the biggest neighbor,
\ben
R_{\rm max}\simeq D\times \left\{\frac{M_{200}^{\rm neighbor}}{M_{200}^{\rm host}}+1\right\}^{-1}\,.
\label{eq:rmax}
\een
We checked in our simulations that applying this simple formula leads to a $\sim$50\% precision in $R_{\rm max}$ with respect to the procedure described above, probably due to a bad estimate of the total mass contained in the direction of the closest neighbor. Even with this uncertainty, however, the resulting error on the escape speed remains at the \% level within the scale radii of halos.

\subsection{Consistency check: reconstructed gravitational potential}
\label{ssec:grav_pot}

\begin{figure}[t!]
\centering
\begin{subfigure}[t]{0.5\textwidth}
  \centering
  \includegraphics[width=1.\textwidth]{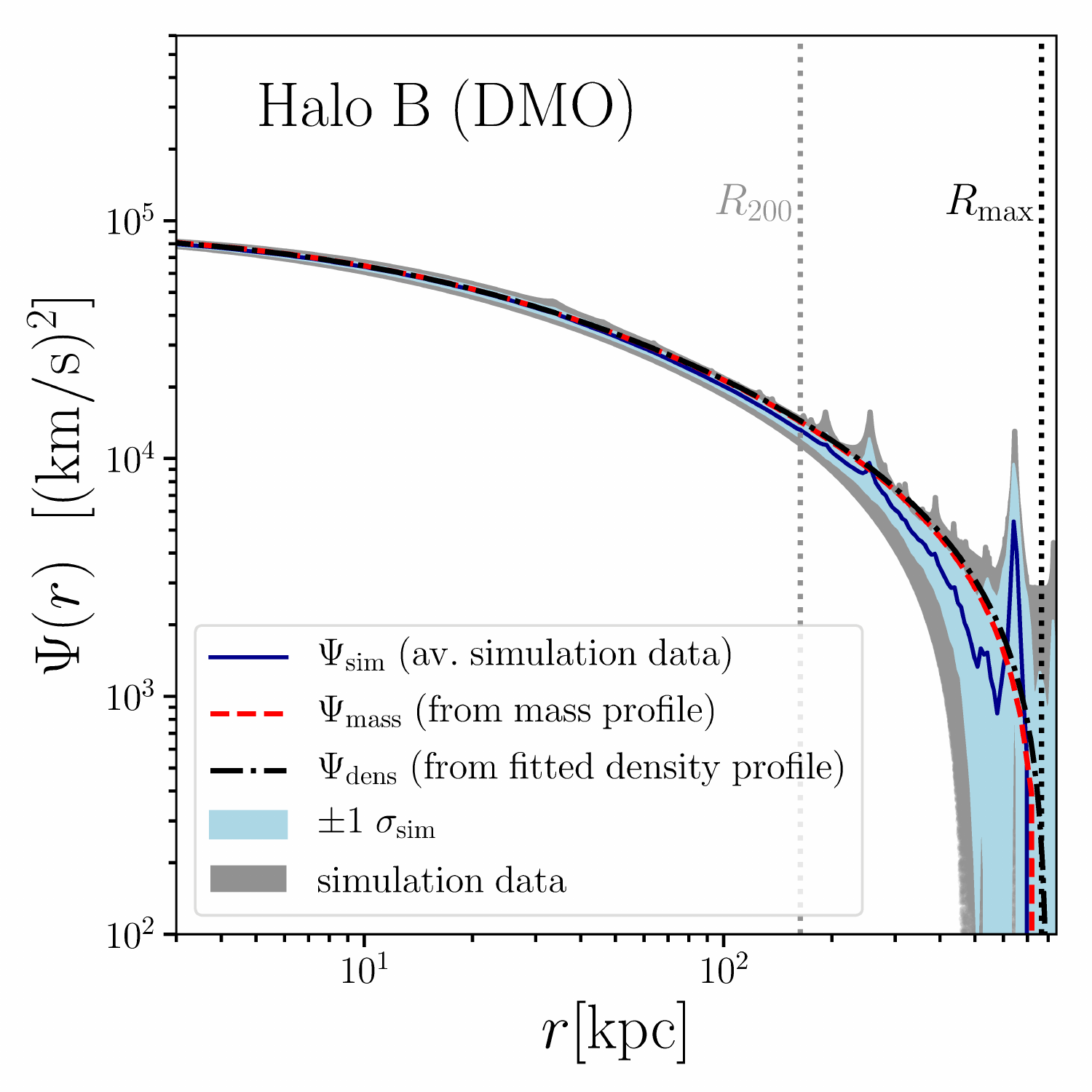}
\end{subfigure}%
\begin{subfigure}[t]{0.5\textwidth}
  \centering
  \includegraphics[width=1.\textwidth]{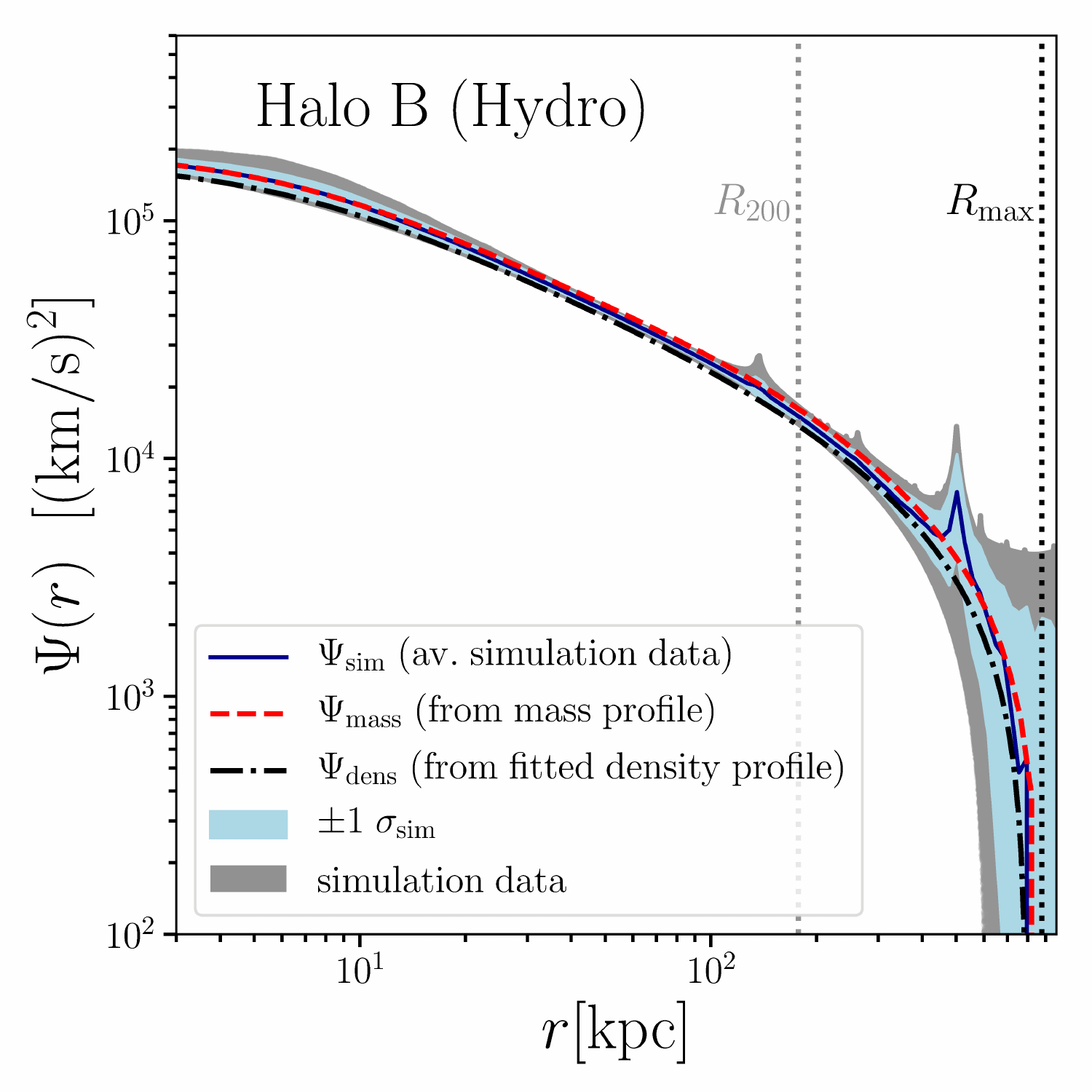}
\end{subfigure}
\caption{\small Relative gravitational potential as a function of galactocentric radius for Halo~B. Comparison between the spherical potential reconstructed from the fitted density profile (black dot-dashed), from the mass profile (red dashed), and averaged from the potentials of particles in the simulation, computed using the method of Ref.~\cite{Dehnen2002} (solid blue), for the DMO (left) and hydro (right) runs. The simulation data are shown as gray bands, and the 1-$\sigma$ standard deviation around the mean is shown as a light blue shaded band in each panel.}
\label{fig:Potential-comparison-HALOB}
\end{figure}

In order to ensure that the relative gravitational potential $\Psi$ derived from the mass models given in \citesec{ssec:fits}---and which is an important input of the Eddington inversion method---is consistent with the simulation, we compare the averaged particle potential computed in the simulation using the method described in Ref.~\cite{Dehnen2002}, (i) with the mean potential calculated from the spherically-averaged mass profile from the simulation (see \citeeqp{eq:relative_potential}), and (ii) with the one calculated from the best-fitting density profile (see \citeeqp{eq:mass}), assuming spherical symmetry again. The results are shown for Halo~B in \citefig{fig:Potential-comparison-HALOB}, for the DMO (left panel) and hydro (right) runs. The three calculations are in excellent agreement up to the virial radius $R_{200}$. In the outskirts of the halos, the reconstruction is somewhat polluted by the presence of DM subhalos. However, external regions are definitely not the cleanest environment to test the Eddington inversion, precisely because of the DM substructure that would require an additional level of analysis beyond the scope of this paper.

\section{Comparison of model predictions with simulation outputs}
\label{sec:comp}

In this section, we want to assess the degree of predictability of the Eddington formalism in terms of DM observables, expressed as velocity moments---see \citesec{ssec:observables}. We proceed as follows: (i) for each simulated galaxy (in both its DMO and hydro configurations), we determine the best-fitting mass models by adjusting the density profiles of all components on the simulation data (see \citesec{ssec:fits}); (ii) from the best-fitting mass density profiles, we compute the total gravitational potential that we further ``sphericize'' to match the spherical symmetry condition imposed by the Eddington inversion; (iii) we derive the full ergodic PSDF for each halo (in both the DMO and hydro configurations), from which we extract the speed distribution (isotropy is assumed) as well as the velocity moments described in \citesec{ssec:observables}, as a function of galactocentric radius.

We will provide and comment on the chain of plots obtained for one synthetic halo, Halo~B, to illustrate our results, which are similar for all the halos of our study. The corresponding figures for the other two halos (Halo~C and Mochima) can be found in \citeapp{app:other_halos}.

\subsection{Comparison of the Eddington PSDF $f({\cal E})$ with the pseudo-PSDF inferred from the simulation data}
\label{ssec:comp_psdfs}
\begin{figure}[t!]
\begin{center}
\includegraphics[width=0.49\textwidth]{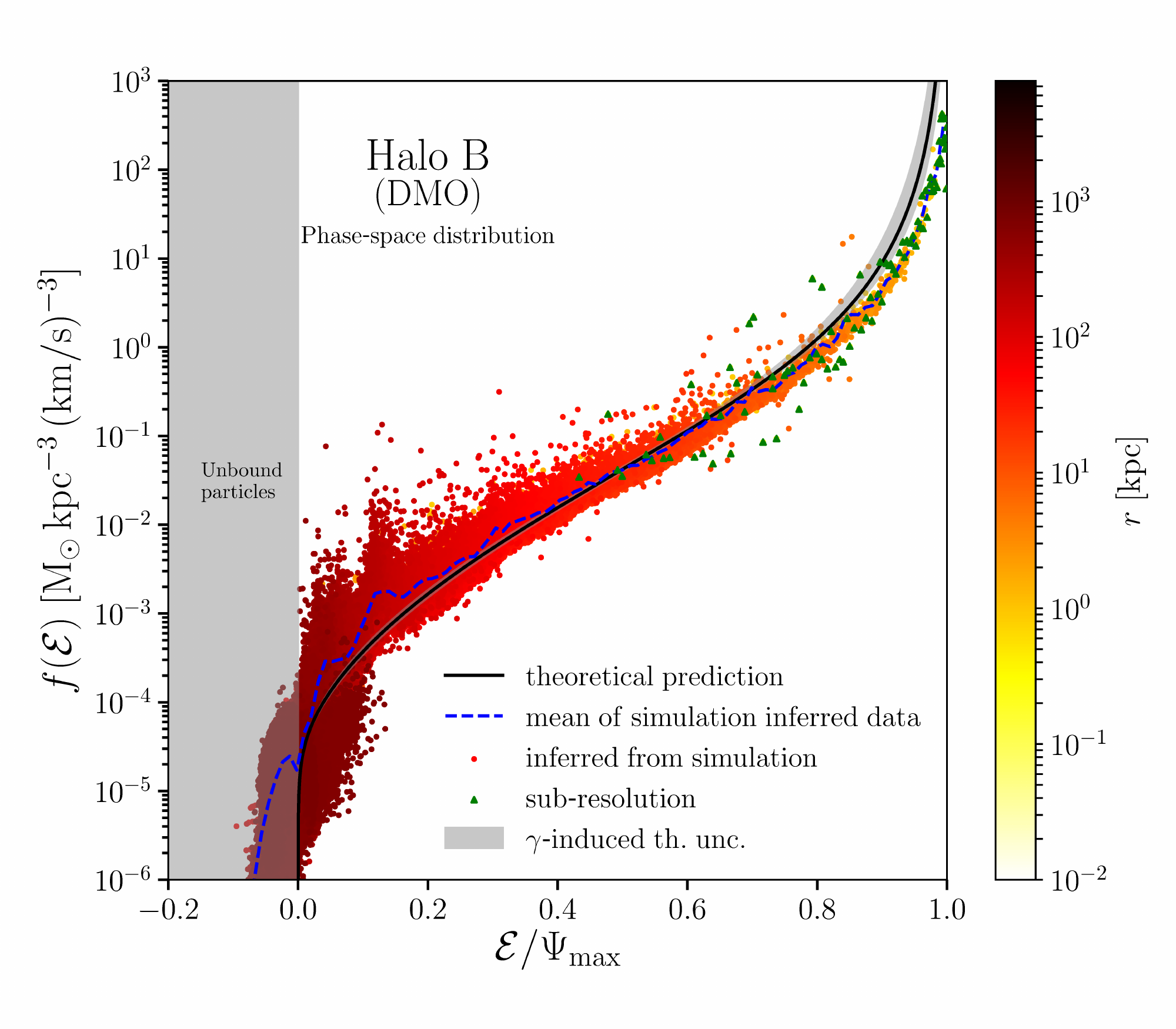} \hfill
\includegraphics[width=0.49\textwidth]{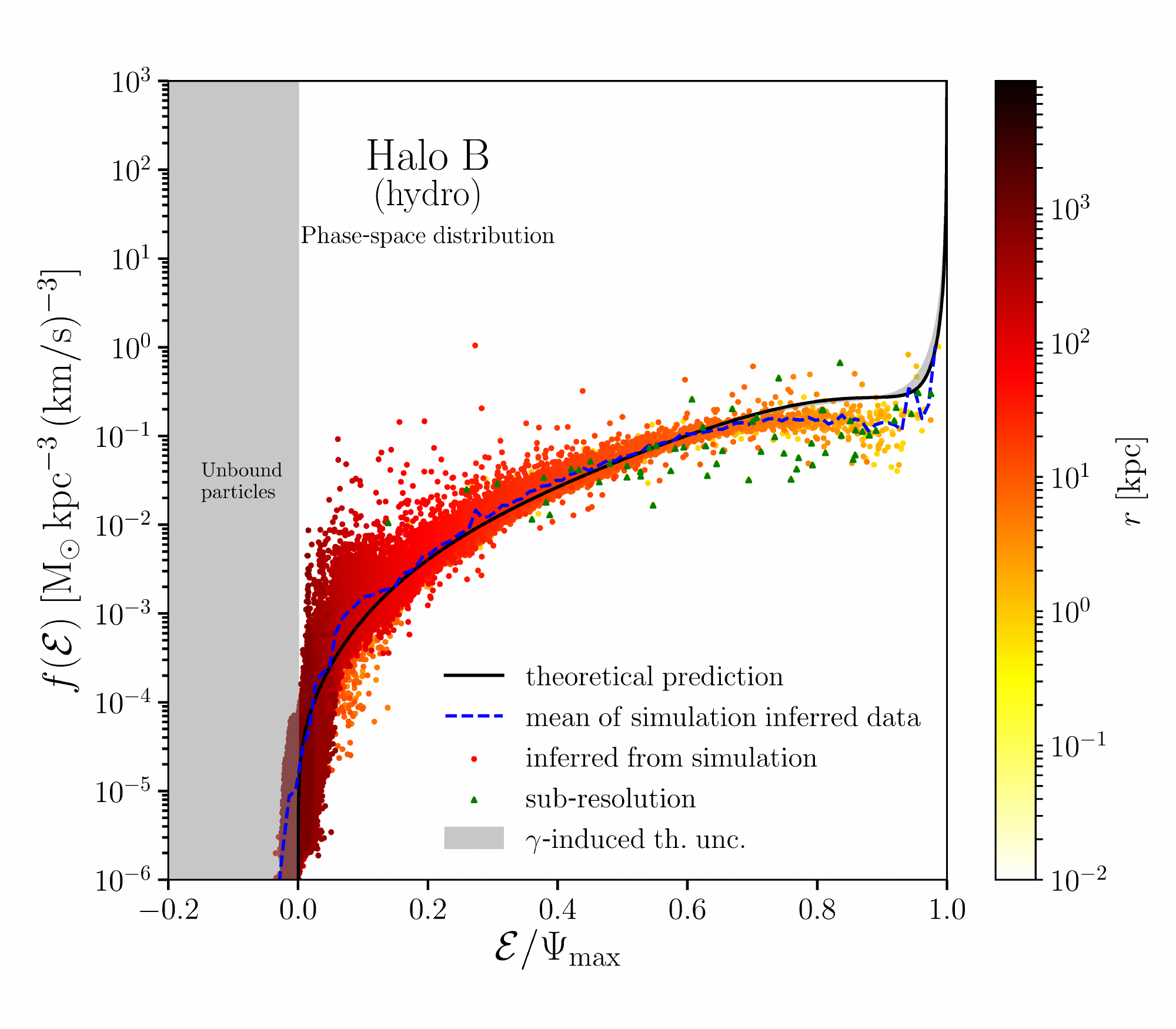}
\caption{\small Ergodic phase-space distribution predicted by the Eddington inversion for Halo~B, for the DMO (left panel) and hydrodynamical (right panel) runs, compared with the pseudo-PSDF derived from the simulation shown as a scatter plot. The mean value of the pseudo-PSDF, averaged in energy bins, is shown as a blue dashed curve. In each panel the black curve represents the Eddington prediction for the ergodic PSDF, calculated from the best-fit density profiles. A color gradient is also used to show the radial origin of each 2D bin. Finally, the gray band represents the 1-$\sigma$ systematic error associated with the fit of the inner slope $\gamma$ of the DM density profile.}
\label{fig:fE_HaloB}
\end{center}
\end{figure}

As a primary consistency check, we want to verify whether the predicted Eddington PSDFs, which only depend on energy, provide reasonable descriptions of DM in the simulated systems. To do so, we need to ``measure'' a quantity consistent with the assumptions made to derive the Eddington PSDF, namely isotropy and spherical symmetry (on top of full equilibrium). However, we have already seen that our numerical galaxies do not exhibit a perfect spherical symmetry, nor do they exhibit full isotropy---see \citesec{sec:sims}. Therefore, what we can compare with an Eddington PSDF is not the real PSDF of a simulated system, but rather a ``pseudo-PSDF''.

The PSDF is defined as the number of particles per unit of phase-space volume, multiplied by the particle mass $m_{\rm p}$ if the distribution is normalized to the mass of the DM halo. This gives in a general context
\ben
f_{\rm sim}(\vec{x},\vec{v}) = m_{\rm p} \dfrac{{\rm d}^{6}N}{{\rm d}^{3}\vec{x} \,{\rm d}^{3}\vec{v}}\,,
\een
\ie~a coarse-grained DF where ${\rm d}N$ denotes the differential number of particles in the simulation, characterized by their position-velocity phase-space coordinates $(\vec{x},\vec{v})$. Since we compare the simulation data with an \textit{ergodic} PSDF $f({\cal E})$ (isoprobability of isoenergy surfaces in phase space) of the DM for the system of interest, we further have to project $f_{\rm sim}(\vec{x},\vec{v})$ on its energy surfaces. This implies ``forcing'' spherical symmetry and isotropy, which transforms the initial coarse-grained PSDF into a pseudo-PSDF. The latter can be expressed as 
\ben
f_{\rm sim}({\cal E}) = m_{\rm p} \dfrac{{\rm d}^{2}N}{(4 \,\pi)^2\, r^{2}\,{\rm d}r \, v^{2} \,{\rm d}v}\,,
\een
where $r$ is the galactocentric distance, and $v$ the speed. Even though this pseudo-PSDF does not a priori contain the full phase-space information of the system, there is no reason whatsoever why this measure should match the Eddington prediction, even after projection on the energy surfaces. Therefore, this provides an interesting test as how departures from dynamical equilibrium, spherical symmetry, or isotropy, may impact on the predictivity of the Eddington inversion method already at the level of the PSDF. Strong differences between the pseudo-PSDF built from the data and the one predicted from the Eddington inversion would already jeopardize the relevance of the latter to describe the simulation. Similar estimates of the pseudo-PSDF were made from DMO runs in Ref.~\cite{VogelsbergerEtAl2009}, with additional assumptions.

In practice, we determine the pseudo-PSDF of each of our halos by dividing the virial sphere embedding the DM halo in linear bins in $r$, associated with index $i$. We set equal size for each radial bin, such that the least populated contains 500 particles. The particle population inside each radial bin is further divided in 100 bins in $v$, associated with index $j$, from the minimal speed to the maximal speed present in the shell selection. Then for each $(r,v)$ bin---or equivalently $(i,j)$---we compute the associated energy as 
\ben
{\cal E}_{ij}=\left\langle \Psi \right\rangle_{ij} - \left\langle \frac{v^2}{2}\right\rangle_{ij} \,,
\een
where $\left\langle\Psi\right\rangle_{ij}$ and $\left\langle v^2/2 \right\rangle_{ij}$ respectively represent the mean potential and the mean kinetic energy per unit mass of the particles inside the corresponding 2-dimensional (2D) bin of indices $(i,j)$. We estimate the mean phase-space density for each 2D bin by dividing the total DM mass inside the bin by its phase-space volume,
\ben
f_{\rm sim}({\cal E}_{ij}) = \frac{m_{\rm p} \, N_{ij}}{(4\,\pi)^2 \, r_{{\rm c},i}^2 \,\Delta r_i \, v_{{\rm c},j}^2\, \Delta v_j}\,,
\een
where $m_{\rm p}$ and $N_{ij}$ are the mass of the DM particles in the simulation and the number of particles inside the bin, respectively. Quantities $r_{{\rm c},i}$ and $v_{{\rm c},j}$ are the central values of the 2D bin $(i,j)$ and $\Delta r_i$ and $\Delta v_j$ are the associated widths. We checked that changing the number of bins and the minimal number of particles per bin does not affect our results.

This coarse-grained pseudo-PSDF is calculated including all particles inside $R_{200}$, and is shown in \citefig{fig:fE_HaloB} for Halo~B as a scatter plot where each data point corresponds to a $(r,v)$ 2D bin. Results are shown for both the DMO and hydro cases (left and right panels, respectively). A color gradient is used to show the radial origin of each 2D bin, illustrating that high energies essentially correspond to the central regions of the halo, as expected. The mean value of the pseudo-PSDF, averaged in energy bins, is shown as a blue dashed curve on top of the scatter plot. In each panel we superimpose as a black curve the Eddington prediction for the ergodic PSDF, calculated from the best-fit density profiles. We also report as a gray band the 1-$\sigma$ systematic error associated with the fit of the inner slope $\gamma$ of the DM density profile.\footnote{It should be noted that the associated uncertainty on the speed distribution and corresponding moments remains at the sub-percent level at all radii accessible in the simulation.} The detailed procedure used to determine that uncertainty band is explained in \citeapp{app:mass_models_params}. This band is quite thin and demonstrates that even exploring the statistical freedom in $\gamma$ does not change the prediction of the model significantly. However, it should be noted that this band does not include variation over the whole available parameter space. Given the large degree of degeneracy in the free parameters characterizing the density profiles, we may wonder whether more freedom could modify this relative stability of the PSDF prediction. Since the Eddington model is fully determined by the detailed density profile (and the full potential), we can argue that for whatever functional form providing a shape very close to the actual density profile, predictions are not likely to change significantly. Therefore, were we to explore the global posterior parameter space within 1~$\sigma$, which would then provide a profile in excellent agreement with the true density profile (see the best-fitting and the true profiles in \citefig{fig:profilesHALOB}), we would likely not see significant changes in the PSDF prediction.

We see from both panels of \citefig{fig:fE_HaloB} that our model predictions are in reasonably good agreement with the averaged pseudo-PSDFs inferred from the simulation data. Some discrepancies appear essentially at high energy, which characterizes low-velocity particles orbiting in the inner parts of the halo. This encapsulates both the lack of resolution of the simulation on sub-kpc scales and the limits of the simplifying assumptions made to apply the Eddington inversion. However, it is already remarkable that such a simple model can match so well the pseudo-PSDF computed from the raw data. Indeed, the model is able to capture the strong differences between the DMO case on the one hand (left panel), which is typical of a cuspy halo PSDF and exhibits an exponential behavior in energy, and the hydro case on the other hand (right panel), which is instead typical of a cored halo PSDF and exhibits a shallower increase as a function of energy---see more discussions on the generic behavior of $f({\cal E})$ as a function of the DM profile for instance in Refs.~\cite{Widrow2000,BinneyTremaine2008,LacroixEtAl2018}, where Ref.~\cite{LacroixEtAl2018} further explains the effect of baryons.

Further comments are in order. We see from \citefig{fig:fE_HaloB} that the model systematically overshoots the data at high energy, while the agreement is better at intermediate energy. Since the high-energy region corresponds to small velocities in the central regions (within $\lesssim 10$-20~kpc), this means that the model is likely to predict a systematic shift in the position of the peak of the speed distribution toward small values. This shift should be less pronounced or even disappear at larger radii. We can also note the existence of some peaky coherent structures at low energy in the data points, and many of them are associated with or close to negative ${\cal E}$, \ie~2D bins of unbound particles (those lying in the gray shaded area on the very left part of the plots). Actually, these features mostly come from DM subhalos or high-speed streams, mostly located in the outskirts ($\gtrsim 100$~kpc), with some of them that are not tied to the host halo. We therefore expect these features to appear in the speed distribution as well, especially in the external regions of the simulated galaxies. Getting rid of these subhalos is a very complicated task, as many of them are tidally stripped by the host potential, some being spatially destroyed while still bound in phase space, some other being relaxed or on the verge of phase mixing with the host halo. We will therefore keep them in the analysis, keeping in mind that they can lead to some features in the speed distribution. Anyway, we will see in \citesec{ssec:comp_fv} that the expected translation of the predicted PSDF and of the measured pseudo-PSDF in terms of speed distribution turns out to be correct. This confirms the physical relevance of comparing the model against data already at the level of the PSDFs.

\subsection{Speed distribution}
\label{ssec:comp_fv}

\begin{figure}[t!]
\centering
\includegraphics[width=0.49\textwidth]{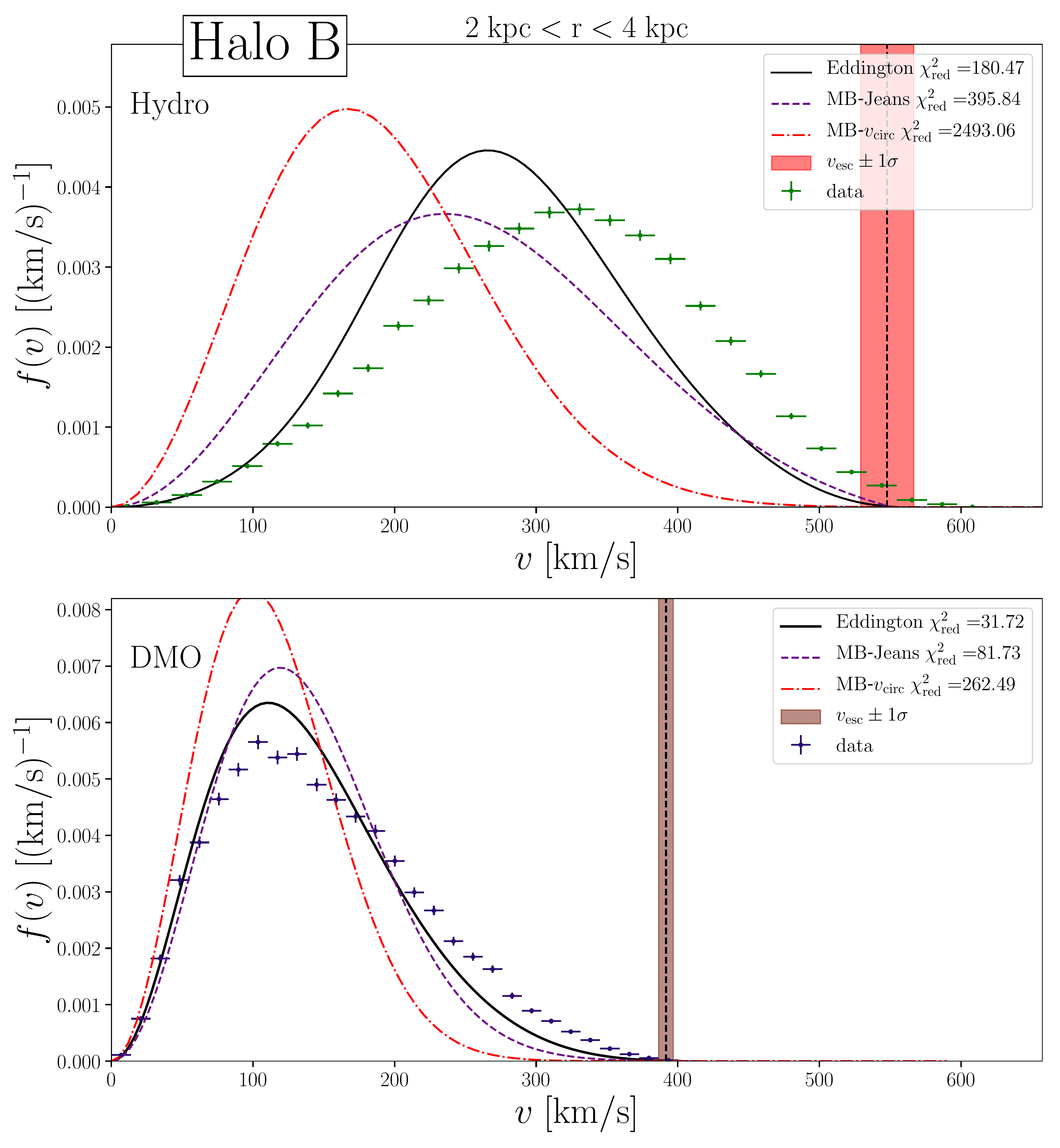}
\hfill 
\includegraphics[width=0.49\textwidth]{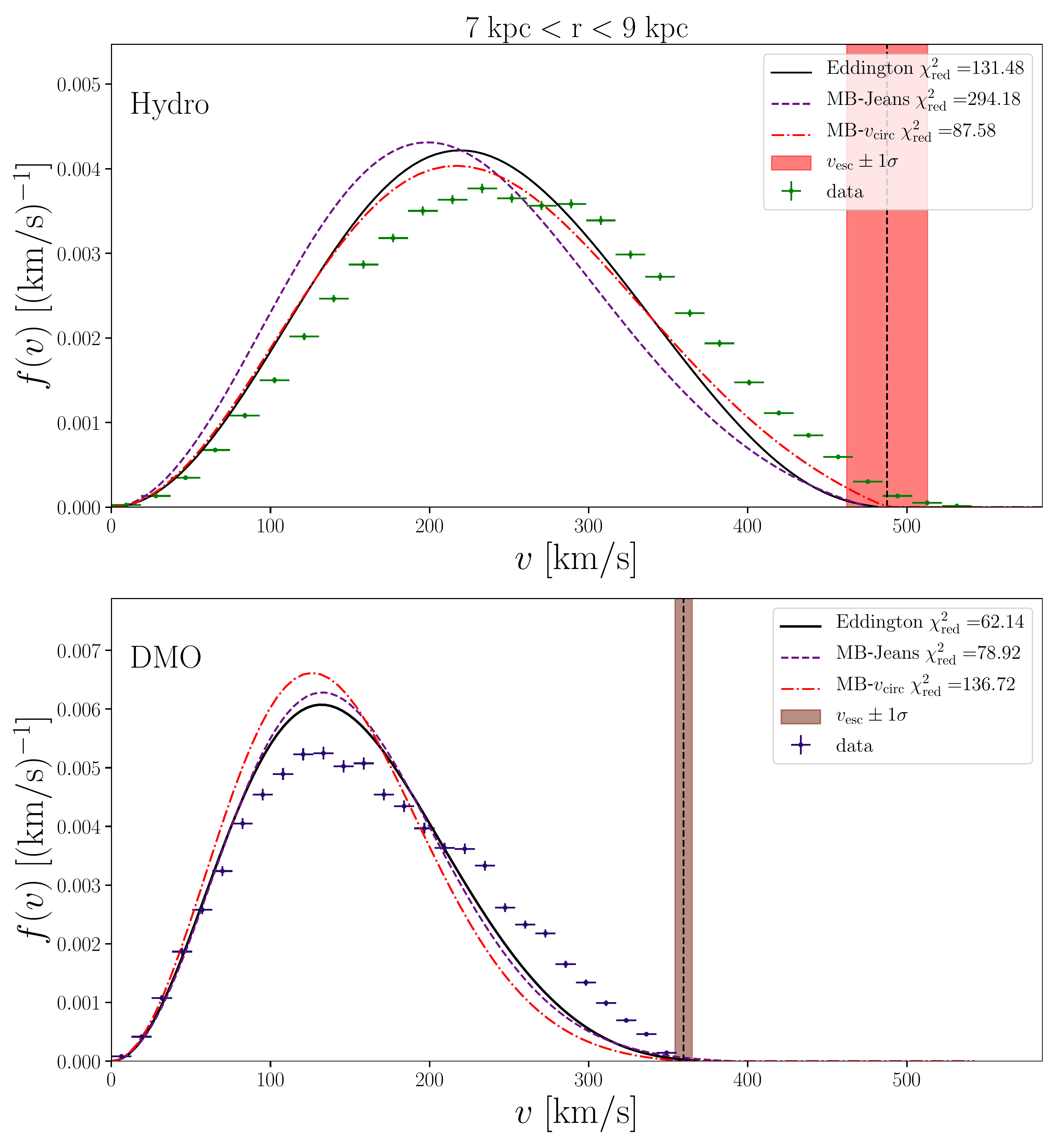}\\
\vspace{0.3cm}
\includegraphics[width=0.49\textwidth]{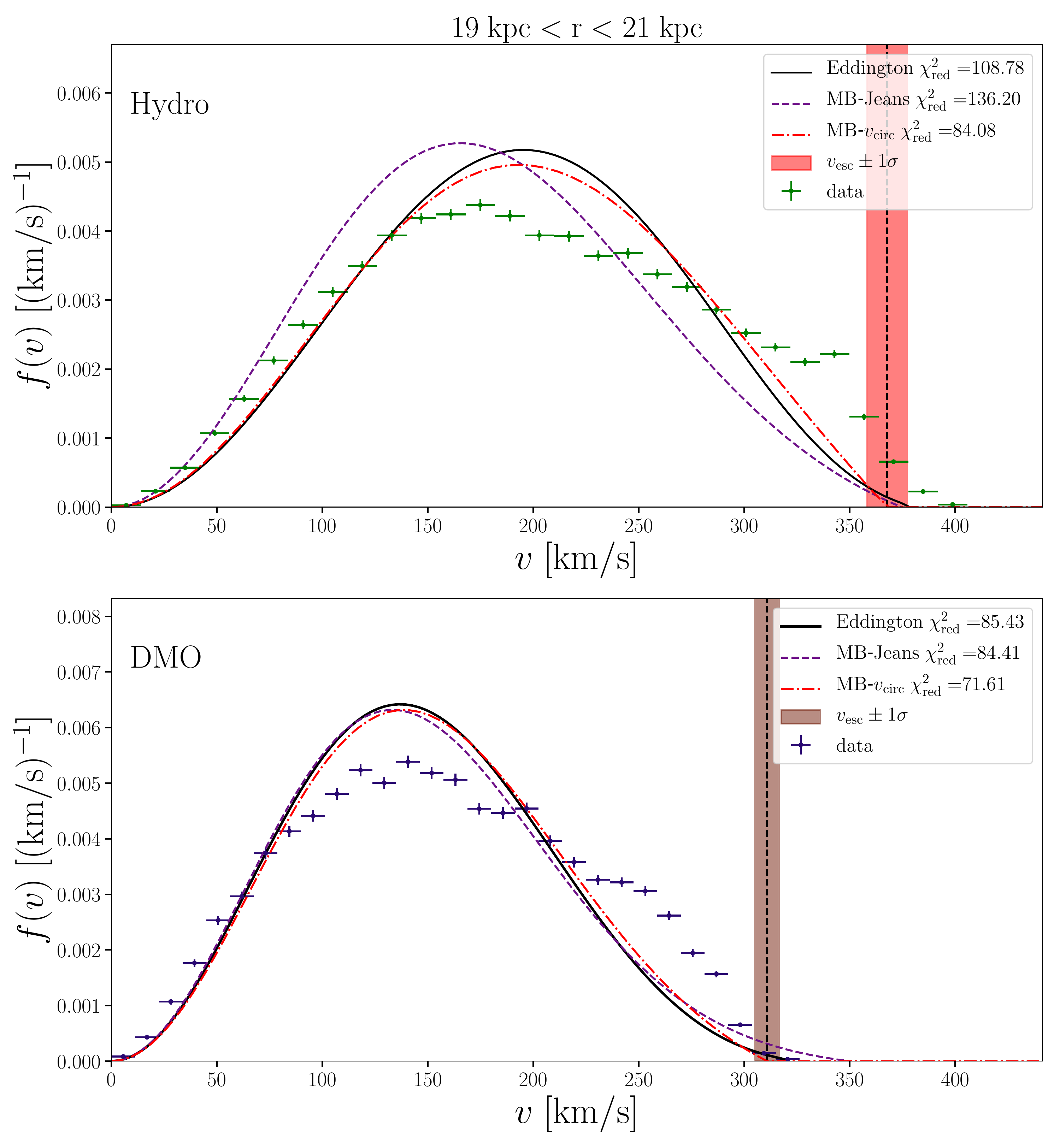}
\hfill 
\includegraphics[width=0.49\textwidth]{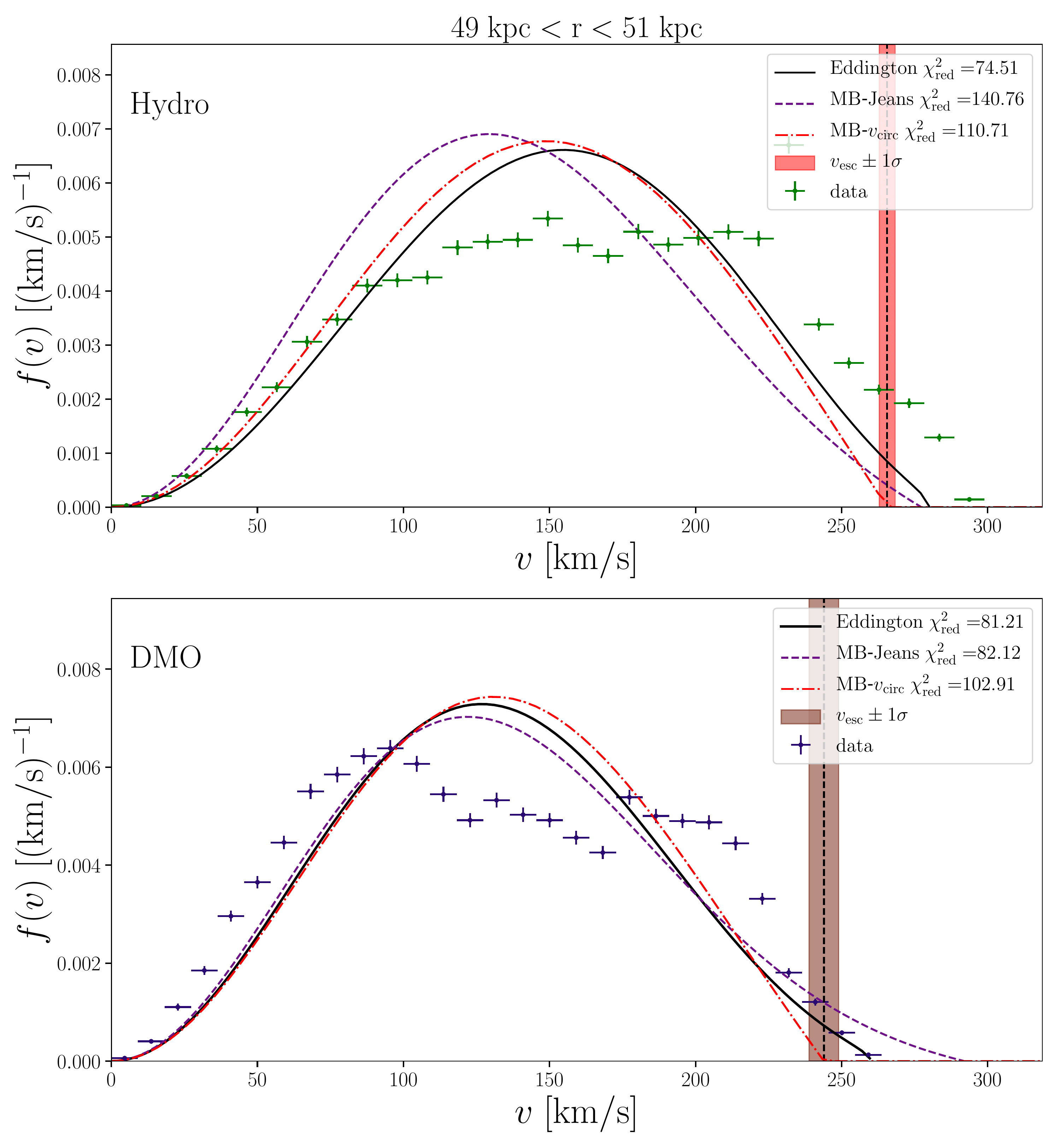}
\caption{\small Comparison between the speed distribution derived with the Eddington method and the simulation data for spherical shells of 2 kpc thickness at different radii (3 kpc, upper left; 8 kpc, upper right; 20 kpc, lower left; and 50 kpc, lower right), for Halo~B, for the DMO (lower subpanels) and hydro (upper subpanels) runs. In each panel, the data points (green for hydro, blue for DMO) correspond to a normalized histogram built from the data. The Eddington predictions are shown as black solid curves, and two models relying on the MB distribution, with different assumptions regarding the peak speed $v_{0}$ are displayed: one in which it is set to the circular velocity (``$v_{\mathrm{circ}}$", red dot-dashed curves), and the other one in which it is given by $\sqrt{2/3}$ times the velocity dispersion calculated by solving the Jeans equation (``Jeans", purple dashed curves). The escape speed $v_{\mathrm{e}}(r)$ from the simulation is shown as black vertical dashed lines, with the associated 1-$\sigma$ error displayed as shaded bands.}
\label{fig:fv-comparison-HALOB}
\end{figure}

Now we compare the Eddington model prediction for the speed distribution $f(v)$ of DM particles with the simulation data. To proceed, we collect the speeds of the DM particles in the various simulated halos in spherical shells of 2~kpc thickness around four benchmark radii, namely 3, 8, 20 and 50~kpc. At each radius, we actually infer the true speed distribution from the normalized histogram built from the data. For each bin in speed, the corresponding error bar on $f(v)$ is simply set to $\sqrt{N_i}/N$, where $N_i$ ($N$) is the number of particles in the $i^{\rm th}$ bin (total number of particles, respectively). This approximates the 1-$\sigma$ Poissonian error.

In \citefig{fig:fv-comparison-HALOB} we confront the Eddington prediction for the speed distribution (black solid curves) to the simulation results (DMO: blue points; hydrodynamical: green points). Upper subpanels are associated with hydrodynamical runs and lower subpanels with DMO runs. For completeness, we also report the predictions obtained using two different models relying on the MB distribution and introduced in \citesec{ssec:MB}, with different assumptions regarding the peak speed $v_{0}$  of the distribution summarized in \citeeq{eq:MB_models}---both models are smoothly truncated at the escape speed $v_{\rm e}(r)$. More specifically, the MB model labeled ``$v_{\rm circ}$'' refers to the isothermal approximation where the velocity dispersion is set to the circular velocity (red dot-dashed curves), and the one labeled ``Jeans'' refers to that in which the velocity dispersion is calculated by solving the Jeans equation (purple dashed curves). These MB models have been widely used in the literature in the context of DM searches.

For each radius we also display the escape speed $v_{\rm e}$ from the simulation as a red shaded vertical band in each subpanel. It is evaluated from the mean gravitational potential in the spherical shell, and the width is set from the associated 1-$\sigma$ statistical uncertainty. As seen from the higher-speed point of the normalized histograms, a few simulation particles are found with speeds beyond $v_{\rm e}$, consistently with \citefig{fig:fE_HaloB}, but these mostly come from unbound subhalos (see for instance the 50~kpc shell in the hydro run). Overall, however, the speed DF inferred from the data is indeed found to vanish as $v\to v_{\rm e}$, which is expected since $v_{\rm e}$ is calculated from the $R_{\rm max}$ measured in the simulation. The predicted speed DF obviously obeys the same behavior because (i) it vanishes at the predicted escape speed by construction, which matches with the endpoint of the tail of the data because $R_{\rm max}$ was inferred from the data\footnote{Should $R_{\rm max}$ be estimated from \citeeq{eq:rmax}, leading to significant error, the prediction of $v_{\rm e}$ would still match at the \% level within the scale radius of the halo, but would more strongly depart from the endpoint of the tail in the outer regions.}, (ii) the gravitational potential calculated from the fits of the density profiles matches very well with the one measured in the simulation data.

In order to better quantify the level of agreement or disagreement between the Eddington prediction and the true speed distribution inferred from the data, we calculate the reduced $\chi^2_{\rm red}=\chi^2/N_{\rm dof}$, where $\chi^2$ is the (usual) total chi-square, and $N_{\rm dof}$ is the number of degrees of freedom (d.o.f). Since there is no free parameter, $N_{\rm dof}$ is equal to the number of bins (29 in the plots). We do the same for the MB models. Given the simplifying assumptions the models are based on, and given the fact that there are local departures from equilibrium in the simulations, we do not expect to find a good $\chi^2_{\rm red}$ in the statistical sense. However, this ``measure'' can still be helpful to define some qualitative hierarchy in the level of predictivity.

Before going to more detailed comparisons between the predictions and the data, it is worth emphasizing again that our MB {\em models} are not {\em fits} to the data. Indeed, it has already been shown that several bell-shaped functions (\eg, generalized Maxwellian or Tsallis functions) with the position and width of the peak inferred from the data themselves would provide reasonable fits to the speed DF at almost all radii (\eg~Refs.~\cite{VogelsbergerEtAl2009,LingEtAl2010,MaoEtAl2013,PillepichEtAl2014a,BozorgniaEtAl2016,NunezCastineyraEtAl2019}). Here, however, we wish to compare the ability of simple models to {\em predict} (or not) the speed DF properties only from a galactic mass model, which is the most simple information we can use to describe the gravitational dynamics of DM in systems like the MW.

By closely inspecting \citefig{fig:fv-comparison-HALOB}, one can first notice that both visually and from the values of the $\chi^{2}_{\rm red}$, all the models considered here are clearly too simple to fully describe the data, as expected. However, the Eddington model matches rather well with the data in the DMO case, much better than the MB models, especially in the inner parts of the virtual galaxy. This trend seems to be rather generic because it is also observed in the other two simulations---see \citeapp{app:other_halos}. This may come from the fact that the central parts of halos are probably the most relaxed ones in the DMO case (in the absence of recent major mergers). This result is interesting because our DMO simulations might actually be representative of what occurs in DM-dominated objects like dwarf galaxies. Moreover, the central regions of galaxies are usually among the best targets for DM searches. At intermediate and larger radii, while the models' predictions get slightly closer to one another, one can see features appearing in the data, which actually come from subhalos or coherent streams, and which can obviously not be captured by the models.

The agreement between the data and the models slightly degrades in the hydro case, and in particular, one can observe a systematic shift of the predicted peak of the speed distribution toward lower values in the central regions, as already expected from the comparisons of the PSDFs---see \citesec{ssec:comp_psdfs}. However, one can note that the shift is much more discrepant for the MB models than for the Eddington prediction, which still provides a much better description of the data in the central regions. Moreover, it is worth mentioning that the observed shift in the peak is globally more pronounced for Halo~B than for the other two halos (see \citeapp{app:other_halos} for Halo~C and Mochima). There might be several explanations: (i) a significant merger took place in the hydro run of Halo~B (not observed in the DMO run), inducing an offset between the DM barycenter and the minimum of the potential (see \citesec{ssec:eq}), which may further induce a combination of departures from equilibrium, spherical symmetry, and isotropy; (ii) baryons are responsible for feedback in the central regions of galaxies, which induces intermittent departure from local equilibrium \cite{PontzenEtAl2012}; (iii) the effective SN feedback is different between Halo~B and Halo~C on the one hand, and Mochima on the other hand, due to the different tuning related to the different resolutions, which may lead to less violent feedback for Mochima, hence a more relaxed system at the center (the peak shift is less salient in the Mochima case). Finally, the same trend as for the DMO case is observed at larger radii, where some features show up in the speed distribution inferred from the data, originating again from subhalos or streams. The models are not able to account for these bumps, but perform reasonably in terms of the global shape of the speed DF.

It is further interesting to compare the overall predictivity of the Eddington model relative to the MB ones. Both by eye and from the $\chi^{2}_{\rm red}$ values, one can reasonably conclude that even though in some cases the MB models seem to provide a satisfactory description of the speed distribution, the Eddington model is in most cases in much better agreement with the simulation at small radii---as represented by the 3~kpc panel in \citefig{fig:fv-comparison-HALOB}. Even in the hydro case, the Eddington model still provides a rather consistent picture of the speed distribution \textit{throughout} the entire halo, down to the resolution radius of the simulation. In particular, it can predict the peak speed of the distribution down to a $\sim$10-20\% precision. This persists in the three simulated halos, both for the DMO runs and in the full hydrodynamical setups. The poorer performance of the MB models comes from their relatively worse estimates of the peak velocity, which are fixed by the choices of $v_0$ (see \citeeqp{eq:MB_models}). The Eddington model will also be shown to perform quite well in terms of velocity moments, much better than at the level of the detailed speed DF---see \citesec{ssec:moments}.

To add up a more theoretical note, it should not come as a surprise that the MB models perform globally worse than the Eddington one, as anticipated in \citesec{ssec:MB}. First of all, it is known that DM halos are highly non-isothermal in their central regions. Moreover, while the MB-Jeans model gives the same velocity dispersion as the Eddington model, the associated position of the peak is overly constrained and not fixed by the internal dynamics of the system (in contrast, the most probable speed is not determined by the dispersion or circular velocity in the Eddington model, but derives from complementary dynamical information). Indeed, even tuned to include some relevant physical information (spatial dependence of peak and width, escape speed), our MB models are still not solutions to the collisionless Boltzmann equations that our simulated systems would obey, should they be perfectly spherically symmetric, isotropic, and in steady state. Even though it was not ensured from the beginning, it is somewhat satisfactory to find that a model that includes physics in a more consistent way provides a better description of a physical system. Along this line, we naturally expect the MB-Jeans model to provide results closer to the Eddington prediction.

Therefore, as a global result, we emphasize that even if based on strongly simplifying assumptions, the Eddington approach has the advantage of giving a satisfactory description of the speed distribution of DM particles in a realistic galactic system, accounting rather well for the dynamical trends as a function of radius, while relying on first-principle grounds. This makes this model a reasonably reliable tool, which does not rely on ad hoc prescriptions, to predict the speed DF of DM throughout galaxies, provided one can constrain the spatial distribution of their components---more reliable than ad hoc MB models. In the next paragraph, we further quantify the level of predictivity of the Eddington model in terms of observables more directly related to DM searches.

\subsection{Moments of the speed distribution}
\label{ssec:moments}
\begin{figure}[t!]
\begin{center}
\includegraphics[width=0.49\textwidth]{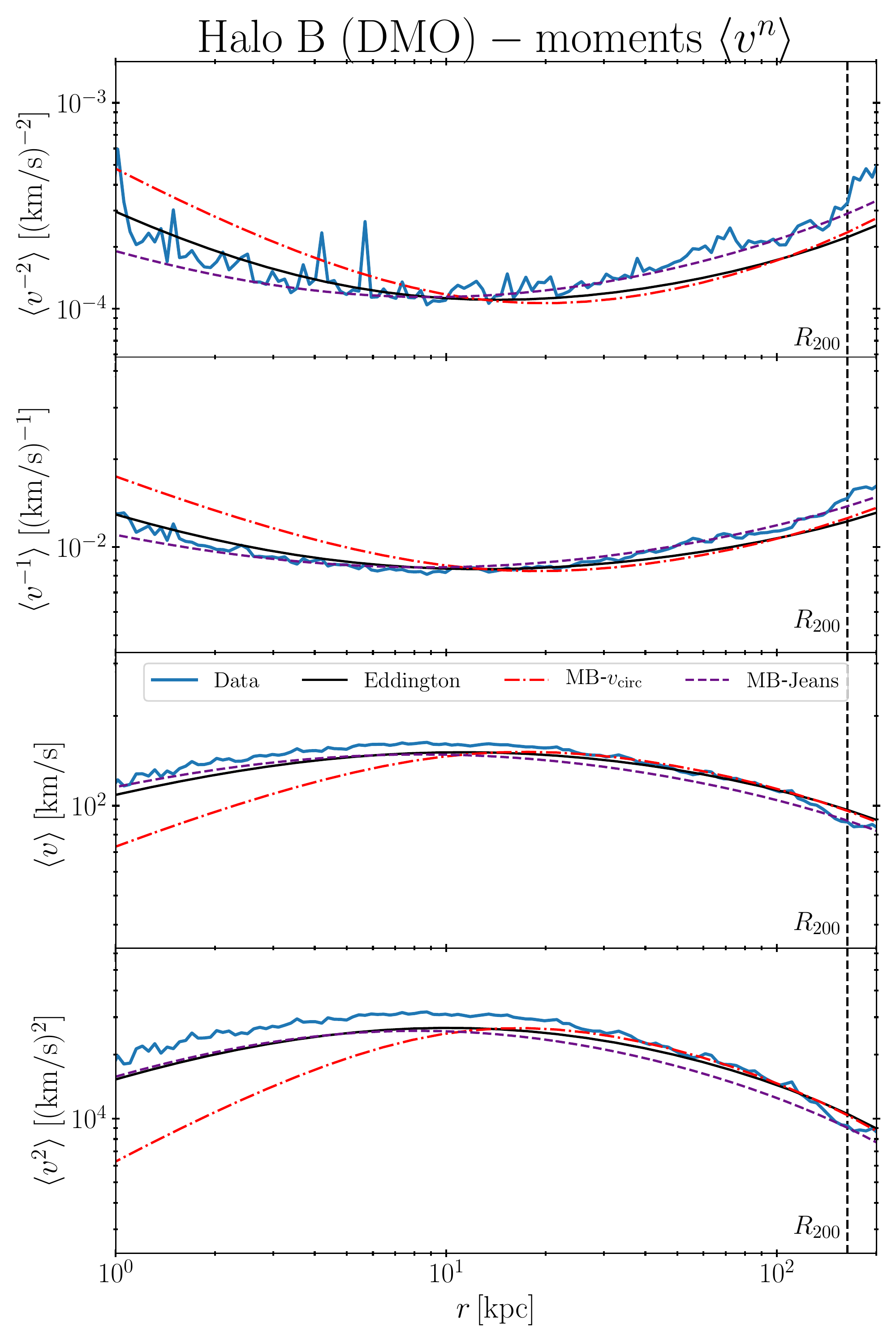} \hfill
\includegraphics[width=0.49\textwidth]{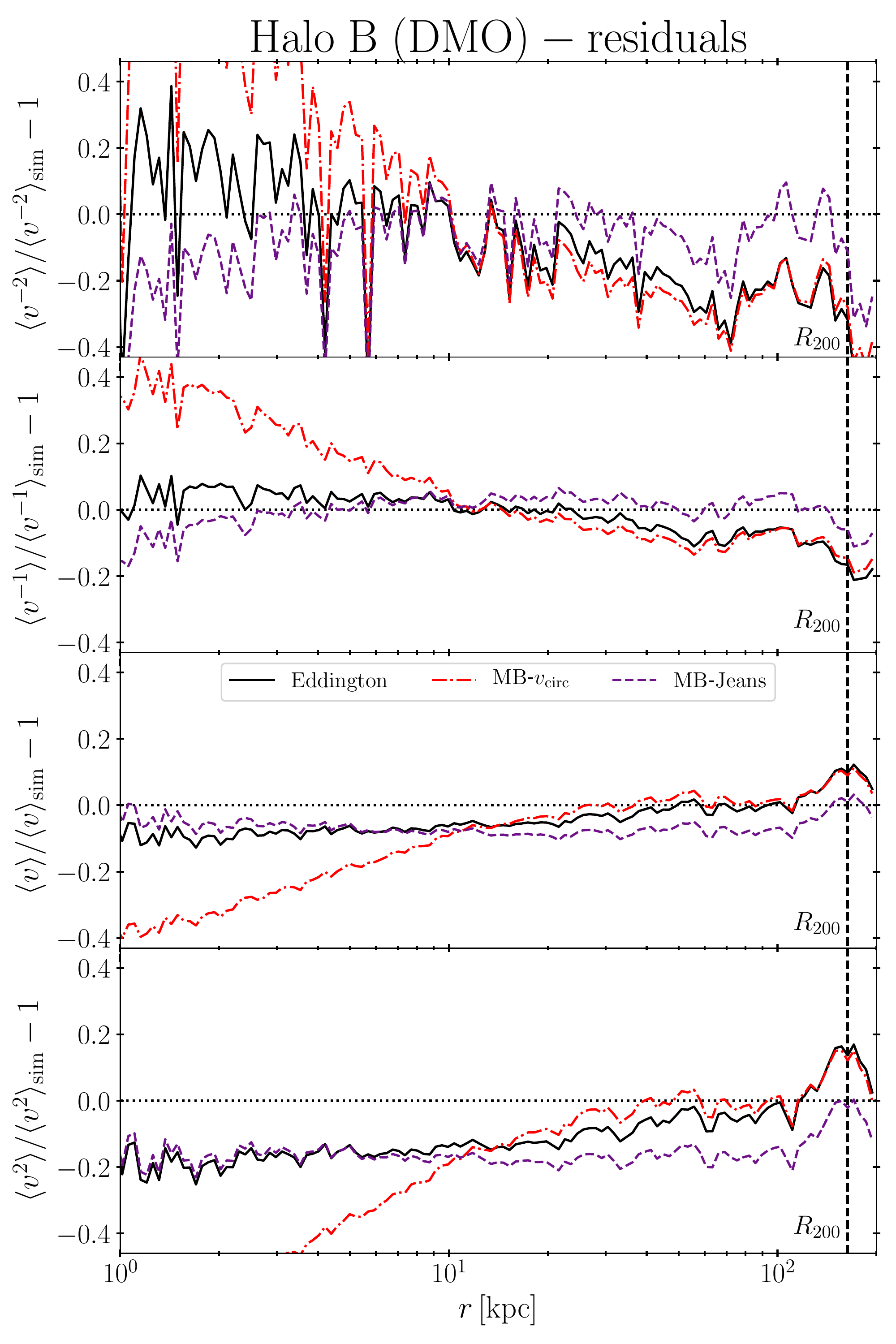}
\caption{\small Comparison of the moments of the speed distribution from the Eddington method and from the simulation for Halo~B (left panel) and associated residuals with respect to the simulation (right panel), for the DMO run. Simulation outputs are shown as blue solid curves. Predictions obtained with the Eddington inversion method are displayed as black solid curves, while the two MB models ``$v_{\mathrm{circ}}$" and ``Jeans" are shown as red dot-dashed and purple dashed, respectively.}
\label{fig:vmoments-comparison-HALOB-DMO}
\end{center}
\end{figure}

\begin{figure}[t!]
\begin{center}
\includegraphics[width=0.49\textwidth]{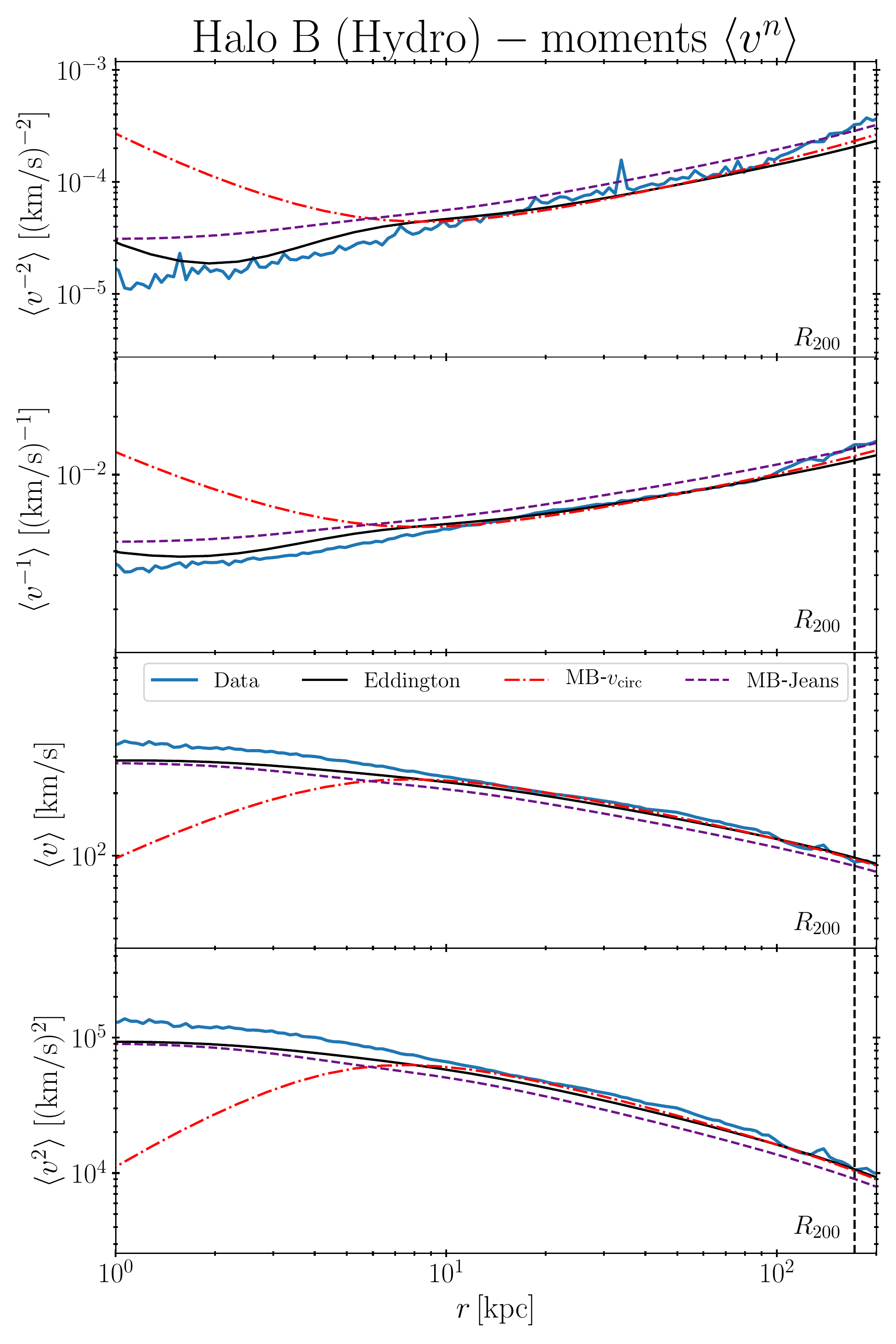} \hfill
\includegraphics[width=0.49\textwidth]{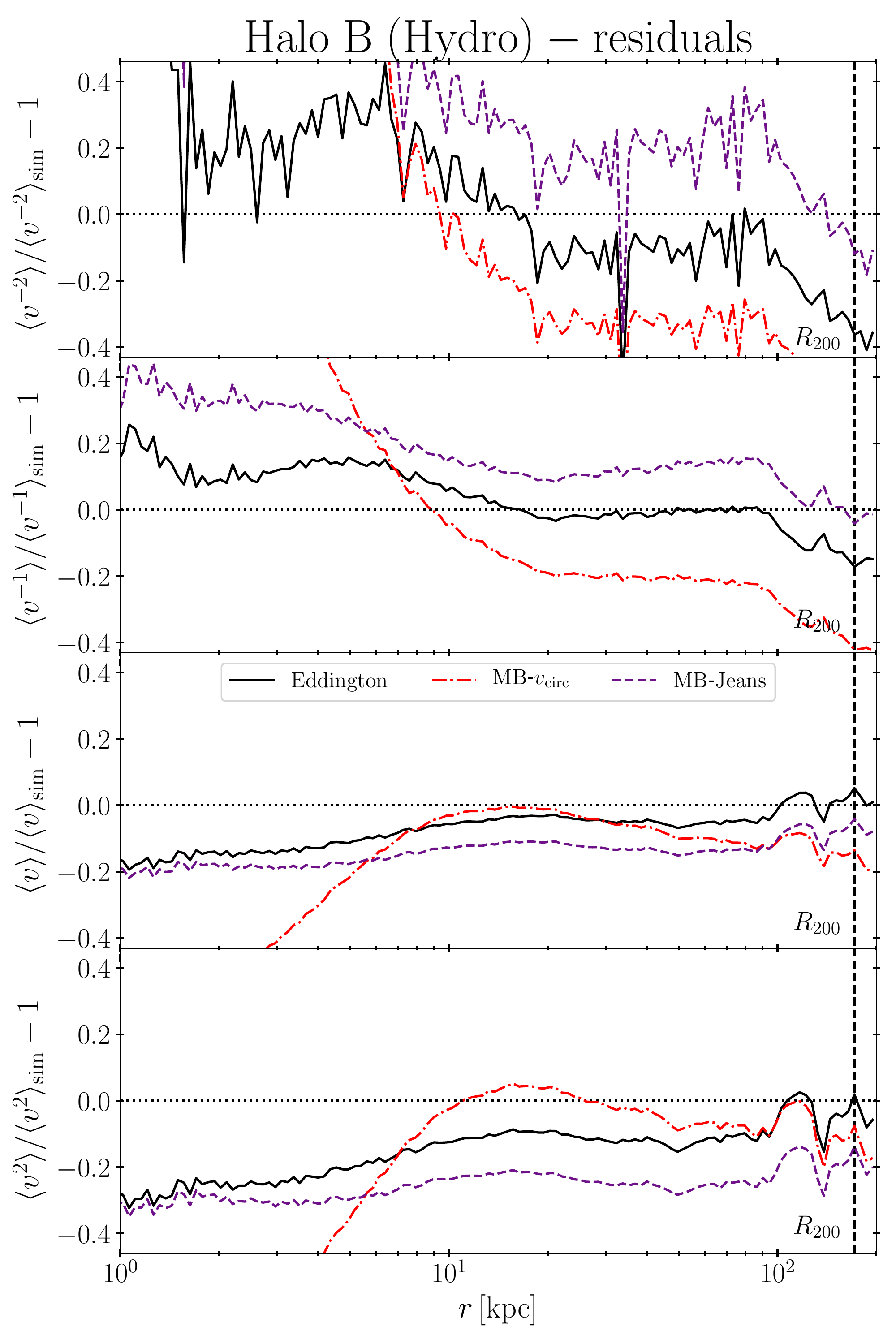}
\caption{\small Comparison of the moments of the speed distribution from the Eddington method and from the simulation for Halo~B (left panel) and associated residuals with respect to the simulation (right panel), for the hydro run. Simulation outputs are shown as blue solid curves. Predictions obtained with the Eddington inversion method are displayed as black solid curves, while the two MB models ``$v_{\mathrm{circ}}$" and ``Jeans" are shown as red dot-dashed and purple dashed, respectively.}
\label{fig:vmoments-comparison-HALOB-hydro}
\end{center}
\end{figure}

Observables in the context of DM searches are actually not directly sensitive to the full speed distribution, but to a few of its moments, as discussed in \citesec{ssec:observables}. Here we make one-to-one comparisons between the Eddington predictions---computed as described in \citesec{ssec:observables}---and the moments extracted from the simulations. More precisely, we compute the mean value of $v^n$ for particles enclosed in spherical shells, for $n=-2,-1,1,2$. The results as a function of radius are shown in \citefig{fig:vmoments-comparison-HALOB-DMO} for the DMO run and in \citefig{fig:vmoments-comparison-HALOB-hydro} for the hydrodynamical run. In each figure, we show the radial profiles of the moments in the left panel, and the relative difference with respect to the simulation in the right panel.

The agreement between the Eddington predictions and the simulations outputs for this sample of moments is particularly good when compared to the simplifying assumptions of the model. In the DMO case (see \citefig{fig:vmoments-comparison-HALOB-DMO}), the model even reaches a precision better than $\sim$10-20\% over a very broad dynamical range. This is consistent with the qualitative statements based on comparisons at the level of the PSDF and of the speed distribution, and discussed in \citesecs{ssec:comp_psdfs} and \ref{ssec:comp_fv}. Close to the resolution limit of the simulation, the relative error increases, but only up to 30\% for the $n=-2$ moment which is more sensitive to the low-speed tail of the distribution. Such a trend is also found in our other test halos---see \citeapp{app:other_halos}. This means that a simple self-consistent equilibrium model based on maximal symmetry is already able to capture the main dynamical properties of a realistic DM-dominated galaxy. This is very promising in the context of DM searches and justifies using the Eddington method to derive predictions for important observables. This somewhat quantifies the theoretical uncertainties one would have when translating a galactic mass model in terms of velocity-dependent observables in the context of DM searches in DM-dominated systems, on top of the uncertainties on the mass model itself.

In the hydro case (see \citefig{fig:vmoments-comparison-HALOB-hydro}), the precision of the model slightly degrades. Although it remains below $\sim 20$\% for the $|n|=1$ moments, the $n=-2$ moment is strongly overestimated, by $\sim 50$\% within the inner 2~kpc. This is consistent with the fact that the peak of the velocity distribution is significantly underestimated, as discussed in \citesec{ssec:comp_fv}. Except for that particular moment, the model is still able to achieve a precision of order $\sim 20$\% over a broad dynamical range. This trend is basically verified in our other test halos---\citeapp{app:other_halos}---except for Mochima, for which the model performs quite well even in the hydro case, with a precision $\lesssim 10$\% over the whole halo for all velocity moments. Again, this somewhat quantifies the theoretical uncertainties associated with the Eddington inversion applied to a realistic galaxy in which baryons dominate the potential at the center. We see that the precision is of order $\lesssim 50$\% for the $n=-2$ moment (down to $\sim 10$\% for one out of three simulations), which is highly sensitive to the low-speed tail of the speed DF, and can reach $\lesssim 20$\% for other moments. This paves the way toward much more controlled predictions on top of defining a next-to-standard type of halo models enabling a decent representation of galactic halos in phase space.

It should be noted that the profiles of the second moment $\left\langle v^2\right\rangle(r)$ obtained with the Eddington method (solid black curve) and the Jeans-based MB model (dashed magenta curve) are different, especially beyond 10~kpc. This may look surprising since the velocity dispersion solution of the Jeans equation is by construction equal in both cases. We recall however that we have used the solution to the Jeans equation to define the \textit{peak speed} rather than the velocity dispersion of the MB distribution, and that these two quantities are different when the MB distribution is truncated at the escape velocity.\footnote{Furthermore, the Eddington DF has been regularized to get rid of the divergence at ${\cal E}=0$, which very slightly modifies the result compared to the non-regularized case---see Ref.~\cite{LacroixEtAl2018} for an exhaustive discussion.}
Regardless, the results of the Jeans-based MB and Eddington models are rather close, with the latter still performing slightly better toward the centers of halos. The Eddington formalism, however, remains the only one fully self-contained and self-consistent, which we argue provides an additional theoretical motivation. Moreover, it encapsulates a full PSDF, while the Jeans-MB model does not.

\subsection{Moments of the relative speed distribution}
\label{ssec:moments_vrel}

\begin{figure}[t!]
\begin{center}
\includegraphics[width=0.49\textwidth]{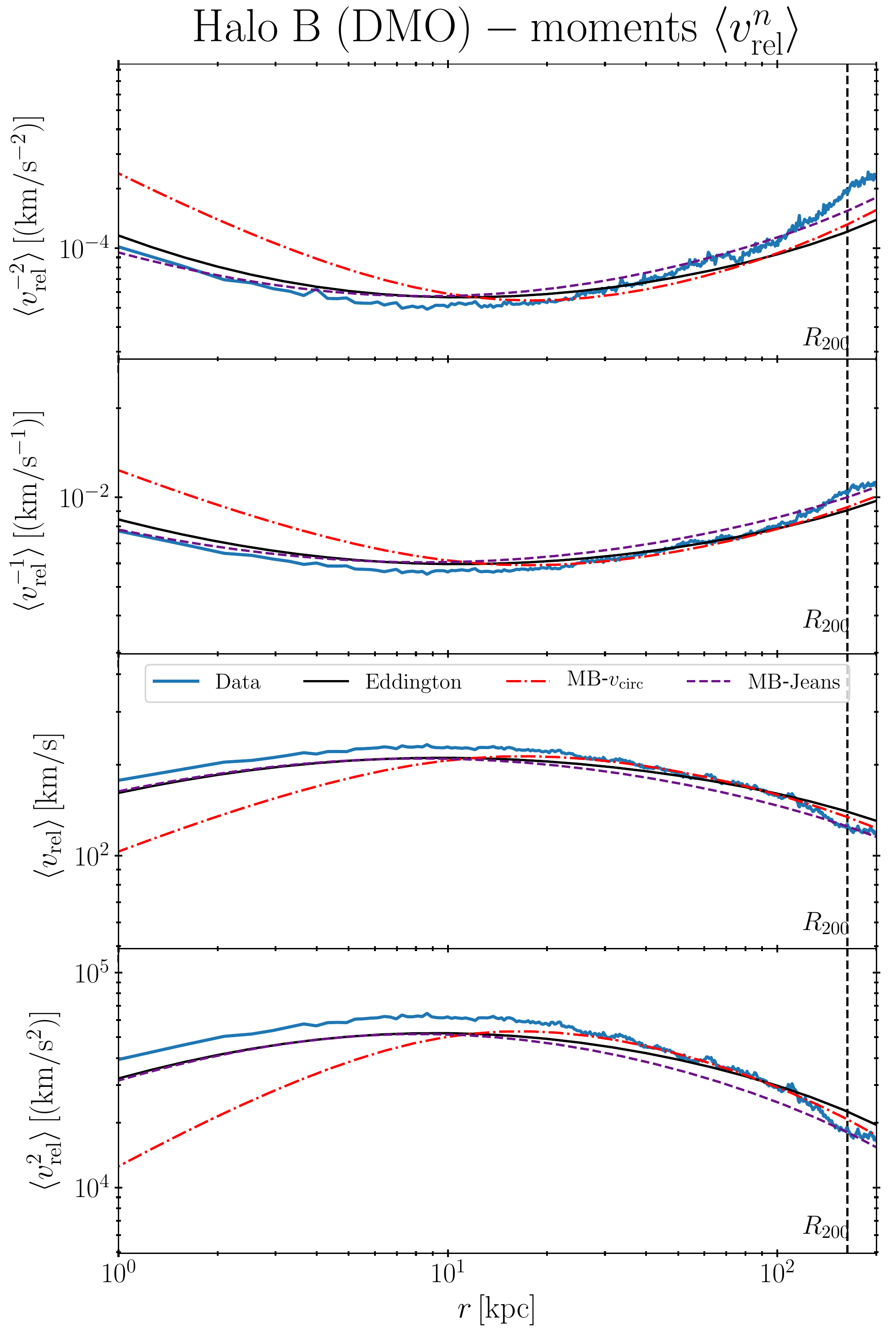}  \hfill  
\includegraphics[width=0.49\textwidth]{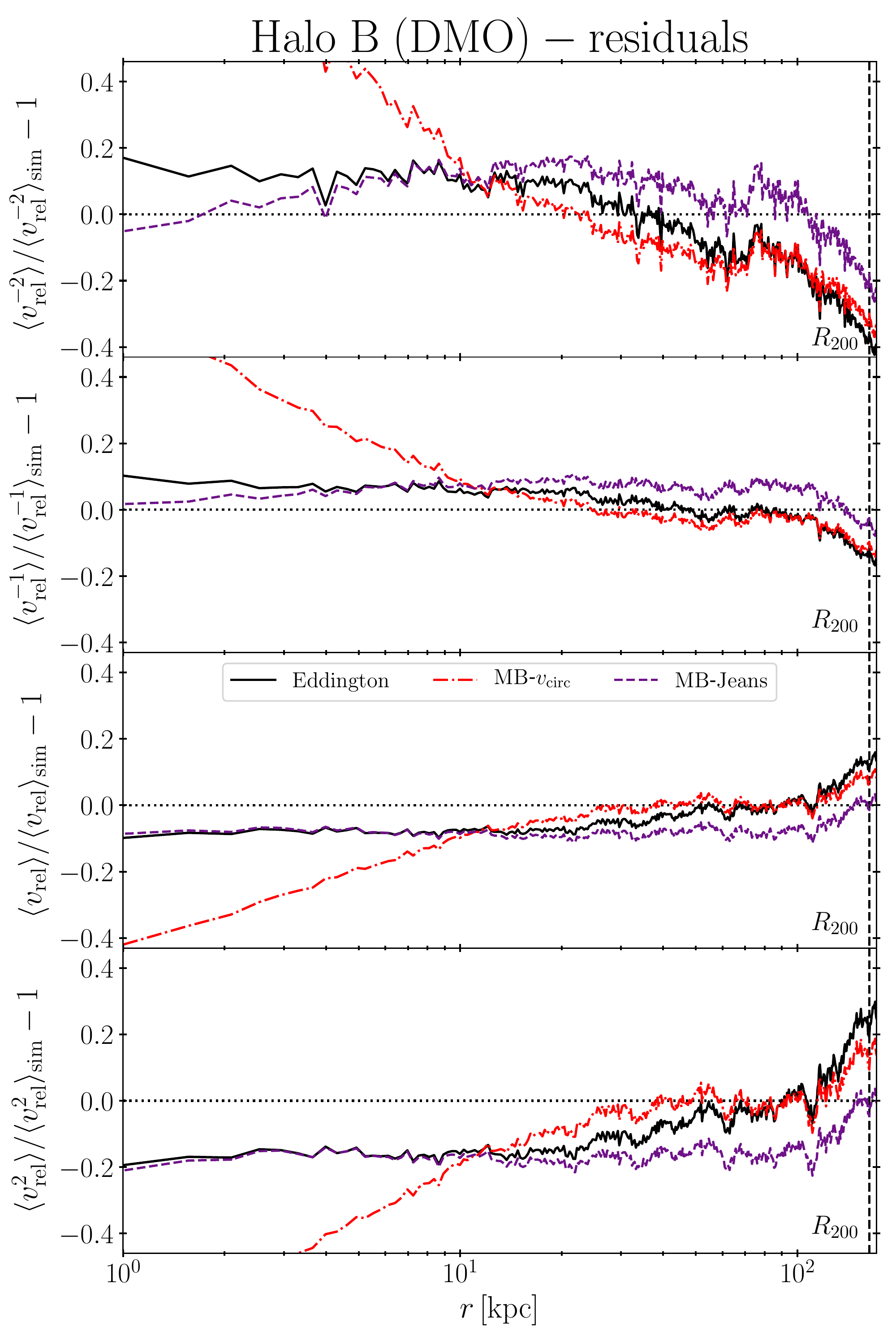}
\caption{\small Comparison of the moments of the \textit{relative} speed distribution from the Eddington method and from the simulation for Halo~B (left panel) and associated residuals with respect to the simulation (right panel), for the DMO run. Simulation outputs are shown as blue solid curves. Predictions obtained with the Eddington inversion method are displayed as black solid curves, while the two MB models ``$v_{\mathrm{circ}}$" and ``Jeans" are shown as red dot-dashed and purple dashed, respectively.}
\label{fig:v-rel-moments-comparison-HALOB-DMO}
\end{center}
\end{figure}

\begin{figure}[t!]
\begin{center}
\includegraphics[width=0.49\textwidth]{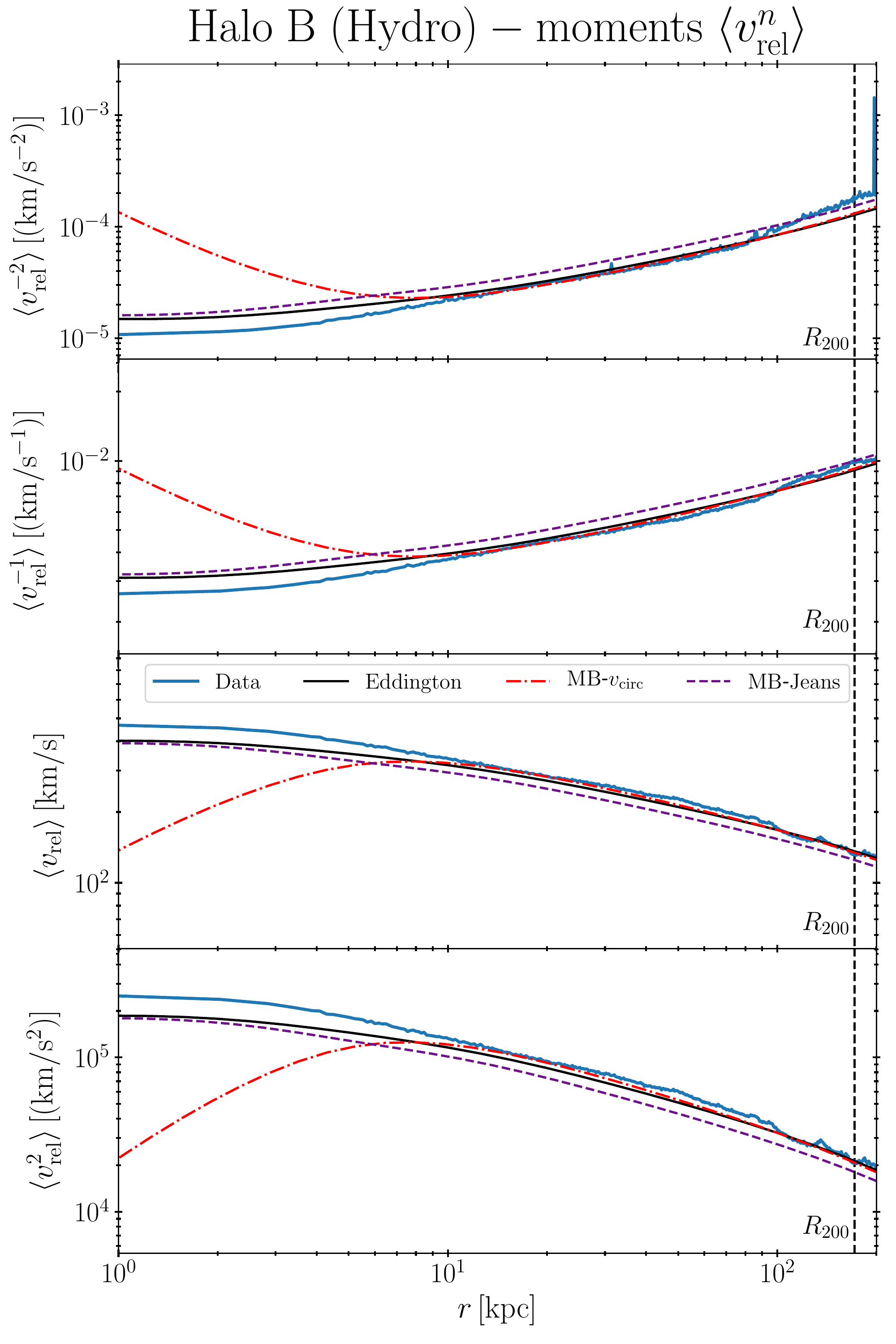}  \hfill
\includegraphics[width=0.49\textwidth]{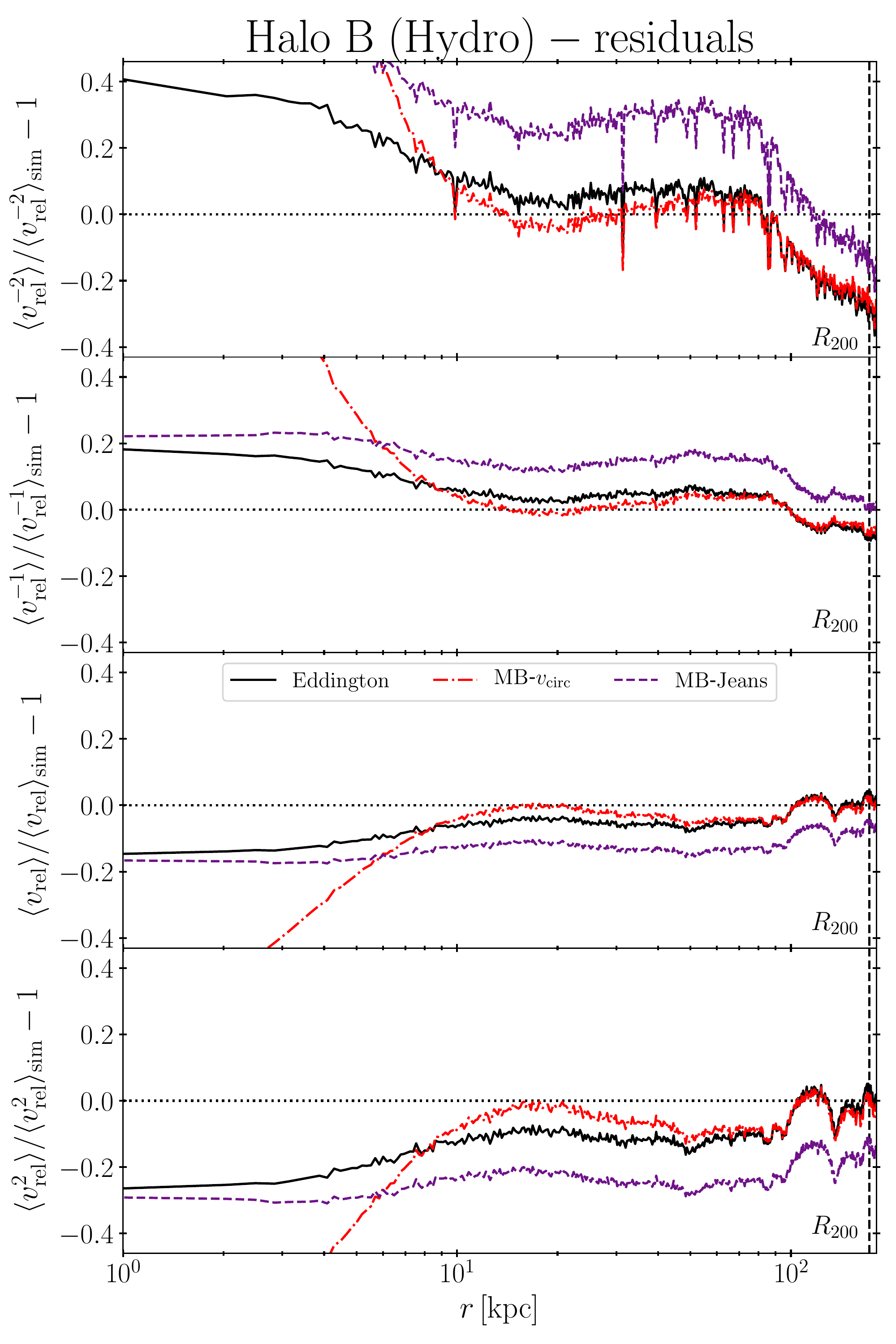}
\caption{\small Comparison of the moments of the \textit{relative} speed distribution from the Eddington method and from the simulation for Halo~B (left panel) and associated residuals with respect to the simulation (right panel), for the hydro run. Simulation outputs are shown as blue solid curves. Predictions obtained with the Eddington inversion method are displayed as black solid curves, while the two MB models ``$v_{\mathrm{circ}}$" and ``Jeans" are shown as red dot-dashed and purple dashed, respectively.}
\label{fig:v-rel-moments-comparison-HALOB-hydro}
\end{center}
\end{figure}

In the context of indirect DM searches involving self-annihilation, the actual physical quantities of interest are the moments of the relative speed distribution. These are derived from the Eddington method as presented in \citesec{ssec:observables}. In the simulations, the moments of the relative speed are computed by evaluating the mean of $v_{\rm rel}^{n}$, where
\ben
v_{\rm rel}=\sqrt{(v_{x,i}-v_{x,j})^2+(v_{y,i}-v_{y,j})^2+(v_{z,i}-v_{z,j})^2}\;\;\; \forall \; i\neq j,
\een   
where indices $x$, $y$, $z$ refer to the Cartesian coordinates in the reference frame, and indices $i$ and $j$ iterate over all the particles inside the radial bin of interest, in the same way as for the moments of the speed distribution. Again we consider $n=-2,-1,1,2$. Our results are shown as a function of radius in \citefig{fig:v-rel-moments-comparison-HALOB-DMO} for the DMO run and in \citefig{fig:v-rel-moments-comparison-HALOB-hydro} for the hydrodynamical run, with the radial profiles in the left panel and the residuals in the right panel.

The agreement between the Eddington prediction and the reconstructed relative moments from the simulation is again better than for the speed distribution---and slightly better than for the moments of the speed distribution---as we consistently obtain $\sim 10$-$20\%$ throughout the halos for all the moments of interest in the DMO case. In the hydro case, as for the speed moments, the precision of the model slightly degrades, especially in the $n=-2$ moment (in the inner regions), which is the most sensitive to the low-speed tail of the speed DF. However, except for that specific moment for which we get a precision $\lesssim 40$\%, the model reaches a precision of $\lesssim 20$\% over the whole halo for the other moments. This trend is confirmed in the other test halos, except for Mochima, for which the model still performs remarkably well (precision $\lesssim 10$\%)---see \citeapp{app:other_halos}.

If we now make comparisons of the Eddington model with the other two MB models, we can notice that in the DMO case, the MB-Jeans model gives slightly better results than the Eddington one in the central part of the halo. However, it fails to predict the moments with a relevant precision in the hydro case (inner parts of halos). In the Mochima simulation, the Eddington model performs better in describing the $v_{\rm rel}$ moments in both the DMO and hydro runs. Overall, we can fairly conclude that the Eddington prediction is more reliable as it applies to strongly different configurations, either with DM or baryon domination, providing similar precision.

\section{Summary and conclusion}
\label{sec:concl}
In this work, we tried to quantify the level of predictivity and the relevance of some isotropic models of velocity distribution functions, by comparing their predictions for several observables with direct measurements in highly resolved cosmological simulations, providing realistic test galaxies where both the dark matter and the baryons are dynamically linked through their mutual gravitational interactions. The main question we addressed is the following: can a reliable, though simplified, galactic mass model be translated into as reliable predictions for the speed distribution and related moments? Answering this question and further quantifying the reliability of the procedure is important in a context in which (i) dark matter searches intensify on galactic scales, and (ii) observational data accumulate which can better constrain the dark matter content of target objects or structures. Moreover, discovery prospects as well as exclusion limits on specific dark matter scenarios would certainly benefit from better estimates or control of theoretical uncertainties.

In particular, we have tested a complete model---the Eddington inversion model (see Refs.~\cite{Eddington1916,BinneyTremaine2008} for a general presentation, and Ref.~\cite{LacroixEtAl2018} for a detailed review)---encapsulating a full phase-space description of the dark matter lying in a self-gravitating object, built from first principles while based on several simplifying assumptions: dynamical equilibrium, spherical symmetry, and isotropy. This model, a generic solution to the collisionless Boltzmann equation, allows one to derive the phase-space distribution function of dark matter from the knowledge of its mass density profile and of the full gravitational potential of the system (both required to be spherically symmetric). Therefore, it can be fully derived from a galactic mass model, where the mass density distributions of all components are specified. We have compared this full phase-space distribution model with more ad hoc models for the velocity distribution only, based on the declension of the Maxwell-Boltzmann approximation; one inspired from the isothermal sphere where the peak velocity is set to the circular velocity, and another one in which the peak velocity derives from the velocity dispersion calculated by consistently solving the Jeans equation. These models were used to predict the speed distribution function of a system and several relevant speed moments, as well as relative speed moments. These models are fully described in \citesec{sec:theory}.

Galactic mass models similar to the one of Ref.~\cite{McMillan2017}, including a dark matter profile and several components for baryons, were fitted on three different highly resolved zoom-in cosmological simulations, described in Ref.~\cite{MollitorEtAl2015} and Ref.~\cite{ArturoSimu}. These simulations were used in both their dark matter-only and their hydrodynamical configurations, the former resembling a would-be giant isolated dwarf spheroidal galaxies, and the latter resembling spiral galaxies similar to the Milky Way (the level of ``Milky Way-likeness'' is not essential in this work). The core of the paper focused on one of them, dubbed Halo~B, but similar results were obtained with the other two, Halo~C and Mochima, which can be found in \citeapp{app:other_halos}. The main features of these simulations are detailed in \citesec{sec:sims}, which show that dynamical equilibrium, spherical symmetry, and isotropy can be considered as reasonable assumptions in the perspective of model building, but that there are also clear departures thereof.

We compared the model predictions for several velocity-dependent observables directly with the simulation data in \citesec{sec:comp}. In particular, we compared the Eddington phase-space distribution function for dark matter with the actual coarse-grained phase-space distribution function of the simulation projected on isoenergy surfaces, that we called pseudo-phase-space distribution function. Although the latter does not carry the full information contained in the system which actually departs from local equilibrium and from the symmetries assumed in the model, there is a priori no reason why it should match the model prediction. However, we found reasonably good agreement, especially in the ability of the Eddington model to capture effects related to strong variations in the shape of the dark matter density profile (from cored to cuspy halos). The Eddington model performs well in the dark matter-only configurations, but overpredicts the abundance of low-kinetic energy particles (equivalently large relative energy ${\cal E}$) in the central regions of the halos in the hydrodynamical configurations, which further imprints the velocity distribution, as we checked afterward. In terms of the latter, we found that:
\bi
\item In the dark matter-only galaxies, the Eddington model gives a satisfactory description of the velocity distribution of dark matter across the full halos, and performs slightly better than models based on the Maxwell-Boltzmann approximation. This was expected because the latter contain less consistent physical information. We stress again that our Maxwell-Boltzmann models are not generalized Maxwellian fits to the data, but models based on simplifying assumptions and used to make predictions. Bell-shaped fitting functions using the peak and width of the speed distribution as free parameters are expected to match the data quite reasonably, but have, by definition, no predictive power.
\item In the hydrodynamical runs, the agreement between the Eddington model and the true measured velocity distributions degrades, especially in the central regions of galaxies, where the peak of the velocity distribution is underpredicted by $\sim 20$\%---this is consistent with the overprediction of the low-kinetic energy particles in the phase-space distribution function. Note, however, that the agreement is better in the Mochima galaxy. Still, overall, the Eddington model performs better than the Maxwell-Boltzmann models, especially in the central regions of the simulated halos (as expected).
\ei
We then derived the predictions of the velocity and relative velocity moments, $\langle v^n\rangle(r)$ and $\langle v_{\rm rel}^n\rangle(r)$, respectively, with $n=\pm 1, \pm 2$, and compared them with the simulation data. These moments are directly related to interesting observables in the context of dark matter searches, and associated theoretical uncertainties translate rather straightforwardly in terms of these observables. To summarize, we found that:
\bi
\item For $\langle v^n\rangle(r)$ in the dark matter-only galaxies, the Eddington model gives a precision of $\lesssim 10$\% for $n=\pm 1$, and $\lesssim 20$\% for $n=\pm 2$. Overall, this is slightly better (much better for $n=-2$) than the Maxwell-Boltzmann model using the Jeans dispersion, which also provides a decent matching of the moments in the centers of halos, but better than the isothermal model. High moments are sensitive to the low tail ($n=-2$) or the high tail ($n=2$) of the velocity distribution, hence the decreased precision for them. As expected, however, $n=-2$ is better reproduced by the Eddington model in the central parts of halos.
\item For $\langle v_{\rm rel}^n\rangle(r)$ still in dark matter-only galaxies, the same trend as for  $\langle v^n\rangle(r)$ is recovered, with only little improvement.
\item Moving to hydrodynamical runs for $\langle v^n\rangle(r)$, the overall precision of all models is degraded especially in the central regions of halos, consistently with what was observed in the velocity distributions. For the Eddington model, the precision decreases down to $\lesssim 20$\% for $n=\pm 1$, and $\lesssim 40$\% for $n=\pm 2$, in the central parts of halos. We note that negative moments are systematically overpredicted, and positive moments underpredicted, consistently with the underprediction of the peak of the velocity distribution. However, we note that for the Mochima galaxy, the agreement is better, with a precision $\lesssim 10$\% for all moments. It is also worth noticing that overall, the Eddington model performs much better than the Maxwell-Boltzmann models, especially in the central parts of halos, even when its precision is smallest ($n=-2$).
\item For $\langle v_{\rm rel}^n\rangle(r)$ in the hydrodynamical runs, we recover the same trend as for  $\langle v^n\rangle(r)$ just above, while still with very slight improvement.
\item We may speculate that the departure of the Eddington model from the data in the hydrodynamical runs is somewhat related to the efficiency of the baryonic feedback in the central parts of halos. We will more deeply investigate this potential correlation in a dedicated work.
\ei

As a general conclusion, we can reasonably estimate that the Eddington model provides a fairly good description of the phase-space distribution function of dark matter in galactic structures, reaching a precision\footnote{\change{We remind that such a precision is only related to the prediction
    of phase-space distribution functions from galactic mass models. It is reached starting from
    mass models that are rather accurate in spite of being smooth and forced to spherical symmetry,
    since they are fitted to the full simulation data. With real stellar kinematic data,
    additional uncertainties related to the mass model reconstruction itself come about, but their
    impact on velocity moments can be easily derived from the Eddington inversion. Current and
    future surveys, like the \textit{Gaia} survey, are expected to decrease these uncertainties, but they
    are presently still at a level comparable to the errors discussed in this paper
    (see \eg~\cite{CautunEtAl2020}).}}
of $\sim 10$-20\% for velocity or relative velocity moments of order $n=\pm 1,2$.
It may perform better in describing dark matter-only systems than those with baryonic domination at their centers, with a precision degrading by $\sim 10$\% for the latter---this is not generic though, as our Mochima galaxy was still very well described. It is rather surprising, and even remarkable, that such a simple model can capture the dark matter dynamics so well, especially when one considers the strong assumptions it is built upon. Indeed, none of our simulated objects exhibits perfect dynamical relaxation, spherical symmetry, nor isotropy. Still, the model is able to capture their main dynamical features. We emphasize that the Eddington model, in this context, provides a better description of realistic systems than typical declensions of the Maxwell-Boltzmann approximation used in the literature (in the sense of models, not Gaussian fitting functions). This is rather satisfactory from the theoretical point of view, since the latter lack of solid theoretical grounds in this particular context (\eg~\cite{KazantzidisEtAl2004}).

This work provides a quantitative estimate of the theoretical uncertainties affecting the Eddington inversion in the context of dark matter searches, both in dark matter-dominated objects and in spiral galaxies similar to the Milky Way. We stress that these uncertainties do account for departures from local equilibrium, which are at play in our virtual galaxies. While these results can be straightforwardly used for Galactic searches,\footnote{In Ref.~\cite{LacroixEtAl2018}, the Eddington model was fully applied to the Galactic mass model fitted on recent kinematic data by Ref.~\cite{McMillan2017}. The associated predictions can therefore be assigned the theoretical uncertainties derived in the present work.} it is also tempting to speculate about the relevance of our results in the dark matter-only configurations to dwarf spheroidal galaxies. The main differences come mostly in terms of total mass, of the different impact of baryons (there are still a few baryons in dwarf galaxies), and from the fact that our objects are rather isolated. Anyway, this still may provide a qualitative estimate of the uncertainties on the dwarf galaxy scale.

Recently, the authors of Ref.~\cite{CallinghamEtAl2020} have studied models based on action-angle variables. The main difficulty resides in the fact that there is no generic theoretical method to predict the forms of the action distributions for dark matter, and the latter are often tuned in a semi-empirical way \cite{PostiEtAl2015,BinneyEtAl2015,ColeEtAl2017}. In Ref.~\cite{CallinghamEtAl2020}, the authors have tabulated these distributions directly from the Auriga cosmological simulations \cite{GrandEtAl2017}, using both dark matter-only and hydrodynamical runs. The advantage of actions is that they are adiabatic invariants, so the action distributions should in principle match between the dark matter-only and hydrodynamical runs, in such a way that the latter's properties could be inferred simply by adding the baryonic gravitational potential to the former. All relevant physical observables, including density profiles and anisotropic velocity distributions, can then be calculated from the action distributions and the total gravitational potential. The hope is then to infer the properties of dark matter in a complex system including baryons from the more simple dark matter-only system. However, the authors still find some residual systematic effects attributed to radial action losses (\ie~departure from adiabatic invariance). They then had to calibrate their action distributions on their hydrodynamical runs, thus potentially inducing some dependence on the treatment of baryonic physics---using other recipes for baryonic feedback might therefore lead to changes in the template distributions. Overall, they still get a good precision in reproducing the velocity distributions in solar-like regions of several test simulated galaxies (run in the same baryonic framework, though), seemingly of the order of $\lesssim 10$\% on the first moments (see their Fig.~9), similar to our model's precision; however, results for more central or more distant regions are not available. They have further applied their calibrated method to the Milky Way mass model derived in Ref.~\cite{CautunEtAl2020}, but do not provide predictions for the velocity moments. This interesting study is complementary to ours, which is more minimal since it only relies on a galaxy mass model, and is not calibrated on simulation outputs.

There is obviously room for improvement. Related to the work presented in this paper, it would be interesting to check the level of improvement gained by using anisotropic models, in particular those that can predict the anisotropy \cite{HunterQian1993,Petac2019axisymmetric}, which we leave for a future study. Ultimately and more generally, it would be useful to make cross-comparisons of the predictions obtained for observables relevant to dark matter searches by all the different methods mentioned above, including those based on action distributions, in order to single out the best compromise between calculation complexity and precision.

\acknowledgments
We thank Pol Mollitor for fruitful discussions and continuous exchanges on this project. We also wish to thank our partners of the GaDaMa project, B. Famaey and the ObAS group in Strasbourg, and P. Salati and the LAPTh group in Annecy. We are also grateful to the participants of the previous editions of the {\em News from the Dark} workshop series, with whom we discussed intensively several aspects of this work. We acknowledge financial support by French programs CNRS-INSU/PNHE-PNCG, ANR project ANR-18-CE31-0006, the OCEVU Labex (ANR-11-LABX-0060), the European Union's Horizon 2020 research and innovation program under Marie Sk\l{}odowska-Curie grant agreements No 690575 and  No. 674896; beside recurrent institutional funding by CNRS, the University of Montpellier, the University of Aix-Marseille, and the University of Savoie-Mont-Blanc. We acknowledge Centre de Calcul Intensif d’Aix-Marseille for granting access to its high performance computing resources. TL has received funding from the European Union's Horizon 2020 research and innovation programme under the Marie Sk\l{}odowska-Curie grant agreement No. 713366. The work of TL has also been supported by the Spanish Agencia Estatal de Investigaci\'{o}n through the grants PGC2018-095161-B-I00, IFT Centro de Excelencia Severo Ochoa SEV-2016-0597, and Red Consolider MultiDark FPA2017-90566-REDC.

\appendix

\section{Fitting functions for the density profiles of dark matter and baryons}
\label{app:mass_models}
Here we recall the functional forms we used to describe the density profiles of the various components of our synthetic halos. Parametric density profiles are indeed easier to use as inputs of the Eddington inversion method than interpolating functions. We fitted the radial profiles of the various components of the synthetic galaxies using the same functional forms as in Ref.~\cite{McMillan2017}. More specifically, we obtained a good description of the simulated data using one stellar bulge, two stellar disks, one gas disk and a DM halo. The DM halo is characterized by a generalized $\alpha\beta\gamma$ profile \cite{Zhao1996}
\ben
 \rho_{\rm DM}(x) = \rho_{\rm s}\,x^{-\gamma}\left(1+x^{\alpha}\right)^{(\gamma-\beta)/\alpha}\,,
 \label{eq:halo}
\een    
where $\rho_{\rm s}$ is the scale density, and $x=r/r_{\mathrm{s}}$, where $r_{\rm s}$ is the scale radius. An NFW profile is recovered
with $(\alpha,\beta,\gamma)=(1,3,1)$. It should be noted that for definiteness we fix $\alpha = 1$ for all DMO fits. The bulge profile reads
\ben
\rho_\mathrm{b}=\frac{\rho_{0,\mathrm{b}}}{(1+r'/r_0)^\alpha}\;
\textrm{exp}\left[-\left(r'/r_{\rm b}\right)^2\right]\,,
\label{eq:bulge}        
\een
where $r' = \sqrt{R^2 + (z/q)^2}$. The variable $q$ determines the oblateness of the bulge, $\rho_{0,\mathrm{b}}$ is a scale density,
and $r_{0}$ and $r_{\rm b}$ are scale lengths. The stellar disks are modeled by exponential
profiles:
\ben
\label{eq:disk}
\rho_{j}(R,z)=\rho_{0,j}\;\textrm{exp}
\left(-\frac{|z|}{z_j}-\frac{R}{R_j}\right),
\een
with scale heights $z_{j}$, scale lengths $R_j$ and central densities
$\rho_{0,j}$, where $j = \mathrm{d, D, g}$ for the thin and thick stellar disks, and gas disk, respectively.

\section{Mass model parameters from the simulation}
\label{app:mass_models_params}

The best-fit parameters associated with the parametric density profiles described above, along with the virial radius $R_{200}$ and the spatial boundary $R_{\mathrm{max}}$, are summarized in \citetab{tab:bf_parameters}.

Here, we also describe the derivation of the uncertainty band on the PSDF from the uncertainty on the inner slope of the DM density profile. We define this uncertainty band on the Eddington prediction for $f(\mathcal{E})$ as follows. We fix the value of $\gamma$ at the $\pm 1\sigma$ confidence limit values from the fit, and we then refit all the other parameters. It should be noted that this is a simplified procedure, and that we do not perform a full scan over all the parameters of the fit. However, the Eddington method is insensitive to the actual functional form used as an input, be it a parametric form or an interpolation of the density profile from the simulation. This is different from the situation in which one fits a functional form for the density profile to kinematic data for instance. Therefore, due to degeneracies between parameters of functional forms of the DM profile, at the level of the PSDF we do not expect strong departures from the result obtained for the best-fit values of the density profile parameters. As a result, the procedure outlined here gives a good approximation of the size of the error on the PSDF predicted with the Eddington method from the estimate of the functional form of the DM density profile. Furthermore, here we mean to highlight the impact of the inner slope of the DM profile, which is in general the main source of theoretical uncertainty for DM observables.

\begin{table}[t!]
\centering
\rotatebox{90}{
\begin{tabular}{|c|c|c|c|c|c|c|}
\hline
                                                  & Mochima (DMO)       & Mochima (hydro)     & Halo B (DMO)      & Halo B (hydro)      & Halo C (DMO)        & HALO C (hydro)      \\ \hline
$R_{200}$ {[}kpc{]}                               & $205.66$            & 199.8               & 163.47            & 177.53              & 176.36              & 182.22              \\ \hline
$R_{\rm max}$ {[}kpc{]}                               & $1696.01$           & 1054.59             & 797.89            & 794.93              & 1674.25             & 1571.92             \\ \hline
$\log_{10}(\rho_{\rm s}\;\mathrm{[M_{\odot}/kpc^3]})$   & $6.352 \pm 0.386$   & $5.539 \pm 0.453$   & $7.143 \pm 0.519$ & $7.570 \pm 0.086$   & $6.812 \pm 0.463$   & $7.586 \pm 0.076$   \\ \hline
$r_s$ {[}kpc{]}                                   & $27.615 \pm 25.663$ & $50.787 \pm 49.471$ & $9.856 \pm 6.148$ & $5.049 \pm 0.524$   & $ 14.916 \pm 8.710$ & $ 4.662 \pm 0.427$  \\ \hline
$\alpha$                                          & 1                   & $1.309 \pm 0.466$   & 1                 & $2.954 \pm 1.001$   & 1                   & $ 2.982 \pm 0.881$  \\ \hline
$\beta$                                           & $2.940 \pm 0.314$   & $3.232 \pm 0.476$   & $2.772 \pm 0.247$ & $2.608 \pm 0.112$   & $2.819 \pm 0.300$   & $2.460 \pm 0.085$   \\ \hline
$\gamma$                                          & $1.287 \pm 0.083$   & $1.718 \pm 0.038$   & $1.082 \pm 0.184$ & $0.194 \pm 0.088$   & $1.078 \pm 0.135$   & $0.380 \pm 0.073$   \\ \hline
$\log_{10}(\rho_{\rm b}\;\mathrm{[M_{\odot}/kpc^3]})$   &                     & $10.504 \pm 0.530$  &                   & $8.205  \pm  2.242$ &                     &                  $8.562 \pm 2.633$   \\ \hline
$r_0$ {[}kpc{]}                                   &                     & 0.275               &                   & 0.044               &                     & 0.016               \\ \hline
$\alpha$                                          &                     & 1.501               &                   & 0.267               &                     & 0.092               \\ \hline
$r_{\rm c}$ {[}kpc{]}                                   &                     & 1.05                &                   & 2.799               &                     & 1.699               \\ \hline
$q$                                               &                     & 0.48                &                   & 0.489               &                     & 0.499               \\ \hline
$\log_{10}(\rho_{0,\rm d}\;\mathrm{[M_{\odot}/kpc^3]})$   &                     & $7.634  \pm  2.159$ &                   & $8.198 \pm 2.152$   &                     & $8.941  \pm  2.668$ \\ \hline
$z_{\rm d}$ {[}kpc{]}                                   &                     & 0.371               &                   & 0.299               &                     & 0.686               \\ \hline
$R_{\rm d}$ {[}kpc{]}                                   &                     & 2.302               &                   & 2.799               &                     & 1.993               \\ \hline
$\log_{10}(\rho_{0,\rm D}\;\mathrm{[M_{\odot}/kpc^3]})$   &                     & $7.830 \pm 1.042$   &                   & $8.518  \pm  0.471$ &                     & $8.158  \pm  0.876$ \\ \hline
$z_{\rm D}$ {[}kpc{]}                                   &                     & 1.461               &                   & 0.99                &                     & 1.288               \\ \hline
$R_{\rm D}$ {[}kpc{]}                                   &                     & 3.901               &                   & 4.1                 &                     & 4.987               \\ \hline
$\log_{10}(\rho_{0,\rm g}\;\mathrm{[M_{\odot}/kpc^3]})$ &                     & $8.594 \pm 0.893$   &                   & $9.043 \pm 3.494$   &                     & $ 9.350 \pm 1.064$  \\ \hline
$z_{\rm g}$ {[}kpc{]}                                   &                     & 0.022               &                   & 0.033               &                     & 0.136               \\ \hline
$R_{\rm g}$ {[}kpc{]}                                   &                     & 8.733               &                   & 5.456               &                     & 1.186               \\ \hline
\end{tabular}
}
\caption{\label{tab:bf_parameters}Values of $R_{200}$, $R_{\mathrm{max}}$, and best-fitting mass model parameters for all runs.}
\end{table}

\begin{table}[t!]
\centering
\begin{tabular}{|c|c|c|c|c|c|c|}
\hline
                                                  & Halo B min       & Halo B max    \\ \hline
$\log_{10}(\rho_{\rm s}\;\mathrm{[M_{\odot}/kpc^3]})$   & $7.5686$   & $6.574$    \\ \hline
$r_s$ {[}kpc{]}                                   & $5.913$ & $19.14$   \\ \hline
$\alpha$                                          &   $1\ (\mathrm{fixed})$             & $1\ (\mathrm{fixed})$     \\ \hline
$\beta$                                           & $2.631$   & $3.005$      \\ \hline
$\gamma$                                          & $1.082 - 0.184 = 0.898\ (\mathrm{fixed})$   & $1.082 + 0.184 = 0.266\ (\mathrm{fixed})$     \\ \hline
\end{tabular}
\caption{\label{tab:parameters_uncertainty_bands_HaloB_DMO} Best-fit parameters of the DM density profile, associated with the $\pm 1 \sigma$ containment region on the inner slope $\gamma$, for the Halo B simulation, for the DMO run.}
\end{table}

\begin{table}[t!]
\centering
\begin{tabular}{|c|c|c|c|c|c|c|}
\hline
                                                  &  Mochima min      &  Mochima max      \\ \hline
$\log_{10}(\rho_{\rm s}\;\mathrm{[M_{\odot}/kpc^3]})$   &  $6.713$ & $5.926$    \\ \hline
$r_s$ {[}kpc{]}                                   &  $17.48$ & $46.68$    \\ \hline
$\alpha$                                          &    $1\ (\mathrm{fixed})$            & $1\ (\mathrm{fixed})$     \\ \hline
$\beta$                                           & $2.7378$ & $3.232$     \\ \hline
$\gamma$                                          &  $ 1.287 - 0.083 = 1.204\ (\mathrm{fixed})$ & $1.287 + 0.083 = 1.37\ (\mathrm{fixed})$     \\ \hline
\end{tabular}
\caption{\label{tab:parameters_uncertainty_bands_Mochima_DMO}Best-fit parameters of the DM density profile, associated with the $\pm 1 \sigma$ containment region on the inner slope $\gamma$, for the Mochima simulation, for the DMO run.}
\end{table}

\begin{table}[t!]
\centering
\begin{tabular}{|c|c|c|c|c|c|c|}
\hline
                                                  & Halo B stable       & Halo B stable max      \\ \hline
$\log_{10}(\rho_{\rm s}\;\mathrm{[M_{\odot}/kpc^3]})$   & $7.691$   & $7.635$      \\ \hline
$r_s$ {[}kpc{]}                                   & $5.106$ & $5.486$    \\ \hline
$\alpha$                                          &   $1.9\ (\mathrm{fixed})$          & $1.9 \ (\mathrm{fixed})$       \\ \hline
$\beta$                                           & $2.728$   & $2.756$      \\ \hline
$\gamma$                                          & $0.080$   & $0.08 + 0.06 = 0.14\ (\mathrm{fixed})$     \\ \hline
\end{tabular}
\caption{\label{tab:parameters_uncertainty_bands_HaloB_hydro}Best-fit parameters of the DM density profile, associated with the $\pm 1 \sigma$ containment region on the inner slope $\gamma$, for the Halo B simulation, for the hydro run. Here it should be noted that due to stability criteria, we cannot decrease the value of $\gamma$ within the $1 \sigma$ error, so we bracket the uncertainty from $\gamma$ with the configuration that gives a stable solution to the collisionless Boltzmann equation (`stable'), and the one where we increase $\gamma$ by $1 \sigma$ (`stable max').}
\end{table}

\begin{table}[t!]
\centering
\begin{tabular}{|c|c|c|c|c|c|c|}
\hline
                                                  & Mochima min      &  Mochima max      \\ \hline
$\log_{10}(\rho_{\rm s}\;\mathrm{[M_{\odot}/kpc^3]})$    & $5.007$ & $5.636$    \\ \hline
$r_s$ {[}kpc{]}                                    & $118.3$ & $40.78$    \\ \hline
$\alpha$                                           & $0.904$            & $1.825$     \\ \hline
$\beta$                                          & $4.019$ & $3.020$     \\ \hline
$\gamma$                                          & $1.718 - 0.038 = 1.680 \ (\mathrm{fixed})$ & $1.718 + 0.038 = 1.756 \ (\mathrm{fixed})$     \\ \hline
\end{tabular}
\caption{\label{tab:parameters_uncertainty_bands_Mochima_hydro}Best-fit parameters of the DM density profile, associated with the $\pm 1 \sigma$ containment region on the inner slope $\gamma$, for the Mochima simulation, for the hydro run.}
\end{table}

\clearpage

In addition, for the hydrodynamical run of halo~B, we implement the stability criteria discussed in Sec.~\ref{ssec:sc_range} and Ref.~\cite{LacroixEtAl2018} and restrict ourselves to configurations of parameters that give stable solutions to the collisionless Boltzmann equation, characterized in the Eddington formalism by monotonically increasing ergodic PSDFs. In practice, following the discussions in Ref.~\cite{LacroixEtAl2018}, we fix $\alpha = 1.9$, which controls the sharpness of the transition of the DM profile at the scale radius and is the critical parameter for stability, and fit the other parameters of the DM profile, in particular the inner slope $\gamma$. This gives what we refer to as the `stable' configuration. Then we define the upper limit of the band by fixing $\gamma$ at the upper 1-$\sigma$ limit found in the `stable' fit, and refit the other parameters. The resulting configuration is referred to as `stable max'. It should be noted that the stable configuration defines the lower limit of the uncertainty band since considering a smaller value of the inner slope in this case would result in an unstable configuration.

The corresponding parameters of the modified DM density profiles for Halo B are given in \citetab{tab:parameters_uncertainty_bands_HaloB_DMO} and \citetab{tab:parameters_uncertainty_bands_HaloB_hydro} for the DMO and hydro runs, respectively, and for Mochima in \citetab{tab:parameters_uncertainty_bands_Mochima_DMO} and \citetab{tab:parameters_uncertainty_bands_Mochima_hydro} for the DMO and hydro runs, respectively. The $\gamma$-induced uncertainty on the PSDF derived with the Eddington method is shown in \citefigs{fig:fE_HaloB} and \ref{fig:fE_Mochima} for the Halo B and Mochima simulations, respectively. It should be noted that due to the strong resemblance between Halo B and Halo C, we do not apply the somewhat cumbersome procedure described above to Halo C, since the results would be exactly the same. However, the Mochima halo has a different morphology, and for that one this discussion is important.

\section{Results for other halos in our set of simulations: Halo C and Mochima}
\label{app:other_halos}

In this section we first present figures that illustrate the discussion of the properties of the gravitational potential of the simulated galaxies for the other two halos in our set of simulations, Halo~C and Mochima. The determination of the spatial boundary of the virtual halos is shown in \citefigs{fig:rmax-HALOC} and \ref{fig:rmax-Mochima} for Halo C and Mochima, respectively, for the hydro runs. The method used to derive $R_{\mathrm{max}}$ is described in \citesec{ssec:rmax}---see \citefig{fig:rmax-HALOB} for the corresponding results for Halo B. Then we show in \citefigs{fig:Potential-comparison-HALOC} and \ref{fig:Potential-comparison-Mochima} for Halo~C and Mochima, respectively, for the DMO runs (left panels) and hydro runs (right panels), the excellent agreement up to the virial radius $R_{200}$ of the averaged particle potential in the simulation with the mean potential calculated from the spherically-averaged mass profile from the simulation, and with the one calculated from the best-fitting density profile. See \citesec{ssec:grav_pot} for more details, and \citefig{fig:Potential-comparison-HALOB} for the corresponding results for Halo B.

Next, we provide the figures corresponding to the comparisons between theoretical predictions and simulation outputs discussed in \citesec{sec:comp}, now for the other two halos in our set of simulations, Halo~C and Mochima, in order to illustrate the similarities of our results for all the halos considered in this work. We compare the Eddington PSDF with a ``pseudo-PSDF'' of the simulated system in \citefigs{fig:fE_HaloC} and \ref{fig:fE_Mochima} for the Halo~C and Mochima simulations, respectively, for the DMO runs (left panels) and hydro runs (right panels). See \citefig{fig:fE_HaloB} for the analogous results for Halo~B, and \citesec{ssec:comp_psdfs} for the discussion of the results.

We then compare the Eddington model prediction for the speed distribution $f(v)$ of DM particles with the simulation data for the Halo~C and Mochima simulations in \citefigs{fig:fv-comparison-HALOC} and \ref{fig:fv-comparison-Mochima}, respectively. See \citefig{fig:fv-comparison-HALOB} for the analogous comparison for Halo~B, and \citesec{ssec:comp_fv} for the discussion of the results.


\begin{figure}[t!]
  \centering
    \includegraphics[width=0.49\linewidth]{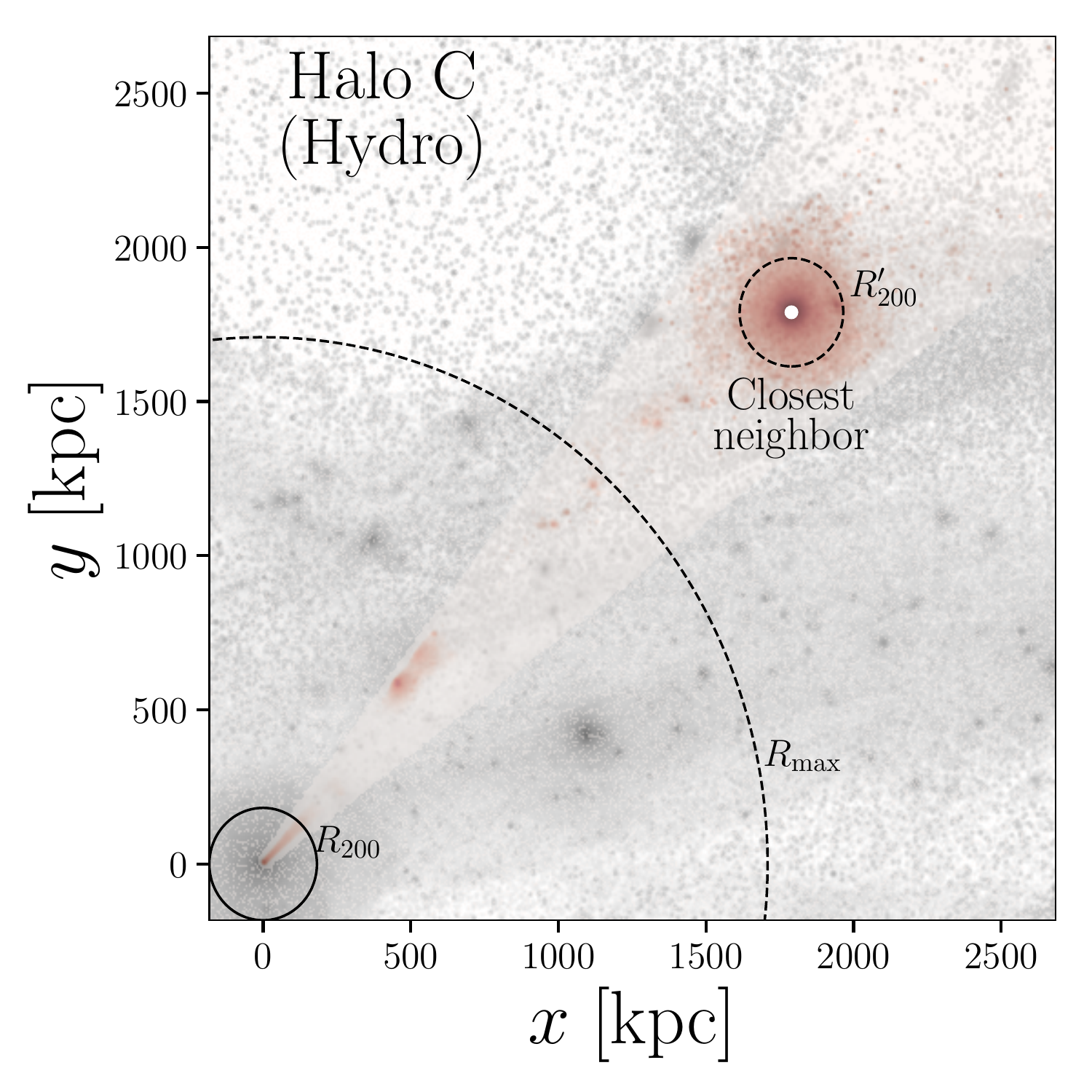} \hfill  \includegraphics[width=0.49\linewidth]{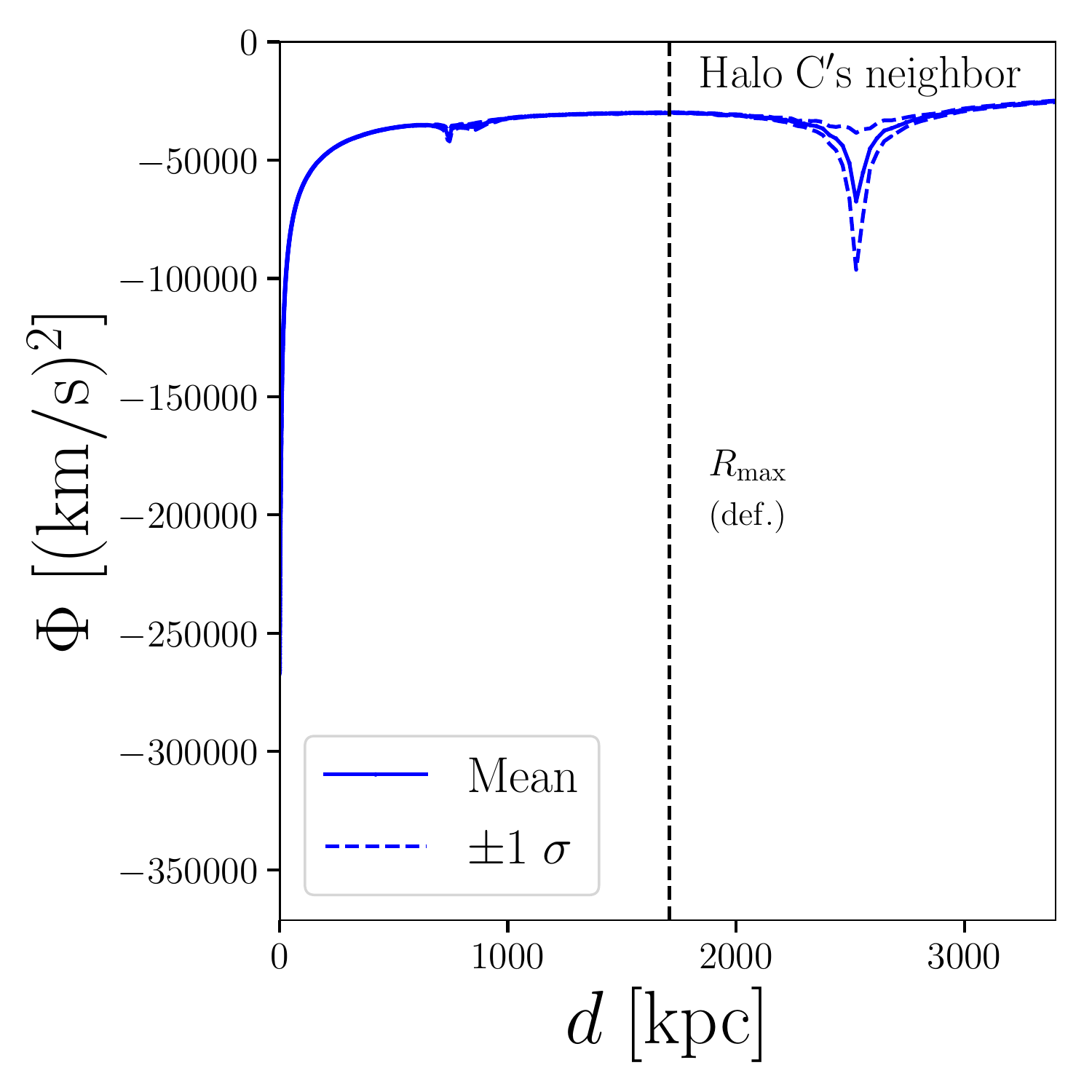}
      \caption{\small Determination of the radial boundary $R_{\mathrm{max}}$ for the Halo C simulation (hydro run). \textbf{Left panel:} Definition of the cone around the imaginary line joining the halo of interest (in the lower left corner) to the next massive neighbor of virial radius $R_{200}'$. The virial radius $R_{200}$ and the boundary radius $R_{\mathrm{max}}$ of the halo of interest are identified by solid and dashed circles, respectively. \textbf{Right panel:} Mean gravitational potential computed in spherical shells inside the cone along the imaginary line between the center of the halo and that of its next massive neighbor, as a function of distance.}
\label{fig:rmax-HALOC} 
\end{figure}

\begin{figure}[t!]
  \centering
    \includegraphics[width=0.49\linewidth]{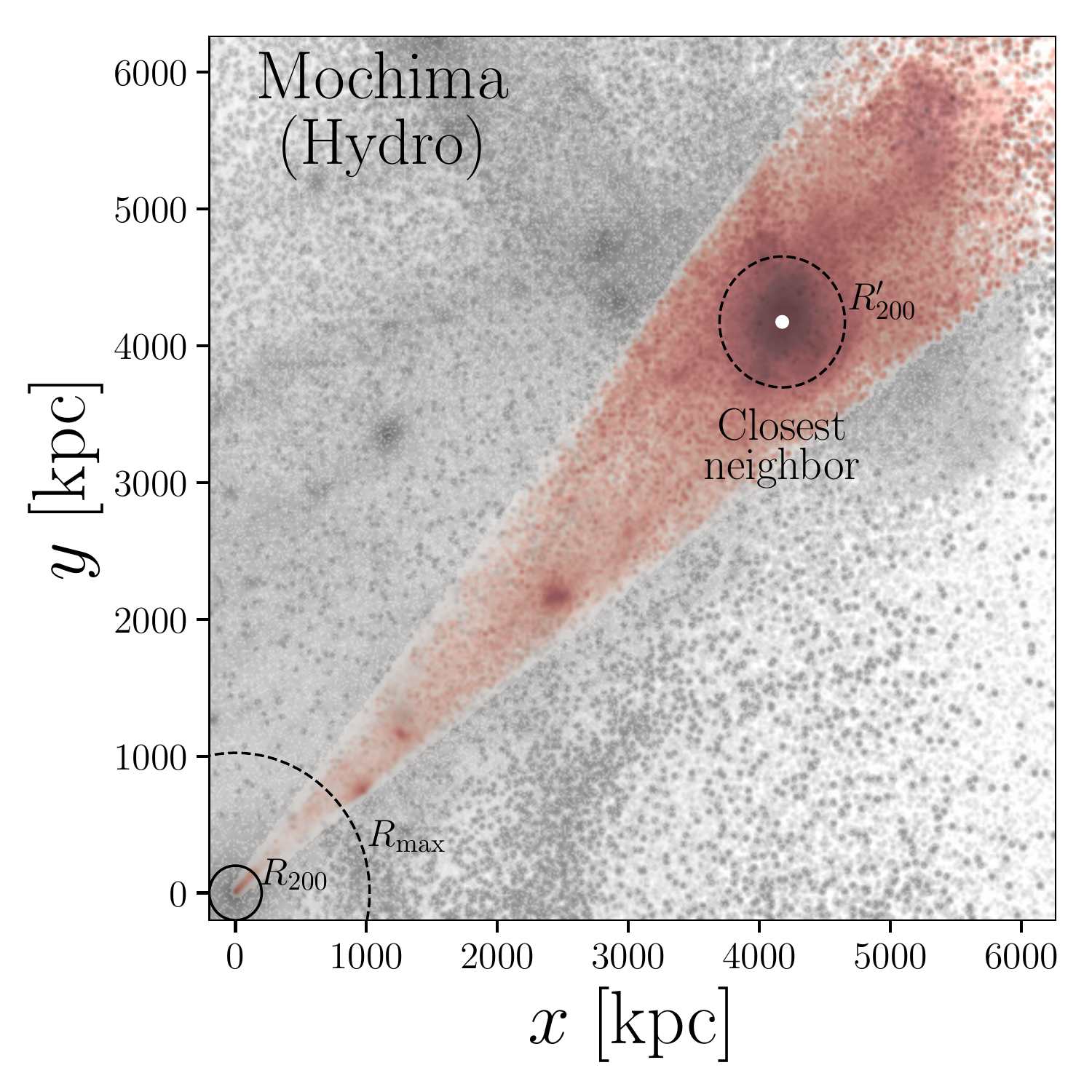} \hfill  \includegraphics[width=0.49\linewidth]{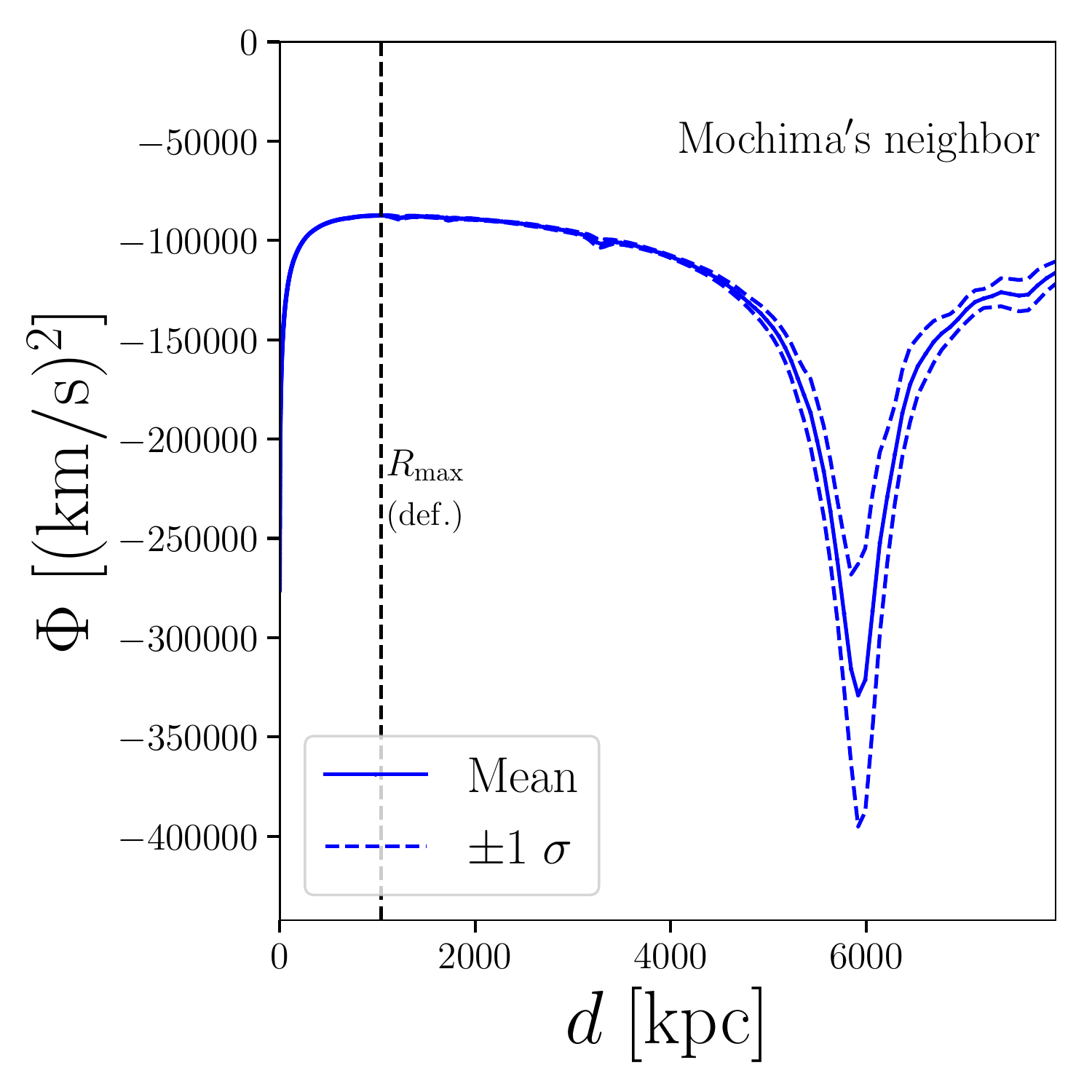}
      \caption{\small Same as \citefig{fig:rmax-HALOC} but for the Mochima simulation.}
\label{fig:rmax-Mochima} 
\end{figure}

\begin{figure}[t!]
\centering
\begin{subfigure}[t]{0.5\textwidth}
  \centering
  \includegraphics[width=1.\linewidth]{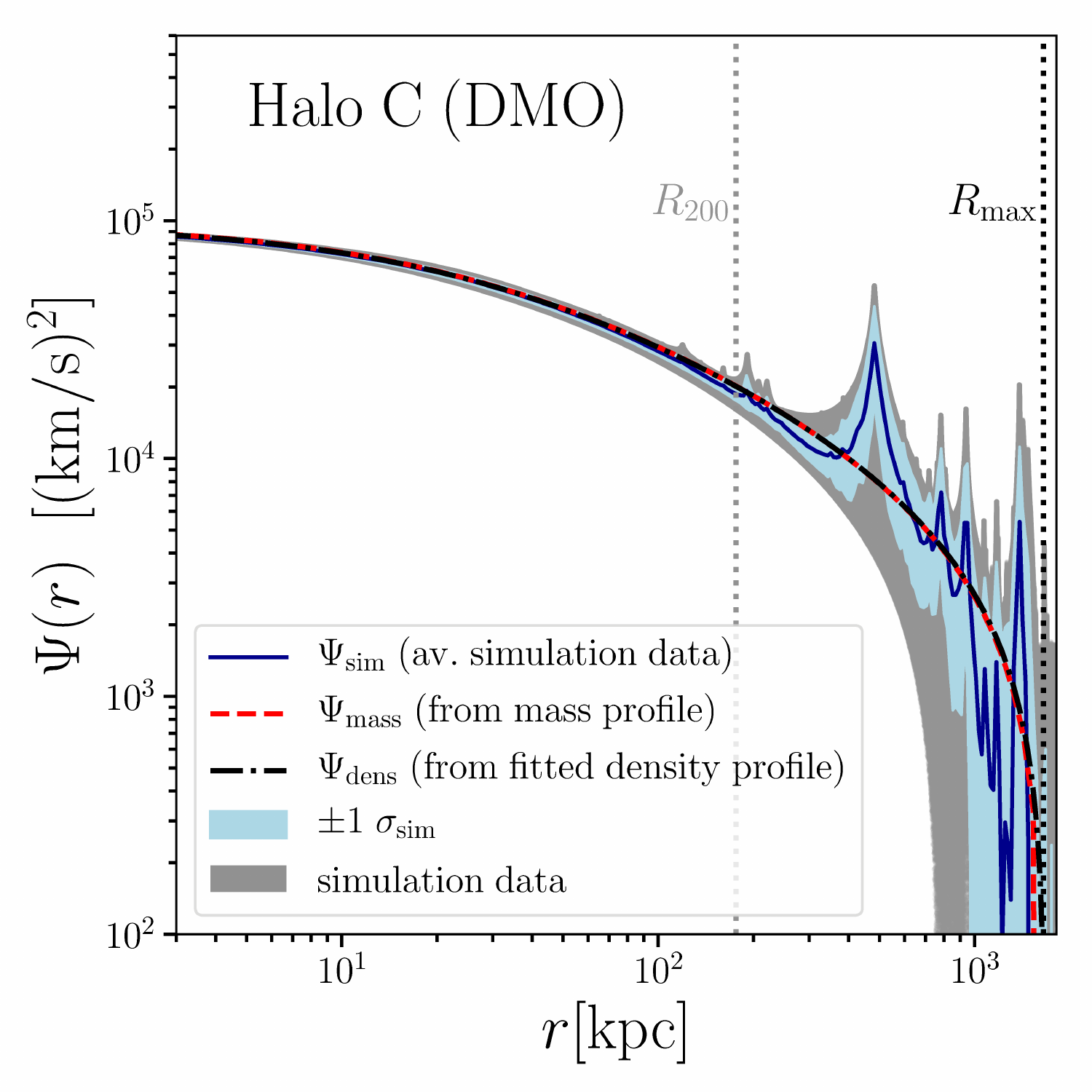}
\end{subfigure}%
\begin{subfigure}[t]{0.5\textwidth}
  \centering
  \includegraphics[width=1.\linewidth]{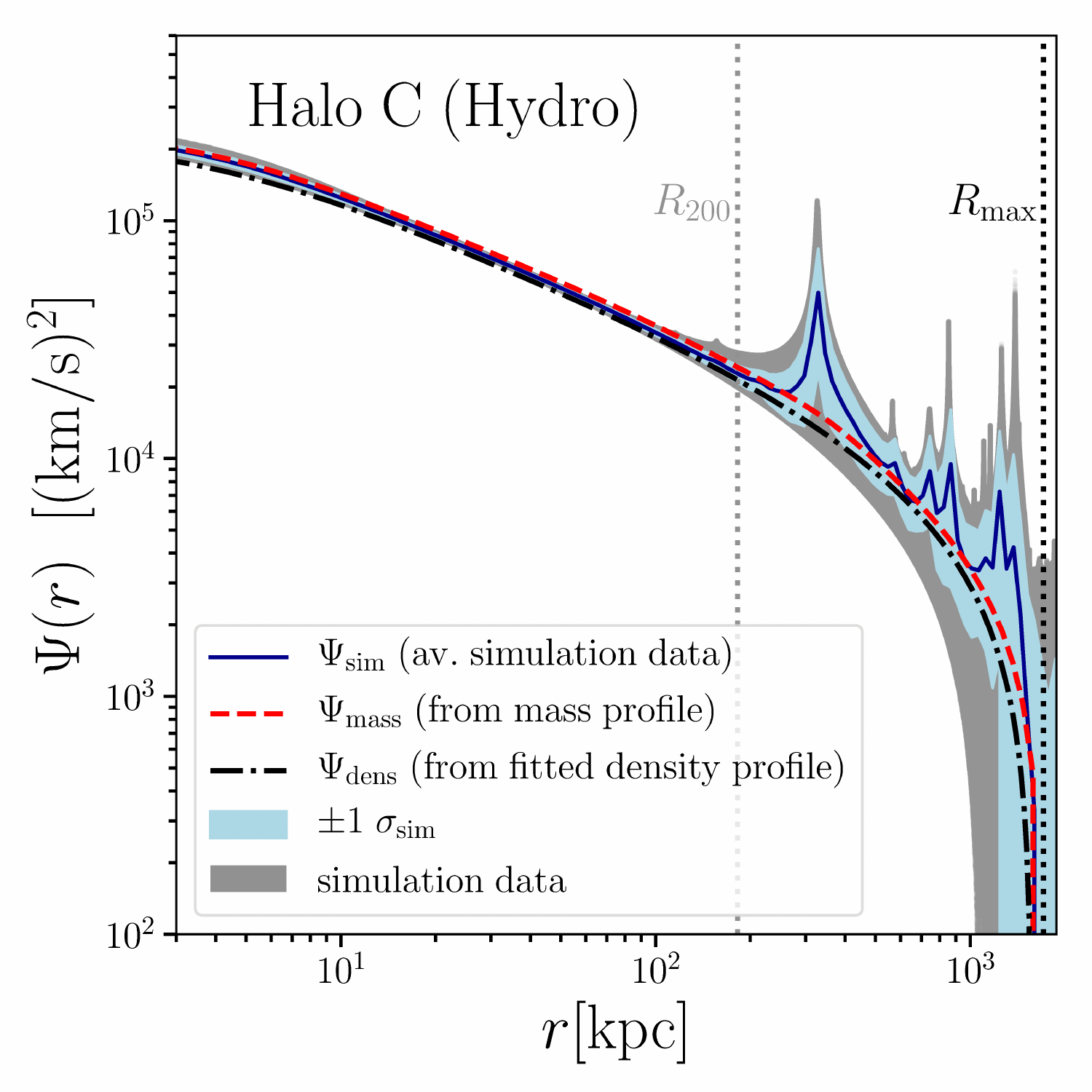}
\end{subfigure}
\caption{\small Relative gravitational potential as a function of galactocentric radius. Comparison between the spherical potential reconstructed from the fitted density profile (black dot-dashed), from the mass profile (red dashed), and averaged from the potentials of particles in the simulation (solid blue), for the DMO (left) and hydro (right) runs, for Halo C. The simulation data are shown as gray bands, and the 1$\sigma$ standard deviation around the mean is shown in light blue.}
\label{fig:Potential-comparison-HALOC}
\end{figure}

\begin{figure}[t!]
\centering
\begin{subfigure}[t]{0.5\textwidth}
  \centering
  \includegraphics[width=1.\linewidth]{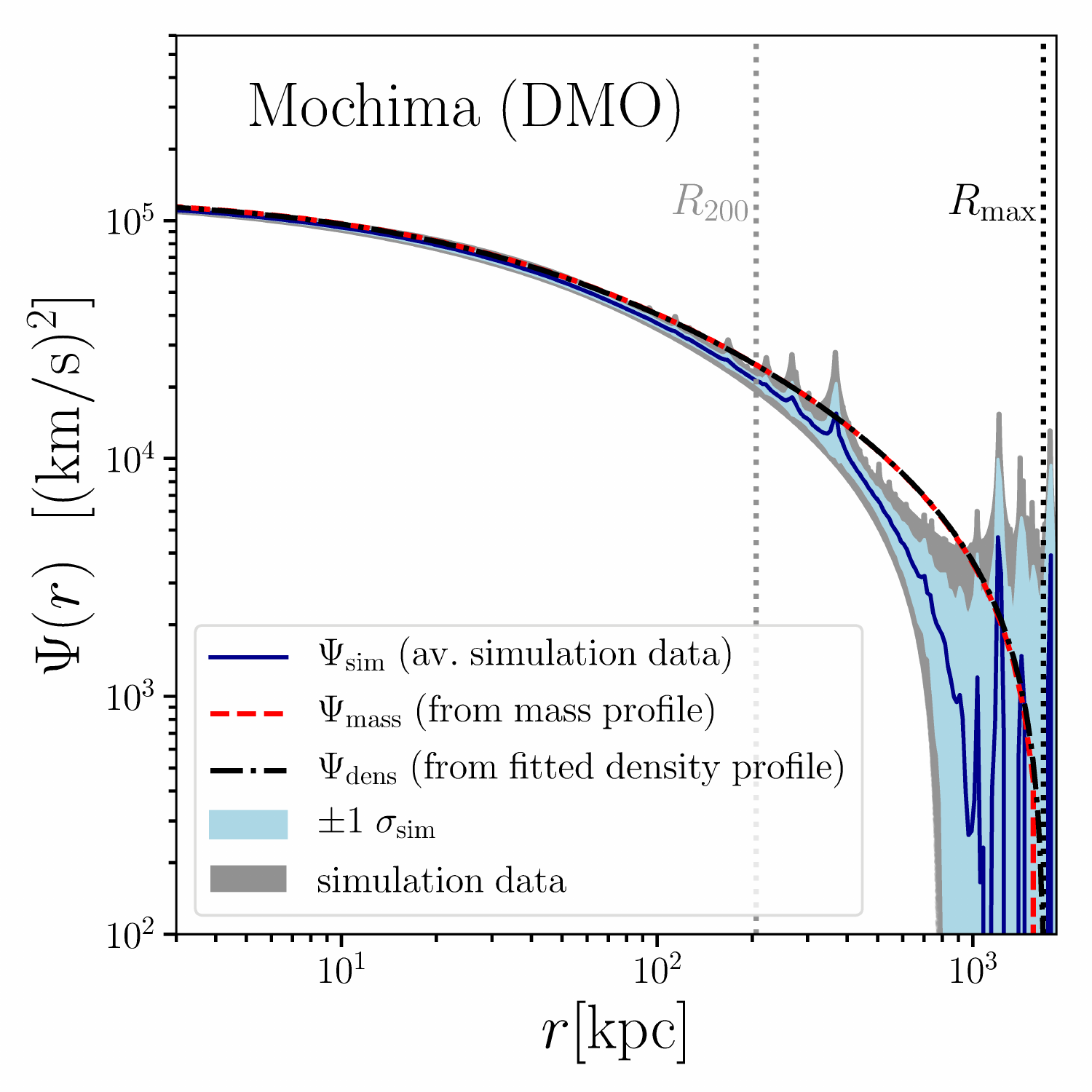}
\end{subfigure}%
\begin{subfigure}[t]{0.5\textwidth}
  \centering
  \includegraphics[width=1.\linewidth]{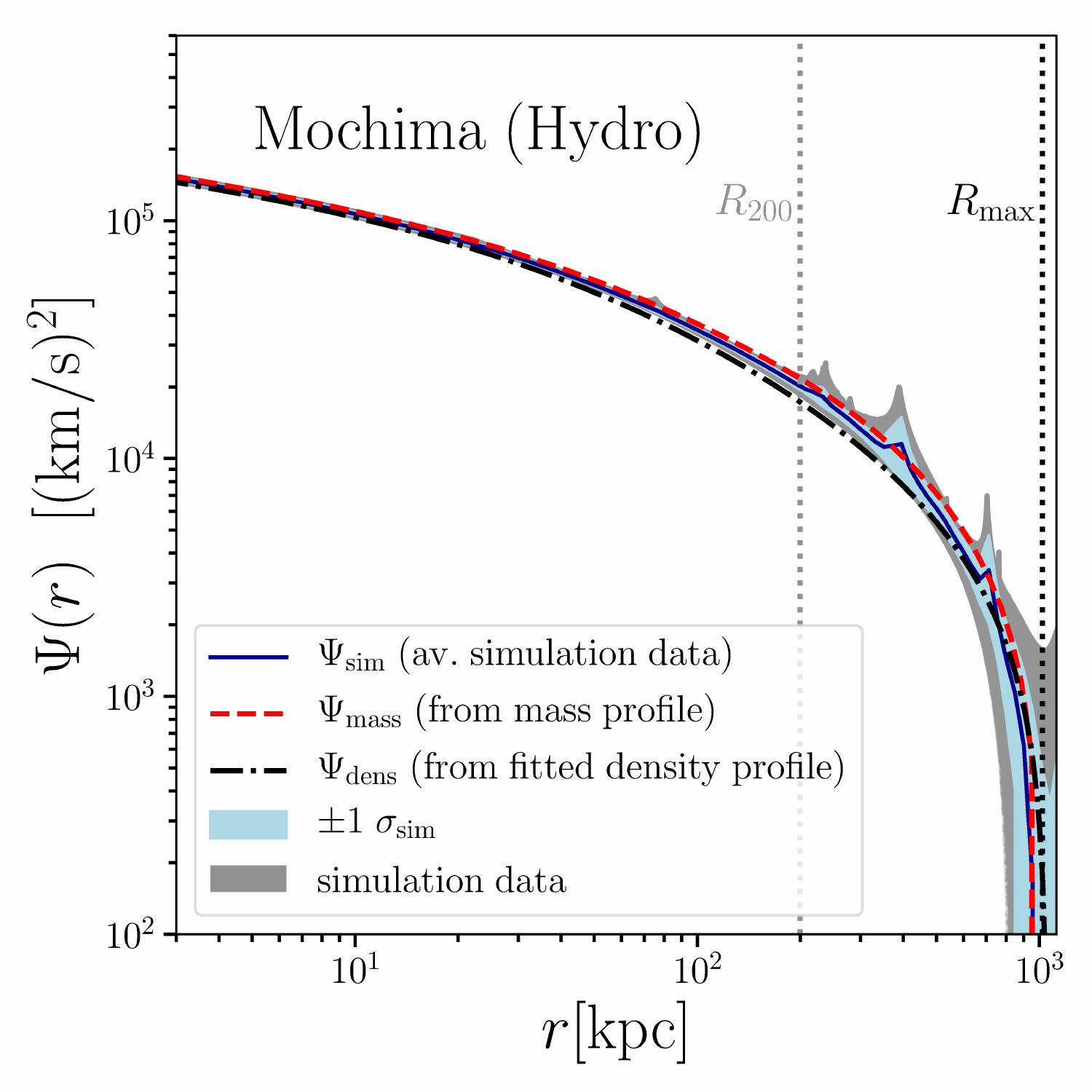}
\end{subfigure}
\caption{\small Same as \citefig{fig:Potential-comparison-HALOC}, but for the Mochima simulation.}
\label{fig:Potential-comparison-Mochima}
\end{figure}

\begin{figure}[t!]
\begin{center}
\includegraphics[width=0.49\textwidth]{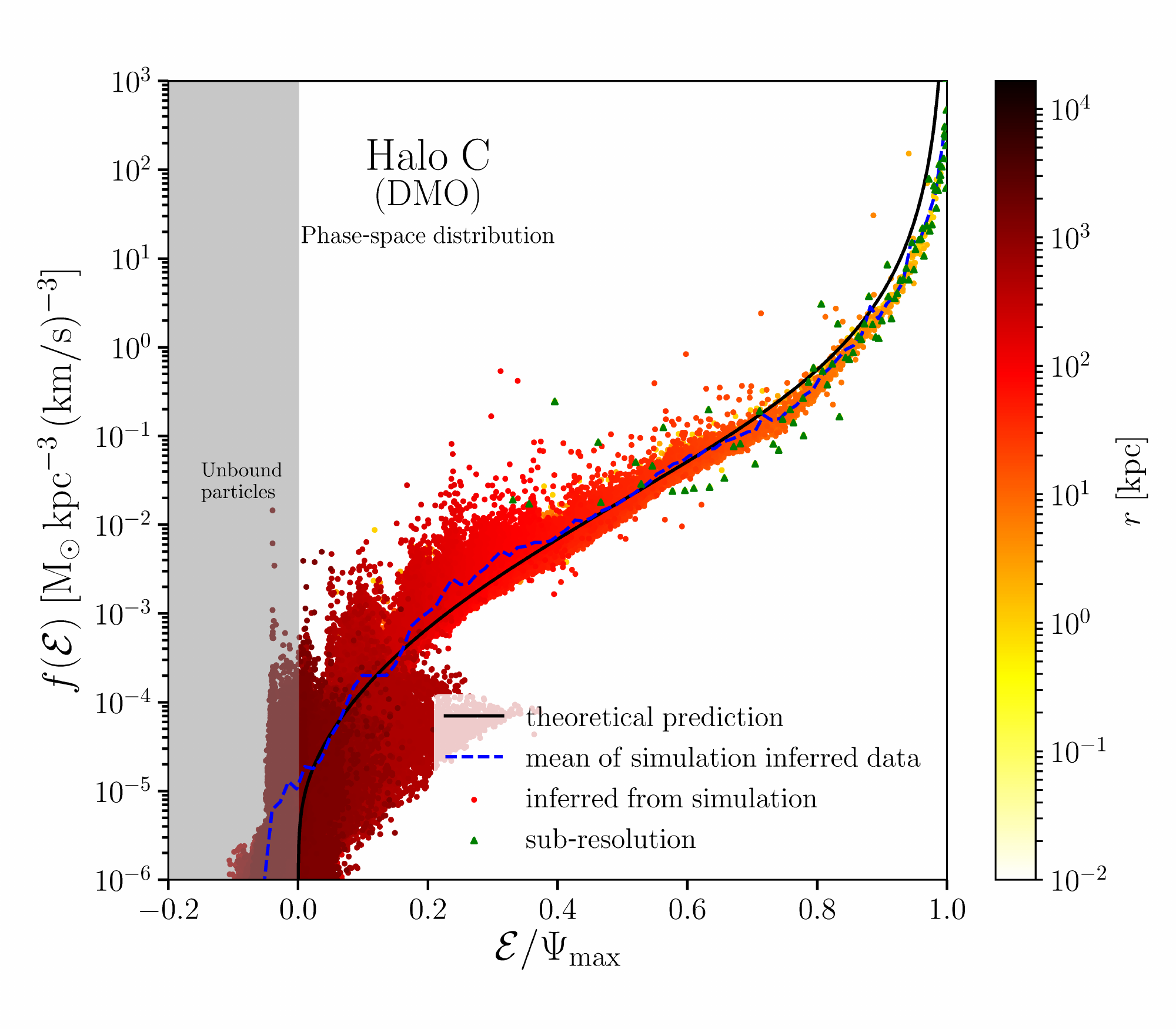} \hfill
\includegraphics[width=0.49\textwidth]{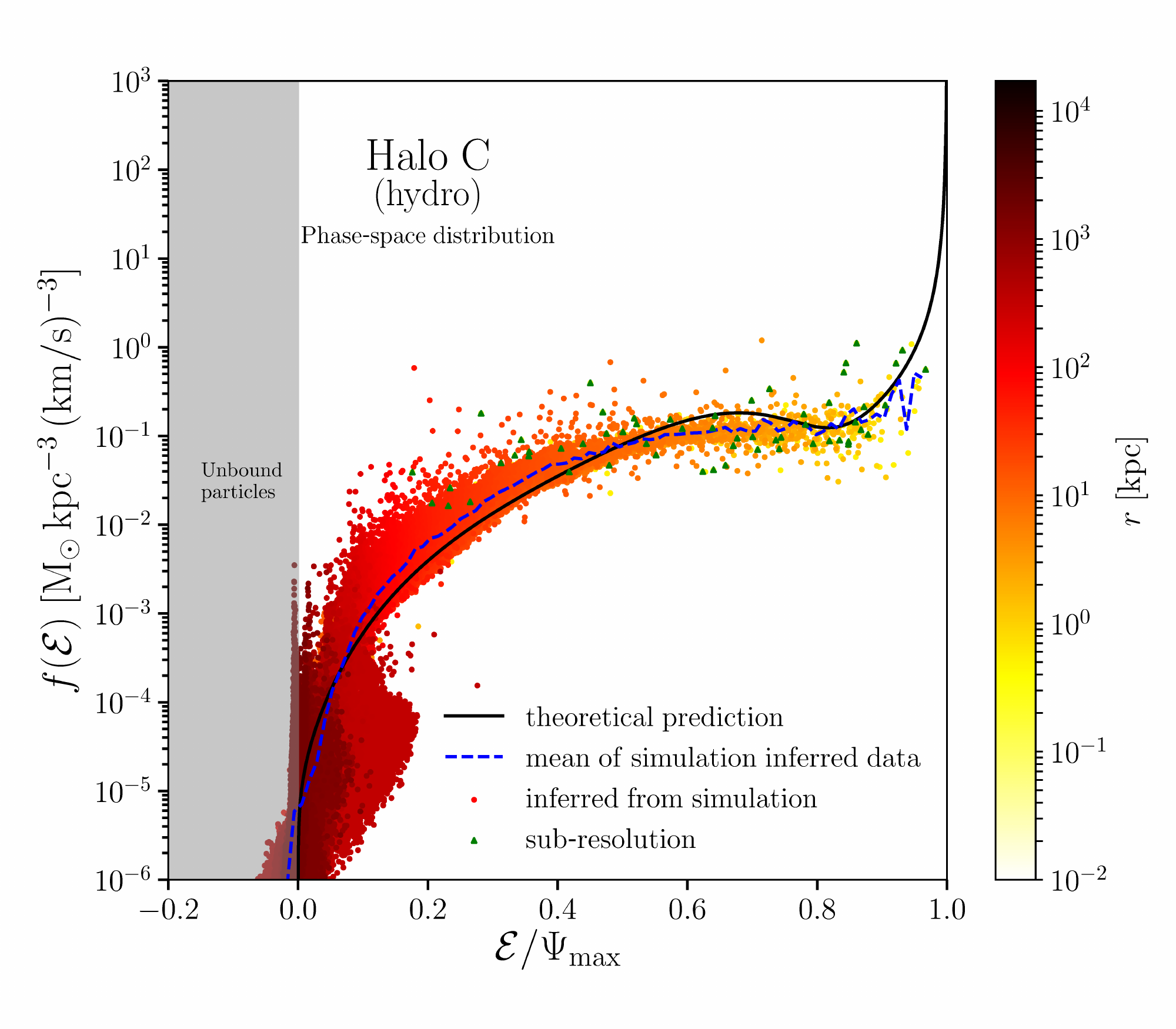}
\caption{\small Ergodic phase-space distribution predicted by the Eddington inversion for Halo C, for the DMO (left panel) and hydrodynamical (right panel) runs, compared with the pseudo-PSDF derived from the simulation shown as a scatter plot. The mean value of the pseudo-PSDF, averaged in energy bins, is shown as a blue dashed curve. In each panel the black curve represents the Eddington prediction for the ergodic PSDF, calculated from the best-fit density profiles. A color gradient is also used to show the radial origin of each 2D bin.
Here, unlike in \citefigs{fig:fE_HaloB} and \ref{fig:fE_Mochima}, we do not include the uncertainty band associated with the fit of the inner slope of the fitted DM density profile since Halo B and Halo C are very similar, and deriving the band requires a somewhat cumbersome calculation.}
\label{fig:fE_HaloC}
\end{center}
\end{figure}

\begin{figure}[t!]
\begin{center}
\includegraphics[width=0.49\textwidth]{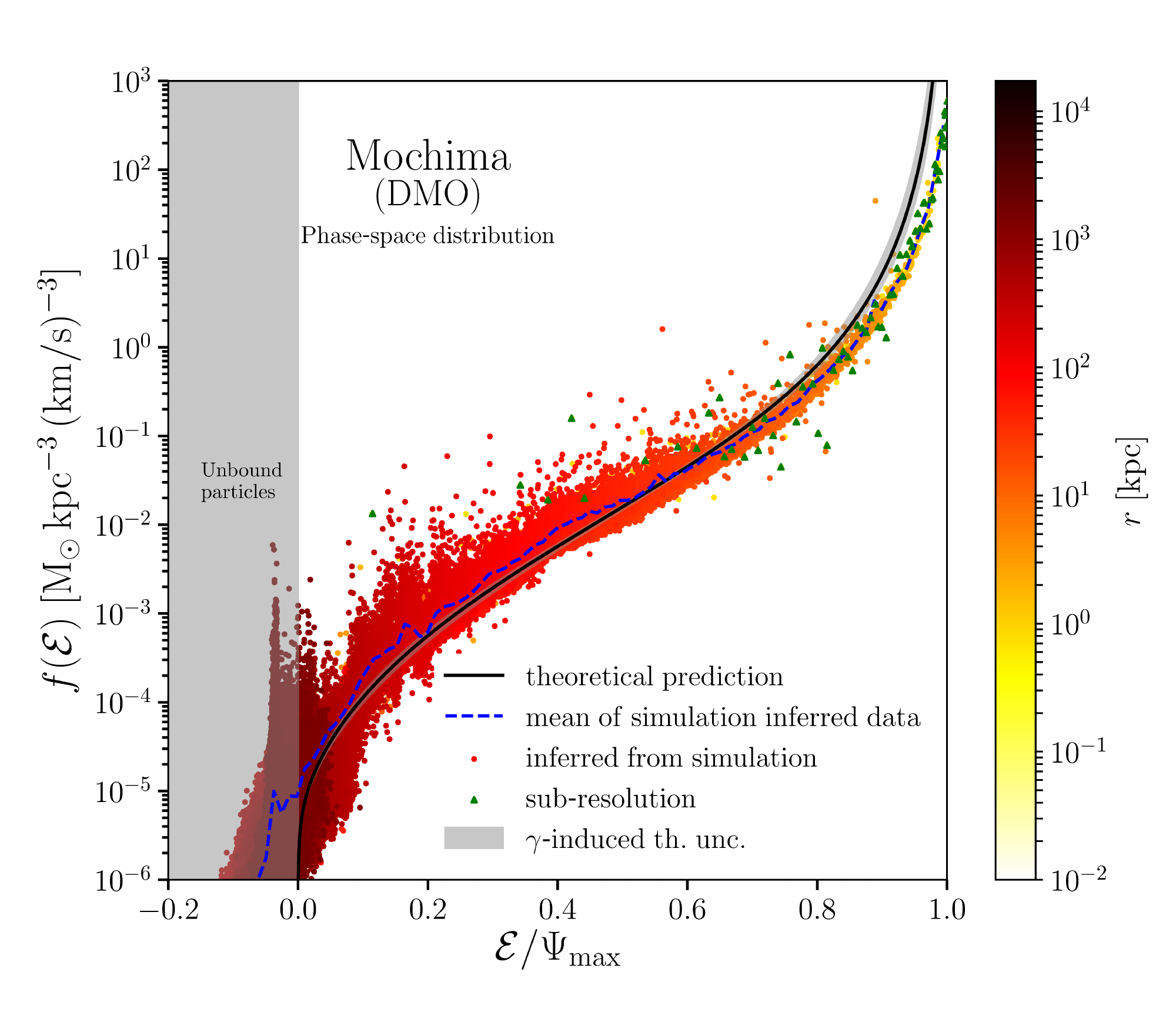} \hfill
\includegraphics[width=0.49\textwidth]{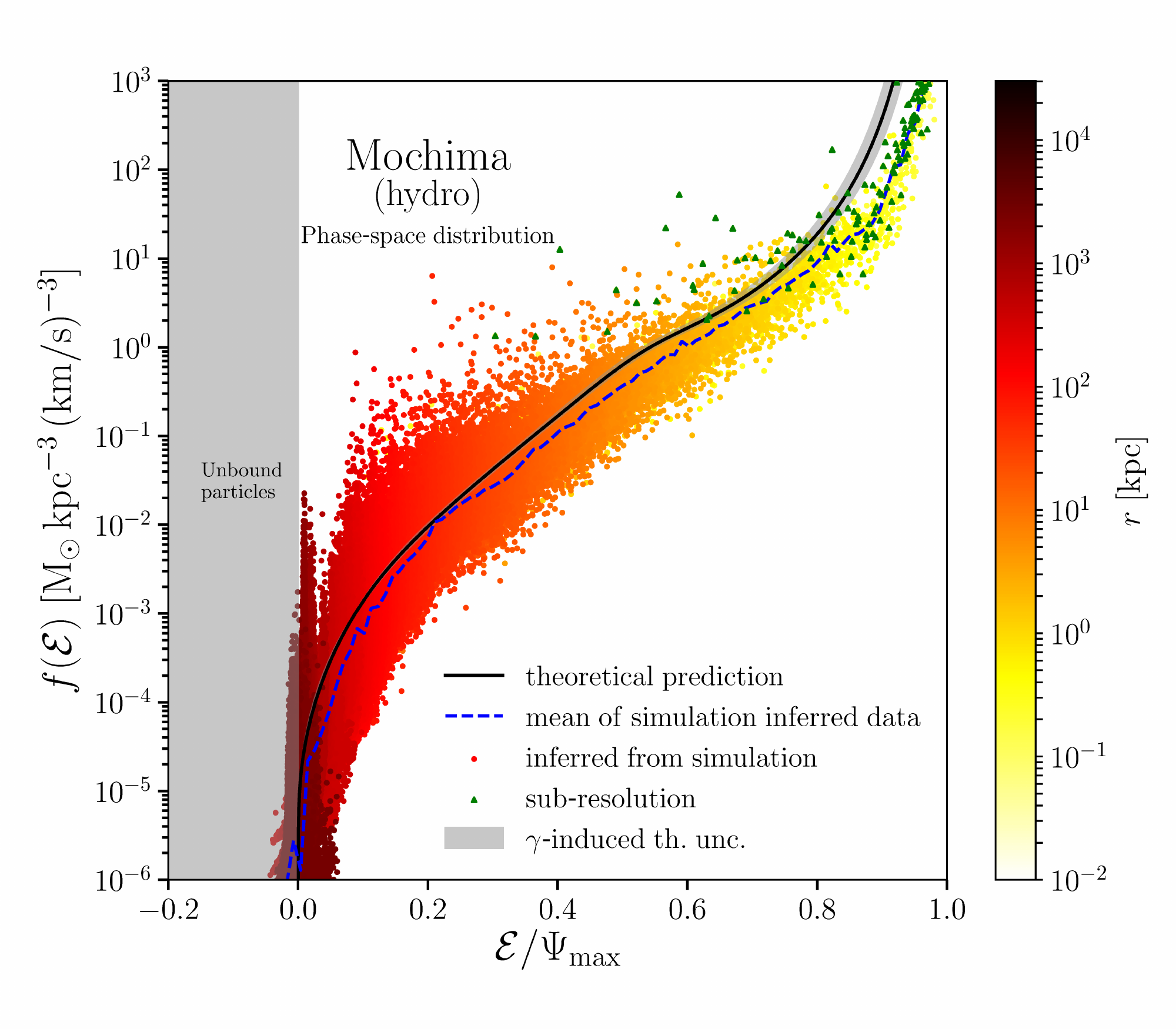}
\caption{\small Same as \citefig{fig:fE_HaloC}, but for the Mochima simulation. The gray band represents the $1 \sigma$ systematic error associated with the fit of the inner slope $\gamma$ of the DM density profile.}
\label{fig:fE_Mochima}
\end{center}
\end{figure}

The results in terms of the moments of the speed distribution are shown in \citefigs{fig:vmoments-comparison-HALOC-DMO} and \ref{fig:vmoments-comparison-Mochima-DMO} for the DMO run of the Halo~C and Mochima simulations, respectively, with $\left\langle v^{n}\right\rangle (r)$ ($n = -2, -1, 1 , 2$) as a function of radius $r$ in the left panels and the residuals with respect to the moments computed from the simulations in the right panels. Shown in \citefigs{fig:vmoments-comparison-HALOC-hydro} and \ref{fig:vmoments-comparison-Mochima-hydro} are the corresponding results for the hydro run of the Halo~C and Mochima simulations, respectively. See \citefigs{fig:vmoments-comparison-HALOB-DMO} and \ref{fig:vmoments-comparison-HALOB-hydro} for the analogous results for Halo~B, and \citesec{ssec:moments} for the discussion of the results.

Finally, the moments of the relative speed distribution, $\left\langle v_{\rm rel}^{n}\right\rangle (r)$, for $n = -2, -1, 1 , 2$, are shown in \citefigs{fig:v-rel-moments-comparison-HALOC-DMO} and \ref{fig:v-rel-moments-comparison-Mochima-DMO} for the DMO run of the Halo~C and Mochima simulations, respectively, and in \citefigs{fig:v-rel-moments-comparison-HALOC-hydro} and \ref{fig:v-rel-moments-comparison-Mochima-hydro} for the hydro run of the Halo~C and Mochima simulations, respectively. In each figure the moments are shown as a function of radius $r$ in the left panel, and the residuals with respect to the corresponding quantities extracted from the simulation, as a function of radius, in the right panel. See \citefigs{fig:v-rel-moments-comparison-HALOB-DMO} and \ref{fig:v-rel-moments-comparison-HALOB-hydro} for the analogous results for Halo~B, and \citesec{ssec:moments_vrel} for the discussion of the results.

\begin{figure}[h!]
\begin{center}
\includegraphics[width=0.49\textwidth]{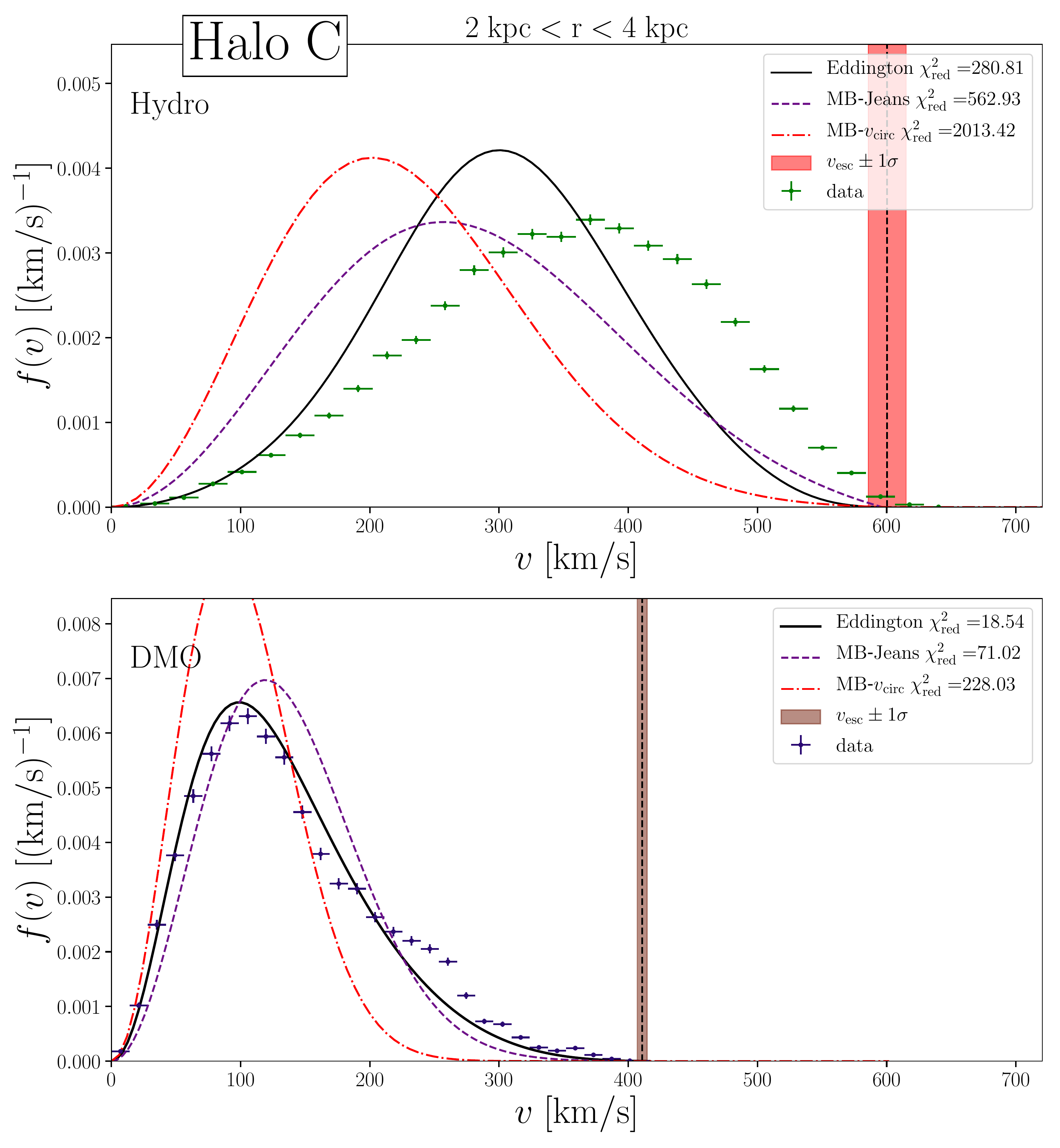} \hfill 
\includegraphics[width=0.49\textwidth]{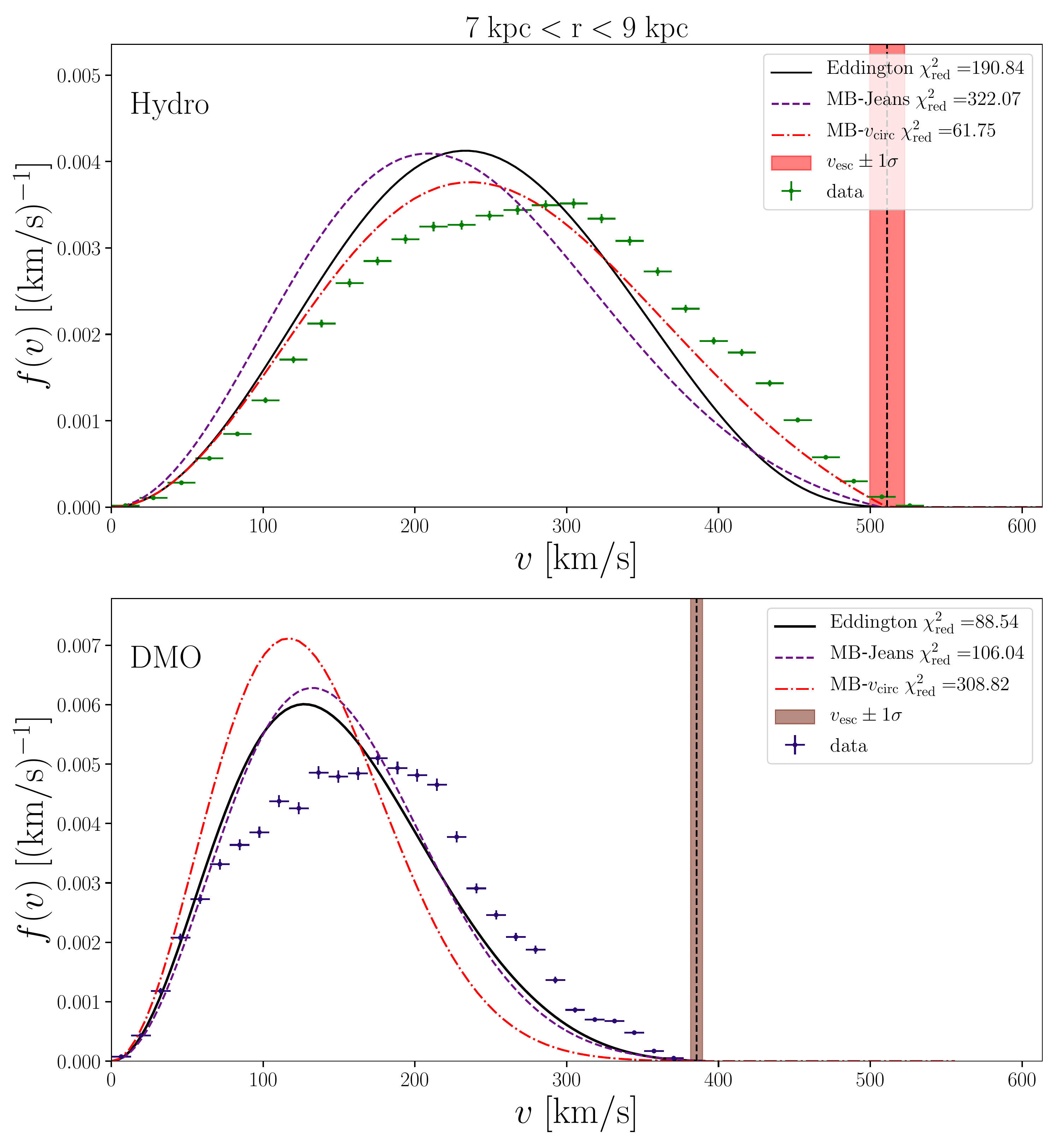}\\
\includegraphics[width=0.49\textwidth]{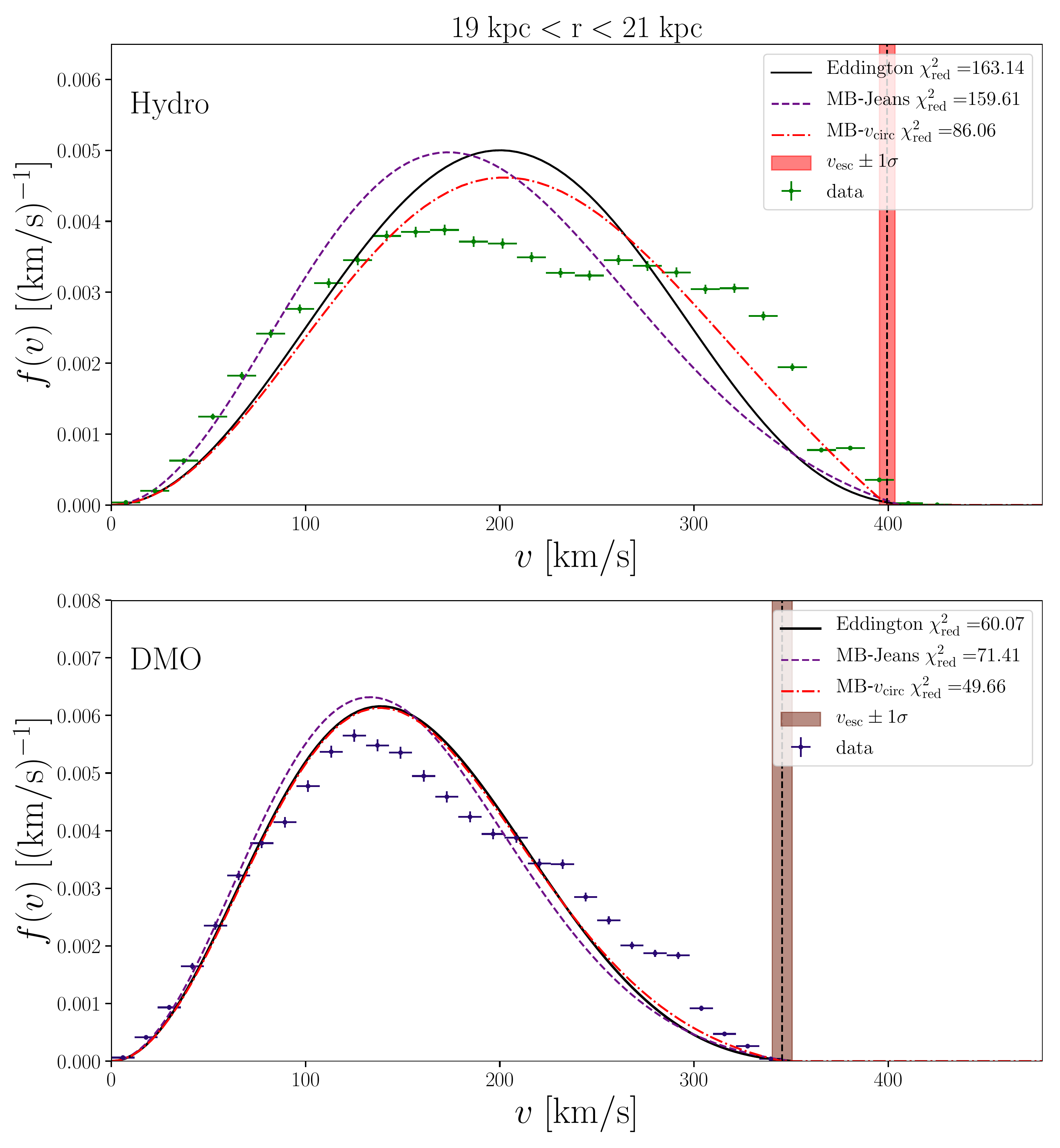} \hfill 
\includegraphics[width=0.49\textwidth]{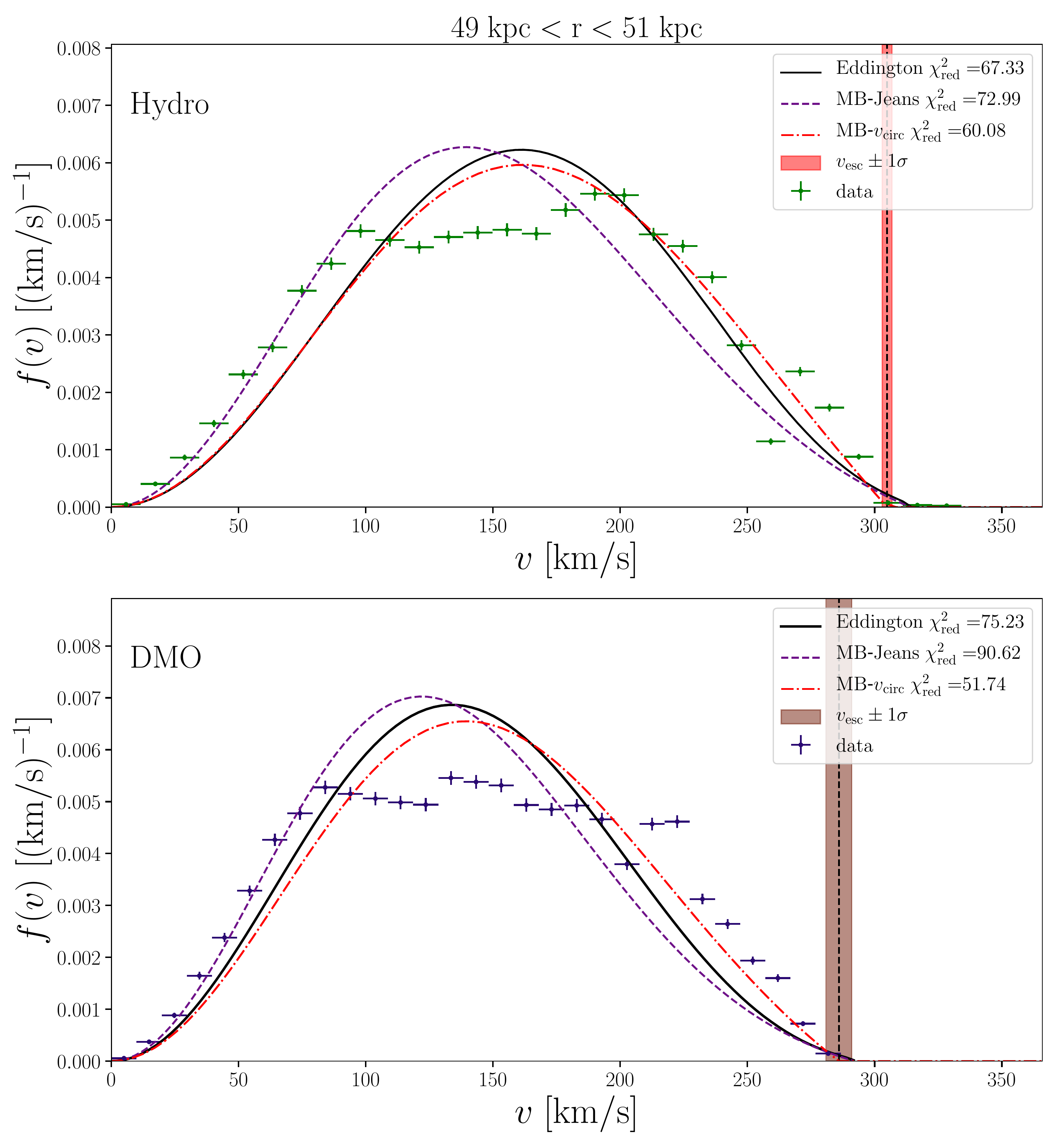}
\caption{\small Comparison between the speed distribution derived with the Eddington method and the simulation data for spherical shells of 2 kpc thickness at different radii (3 kpc, upper left; 8 kpc, upper right; 20 kpc, lower left; and 50 kpc, lower right), for the Halo~C simulation, for the DMO (lower subpanels) and hydro (upper subpanels) runs. In each panel, the data points (green for hydro, blue for DMO) correspond to a normalized histogram built from the data. The Eddington predictions are shown as black solid curves, and two models relying on the MB distribution, with different assumptions regarding the peak speed $v_{0}$ are displayed: one in which it is set to the circular velocity (``$v_{\mathrm{circ}}$", red dot-dashed curves), and the other one in which it is given by $\sqrt{2/3}$ times the velocity dispersion calculated by solving the Jeans equation (``Jeans", purple dashed curves). The escape speed $v_{\mathrm{e}}(r)$ from the simulation is shown as black vertical dashed lines, with the associated 1-$\sigma$ error displayed as shaded bands.} 
\label{fig:fv-comparison-HALOC}
\end{center}
\end{figure}

\begin{figure}[h!]
\begin{center}
\includegraphics[width=0.49\textwidth]{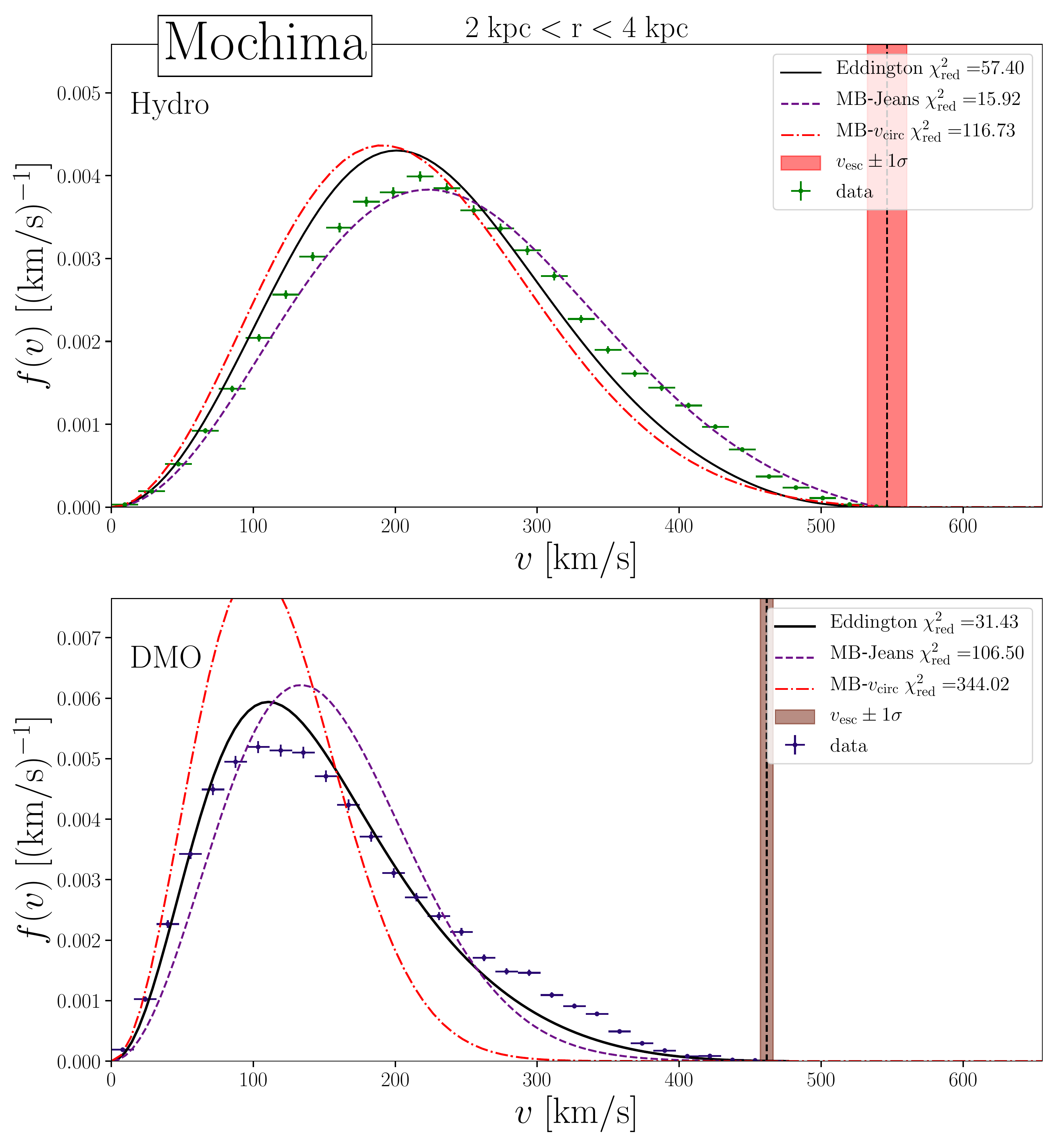} \hfill 
\includegraphics[width=0.49\textwidth]{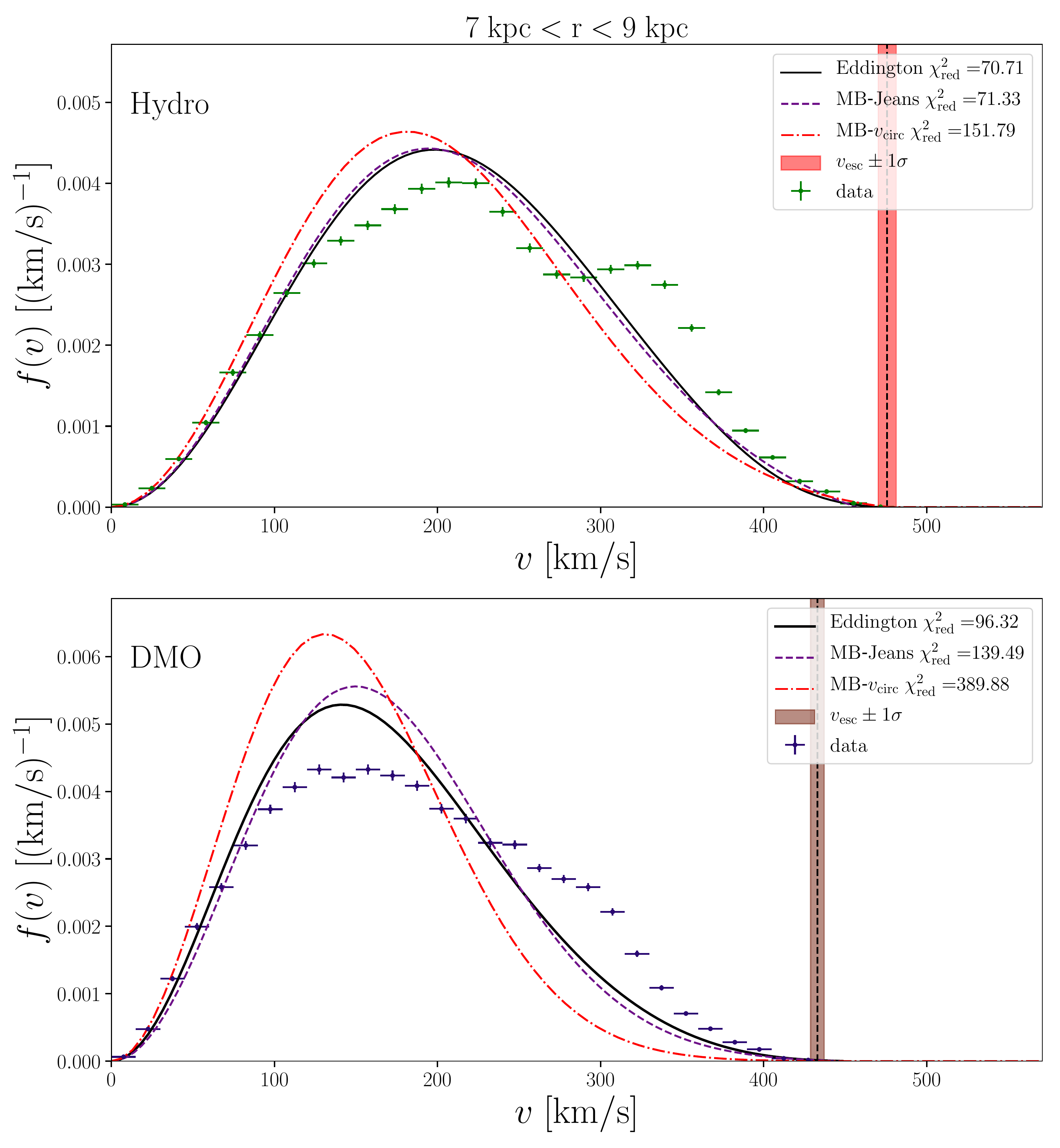}\\
\includegraphics[width=0.49\textwidth]{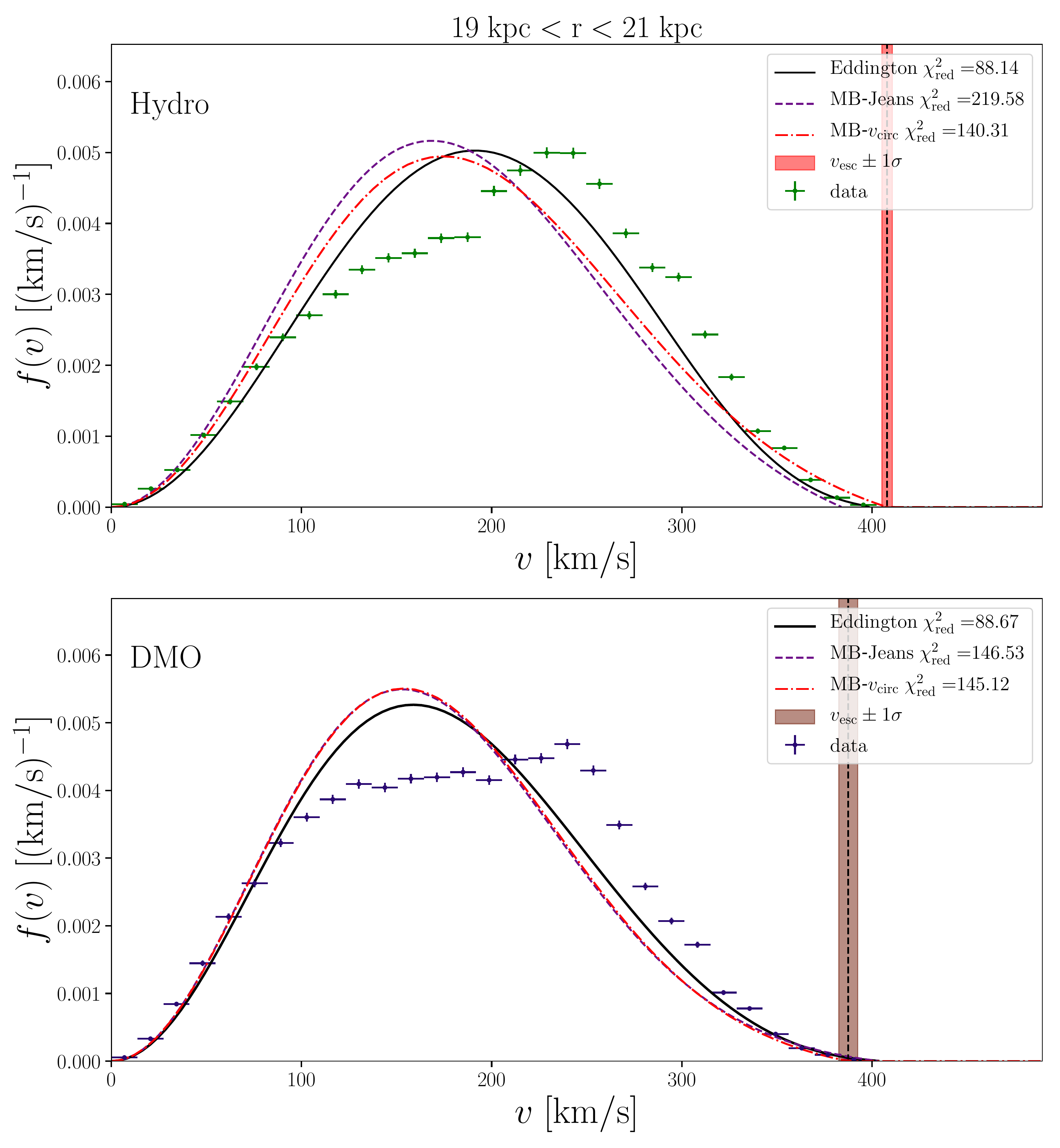} \hfill 
\includegraphics[width=0.49\textwidth]{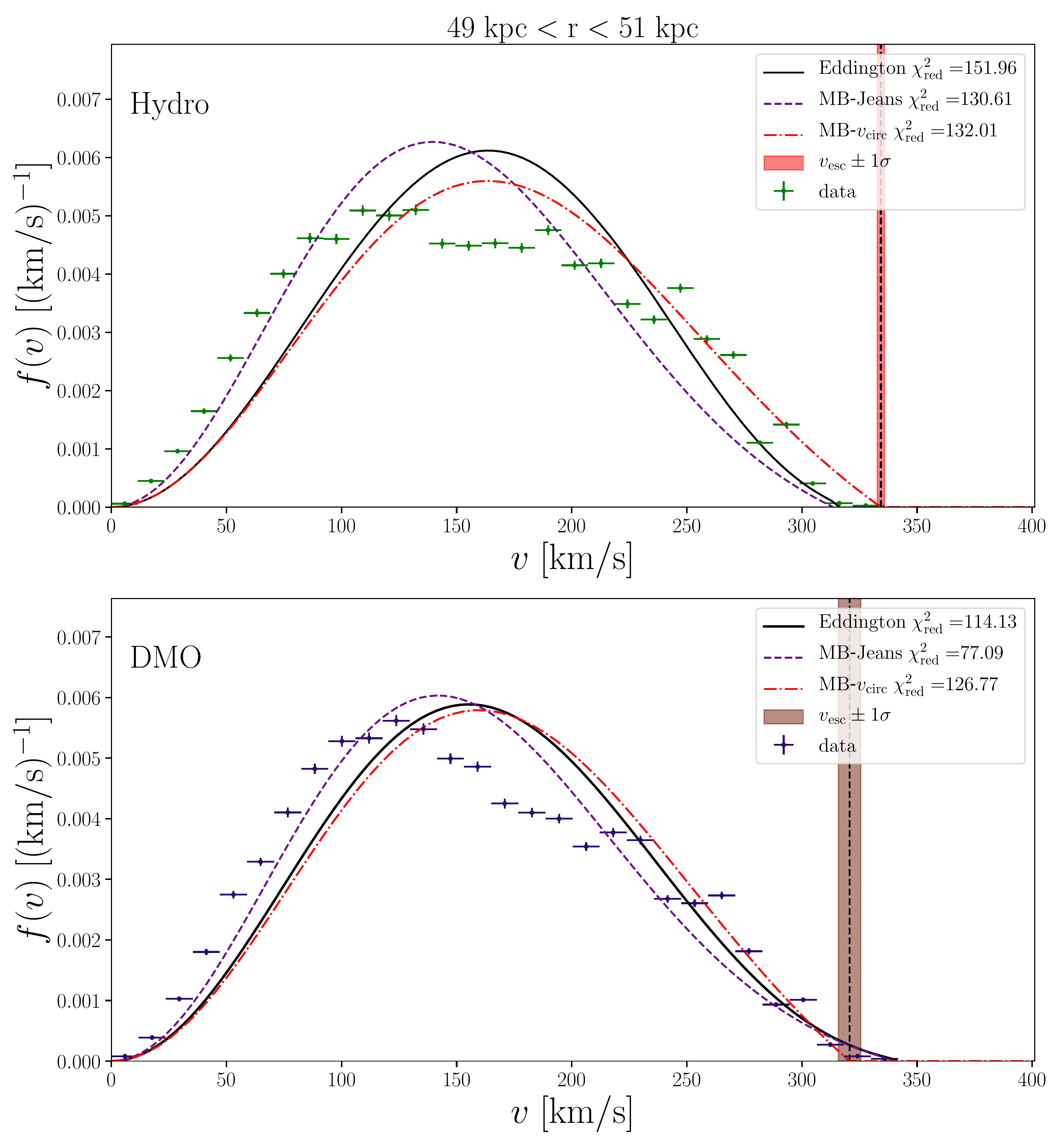}
\caption{\small Same as \citefig{fig:fv-comparison-HALOC}, but for the Mochima simulation.}
\label{fig:fv-comparison-Mochima}
\end{center}
\end{figure}

\begin{figure}[t!]
\begin{center}
\includegraphics[width=0.49\textwidth]{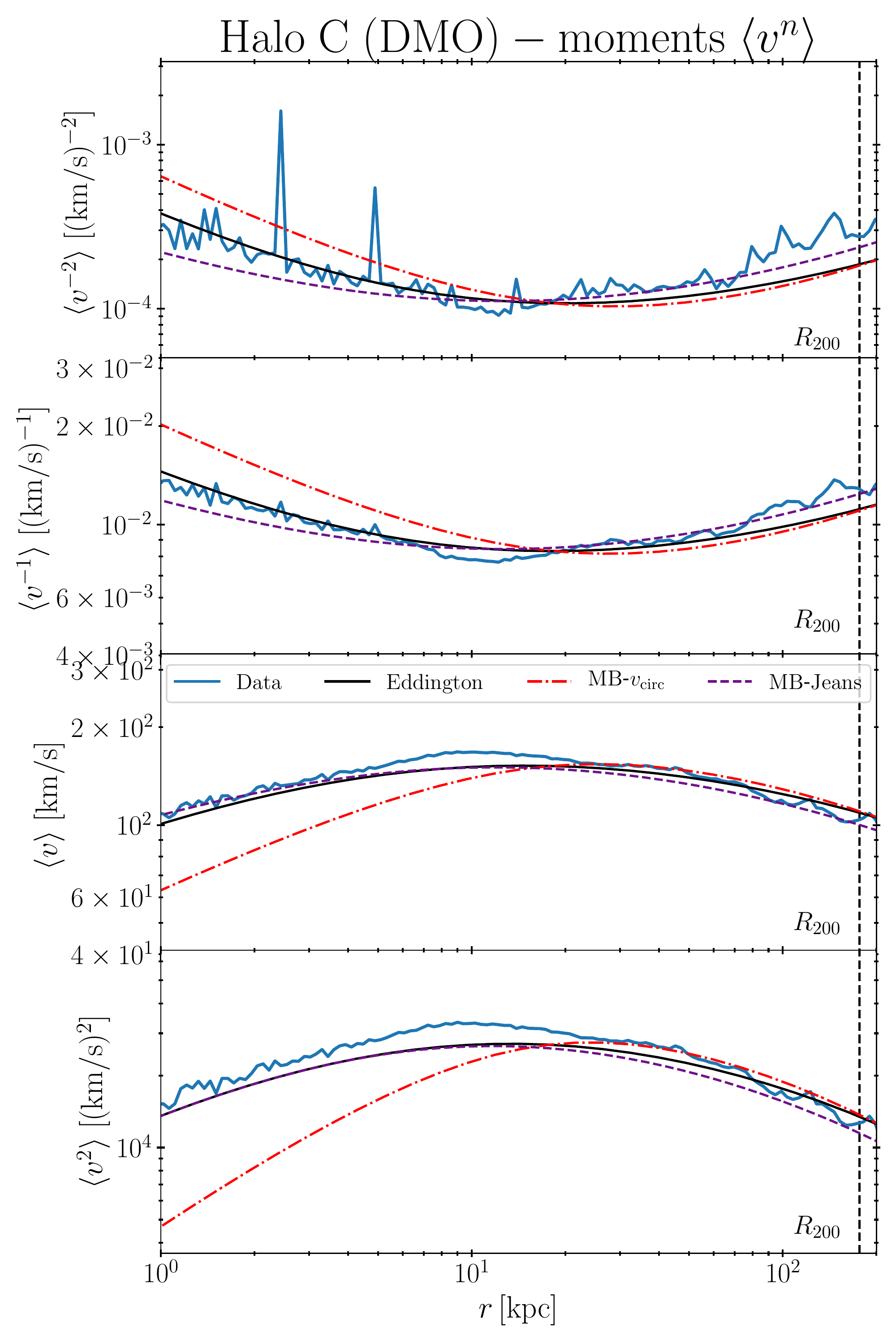} \hfill
\includegraphics[width=0.49\textwidth]{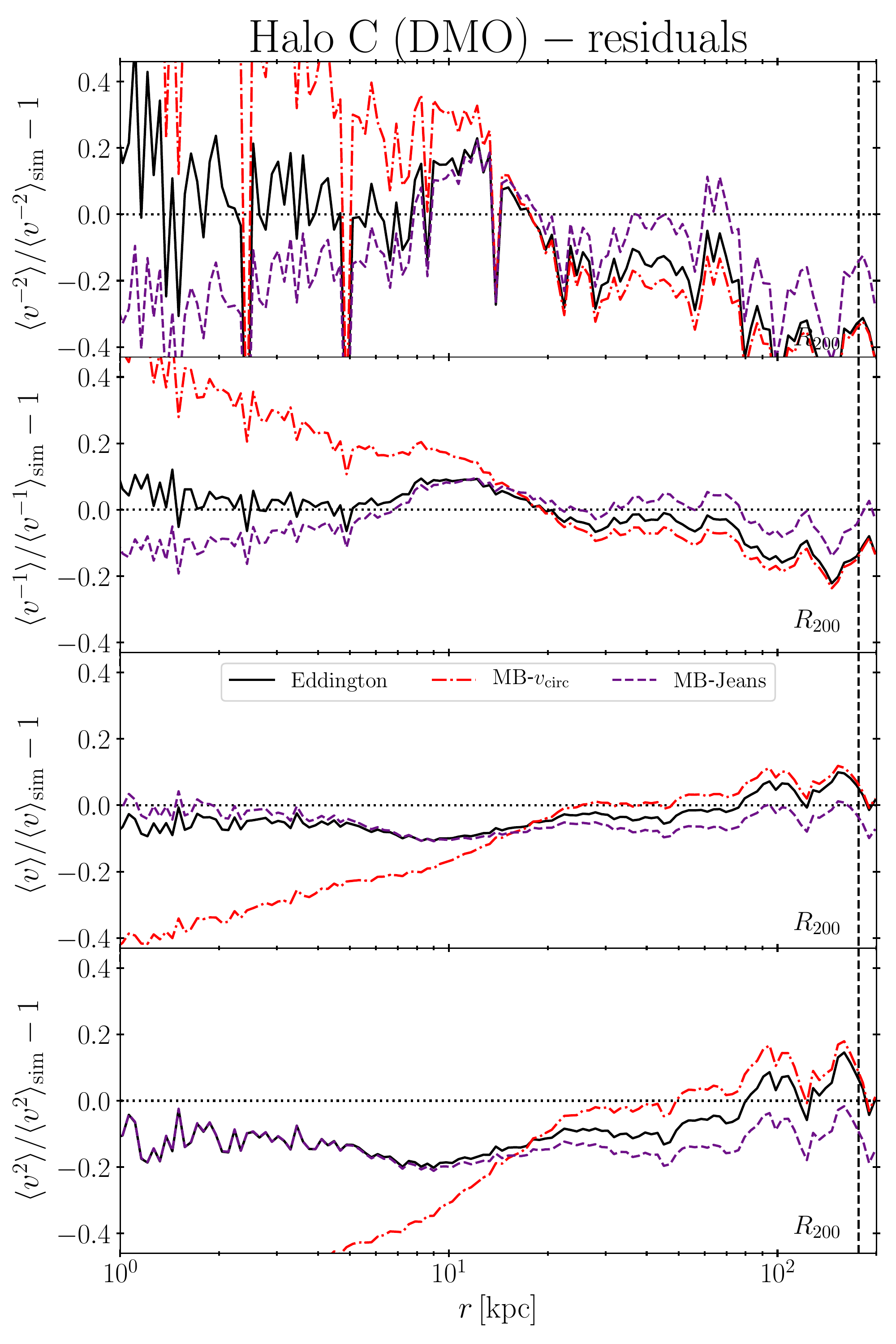}
\caption{\small Comparison of the moments of the speed distribution from the Eddington method and from the simulation for Halo~C (left panel), and associated residuals with respect to the simulation (right panel), for the DMO run. Simulation outputs are shown as blue solid curves. Predictions obtained with the Eddington inversion method are displayed as black solid curves, while the two MB models ``$v_{\mathrm{circ}}$" and ``Jeans" are shown as red dot-dashed and purple dashed, respectively. These line styles match those in the speed distribution figures, \citefigs{fig:fv-comparison-HALOB}, \ref{fig:fv-comparison-HALOC}, and \ref{fig:fv-comparison-Mochima}.}
\label{fig:vmoments-comparison-HALOC-DMO}
\end{center}
\end{figure}

\begin{figure}[t]
\begin{center}
\includegraphics[width=0.49\textwidth]{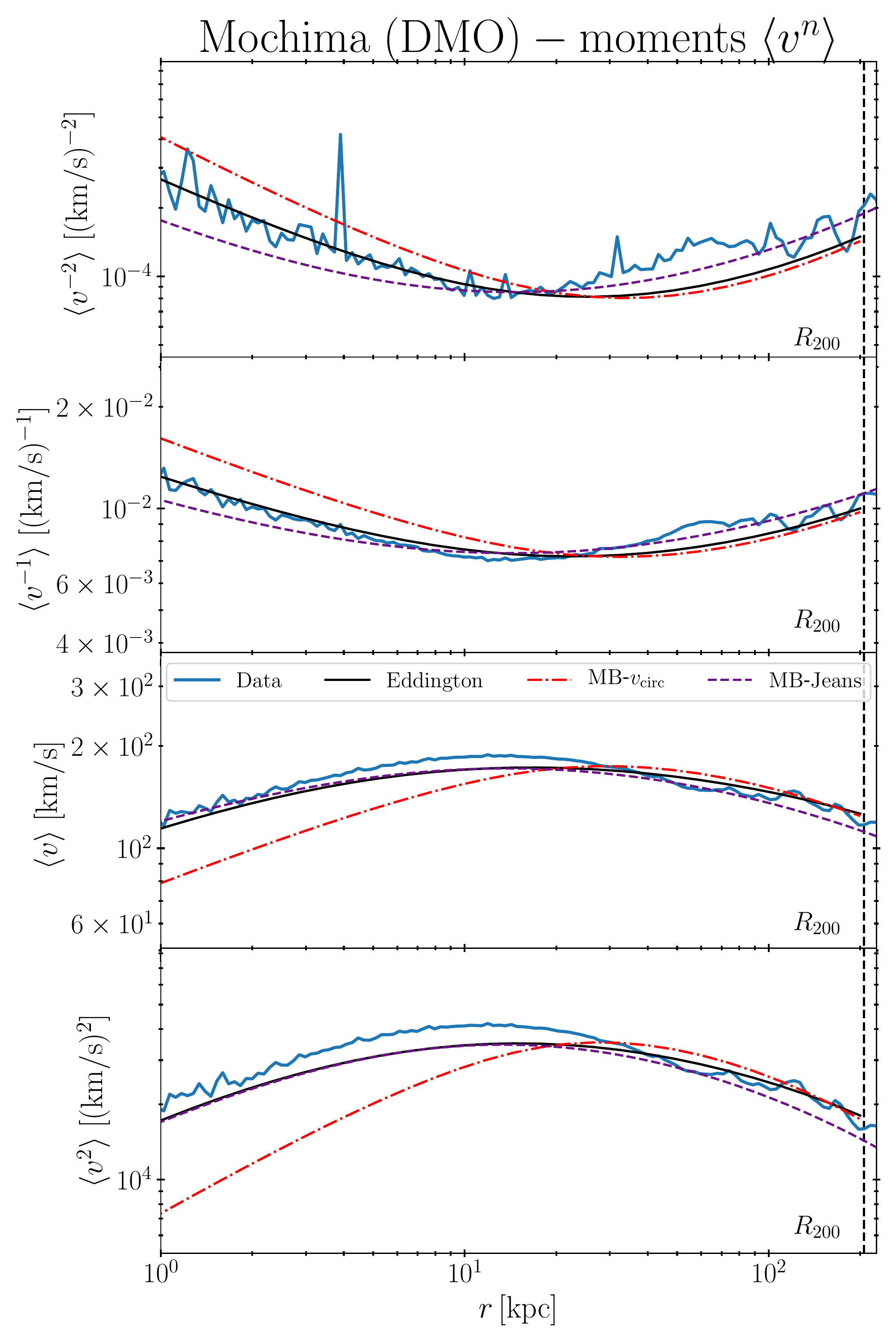} \hfill
\includegraphics[width=0.49\textwidth]{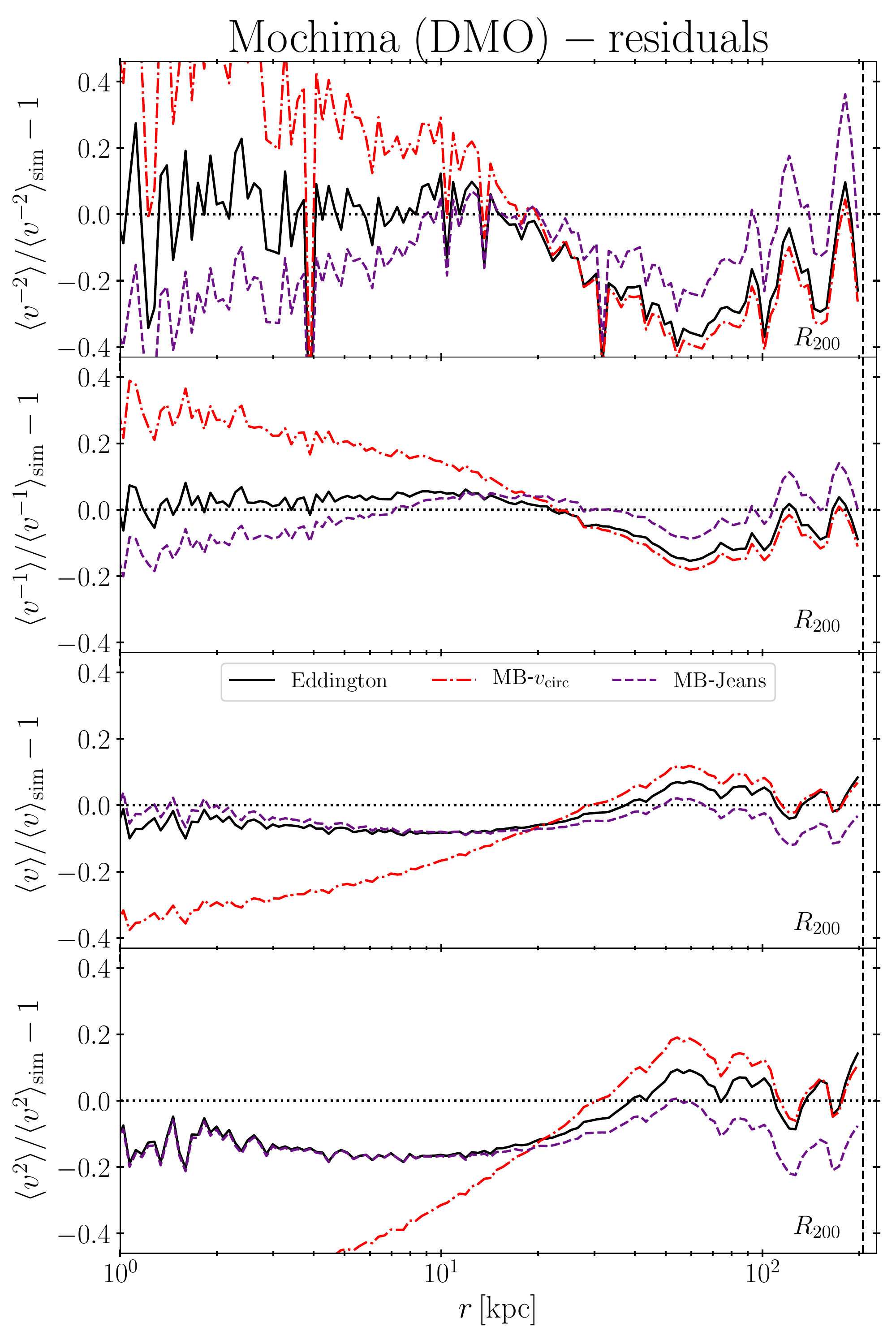}
\caption{\small Same as \citefig{fig:vmoments-comparison-HALOC-DMO}, but for the Mochima simulation.}
\label{fig:vmoments-comparison-Mochima-DMO}
\end{center}
\end{figure}

\begin{figure}[t]
\begin{center}
\includegraphics[width=0.49\textwidth]{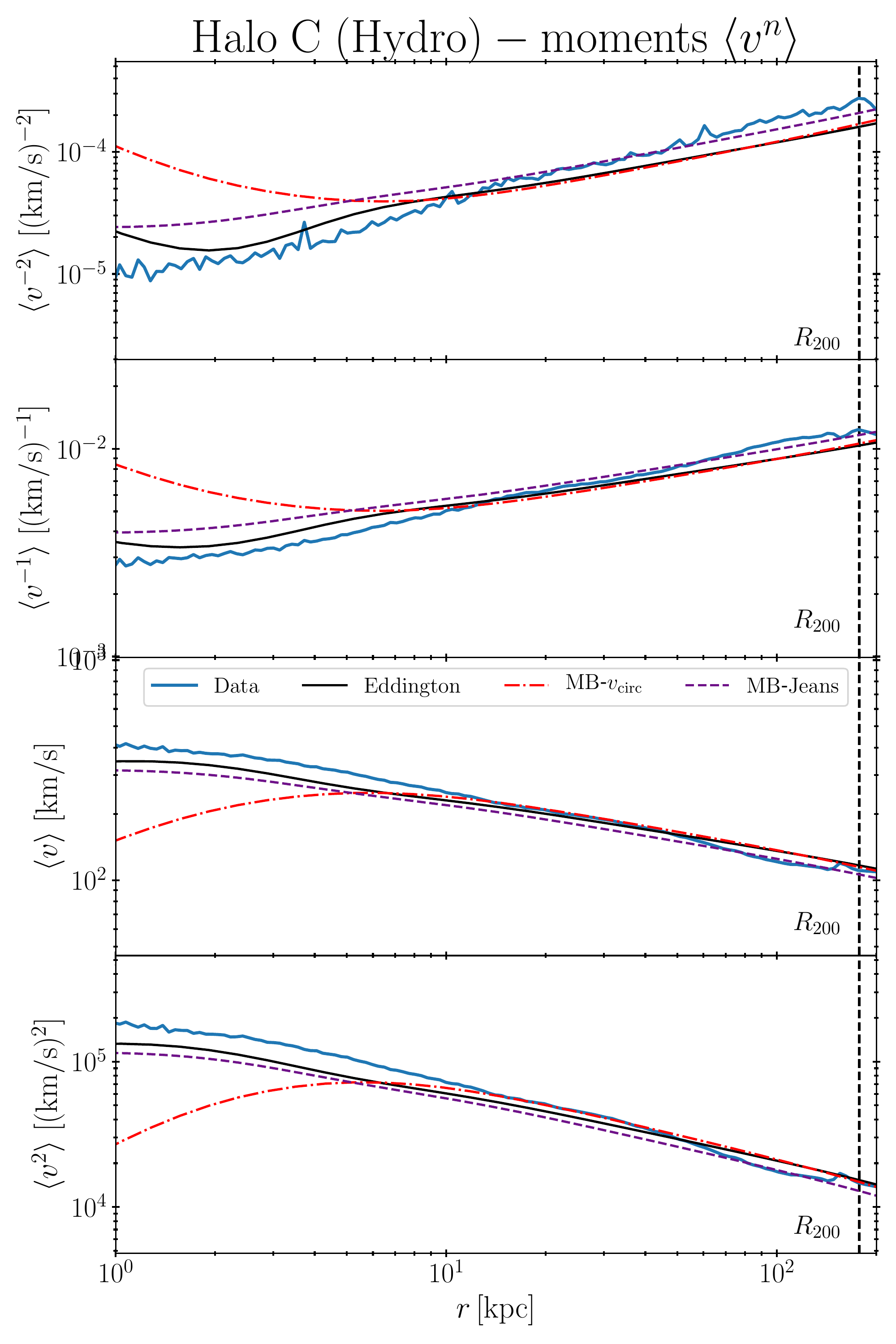} \hfill
\includegraphics[width=0.49\textwidth]{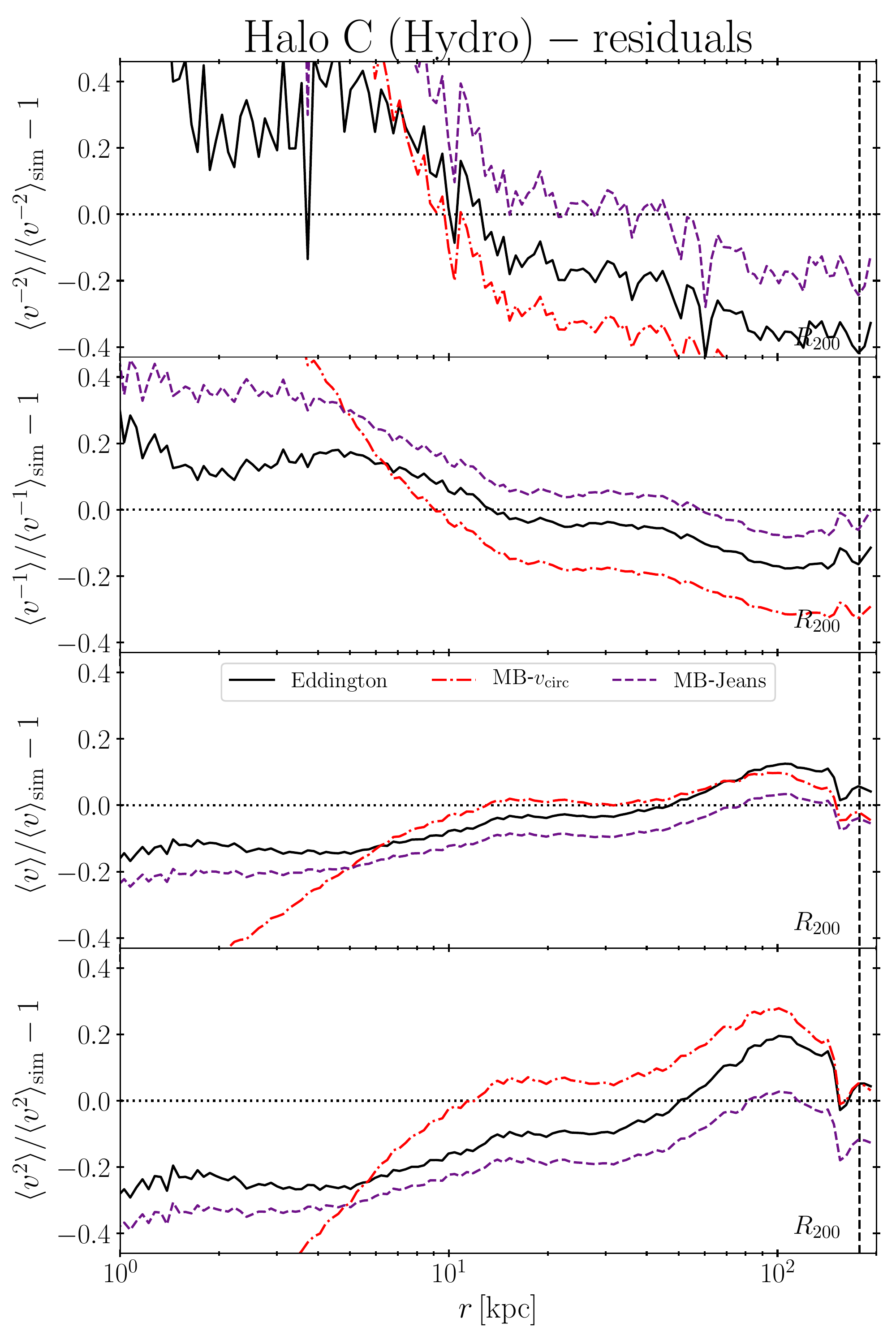}
\caption{\small Comparison of the moments of the speed distribution from the Eddington method and from the simulation for Halo~C (left panel), and associated residuals with respect to the simulation (right panel), for the hydro run. Simulation outputs are shown as blue solid curves. Predictions obtained with the Eddington inversion method are displayed as black solid curves, while the two MB models ``$v_{\mathrm{circ}}$" and ``Jeans" are shown as red dot-dashed and purple dashed, respectively. These line styles match those in the speed distribution figures, \citefigs{fig:fv-comparison-HALOB}, \ref{fig:fv-comparison-HALOC}, and \ref{fig:fv-comparison-Mochima}.}
\label{fig:vmoments-comparison-HALOC-hydro}
\end{center}
\end{figure}

\begin{figure}[t]
\begin{center}
\includegraphics[width=0.49\textwidth]{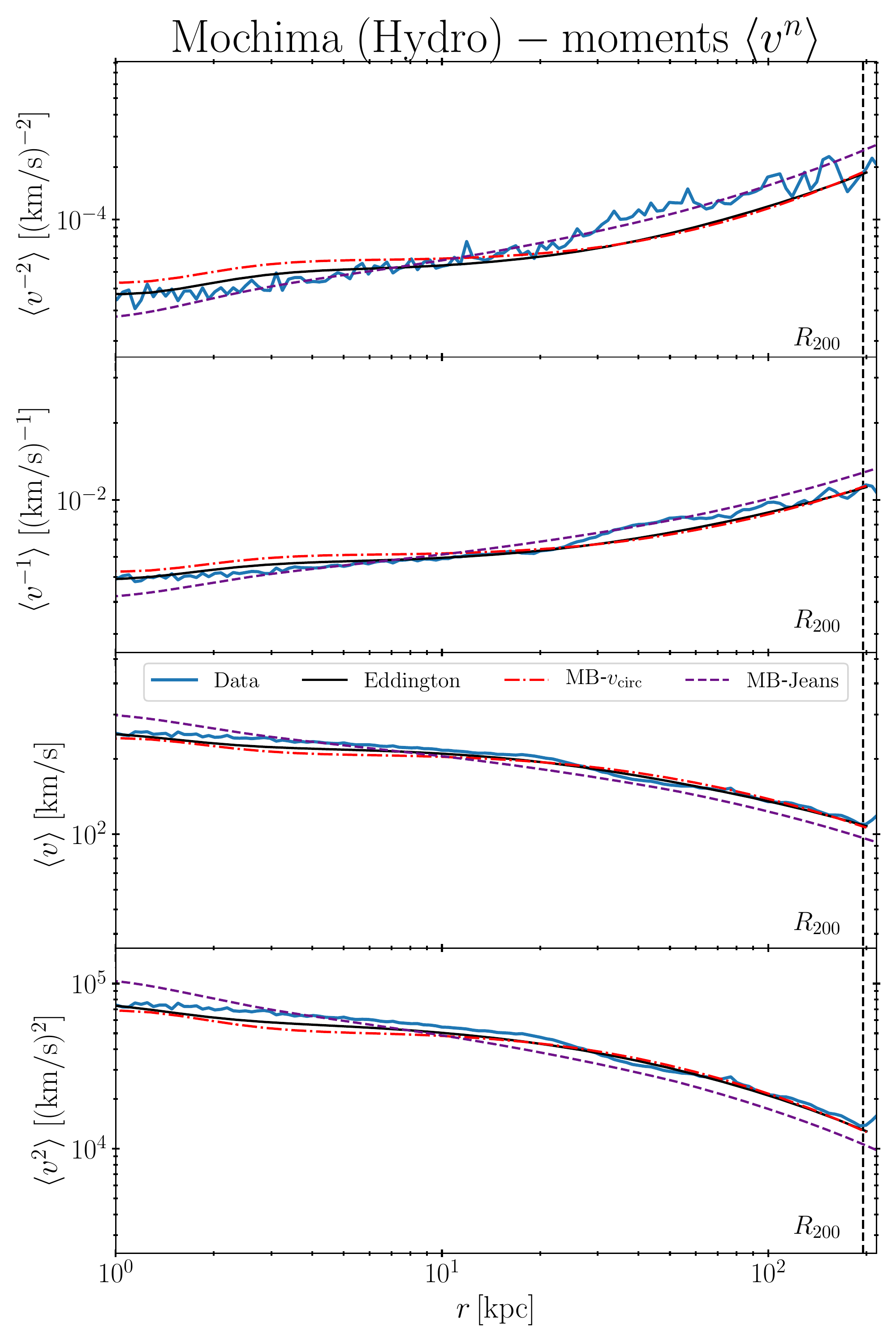} \hfill
\includegraphics[width=0.49\textwidth]{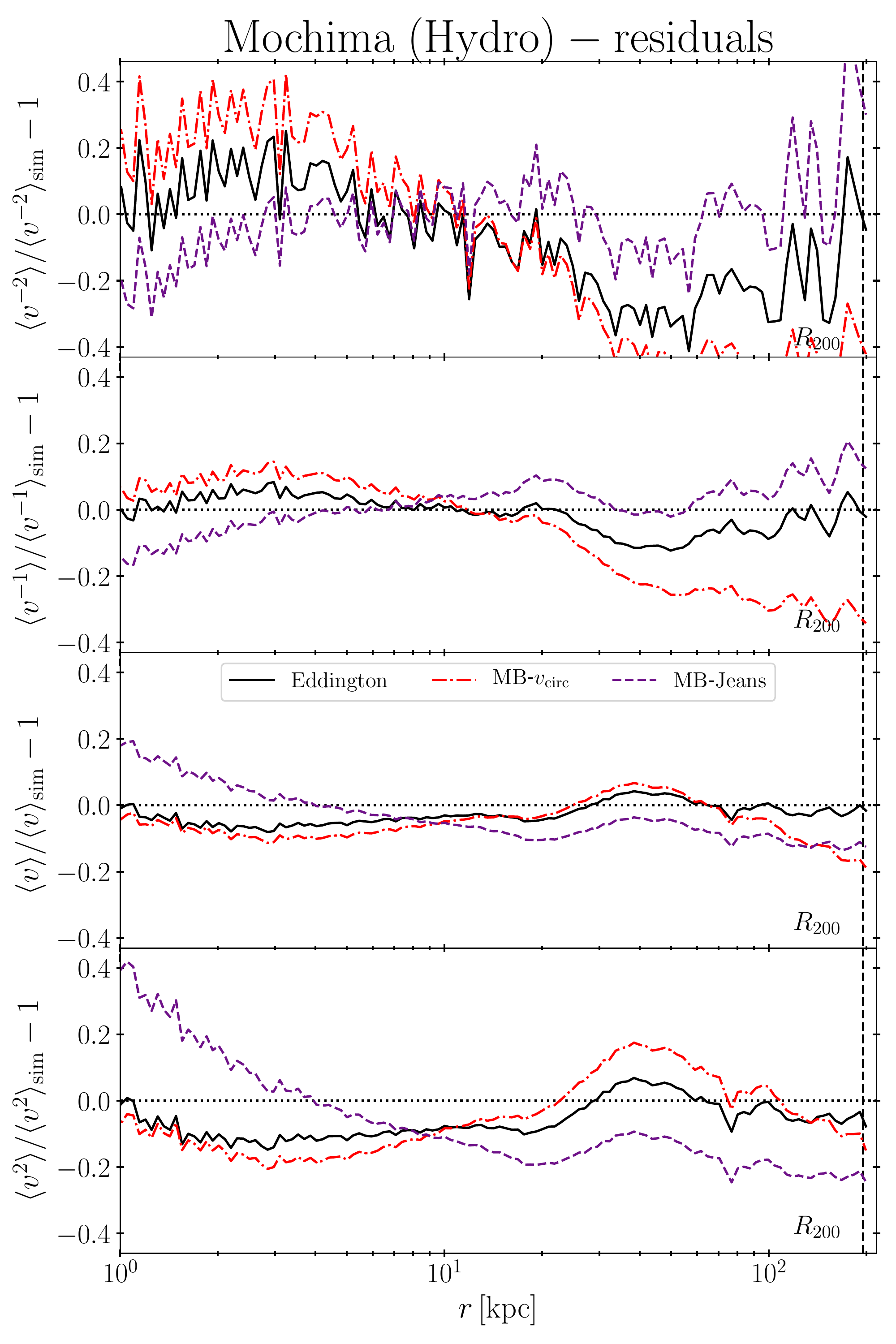}
\caption{\small Same as \citefig{fig:vmoments-comparison-HALOC-hydro}, but for the Mochima simulation.}
\label{fig:vmoments-comparison-Mochima-hydro}
\end{center}
\end{figure}


\begin{figure}[t]
\begin{center}
\includegraphics[width=0.49\textwidth]{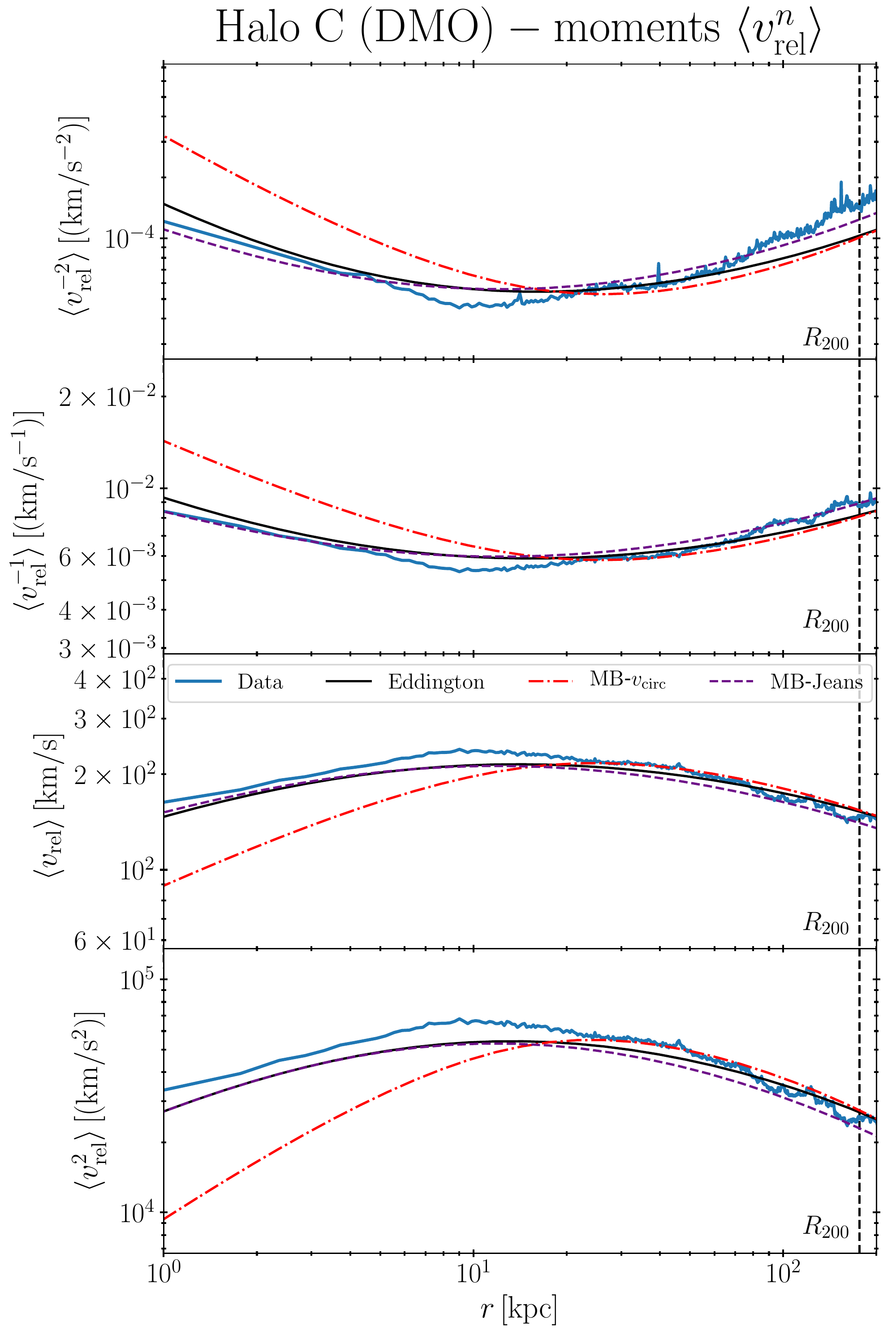} \hfill
\includegraphics[width=0.49\textwidth]{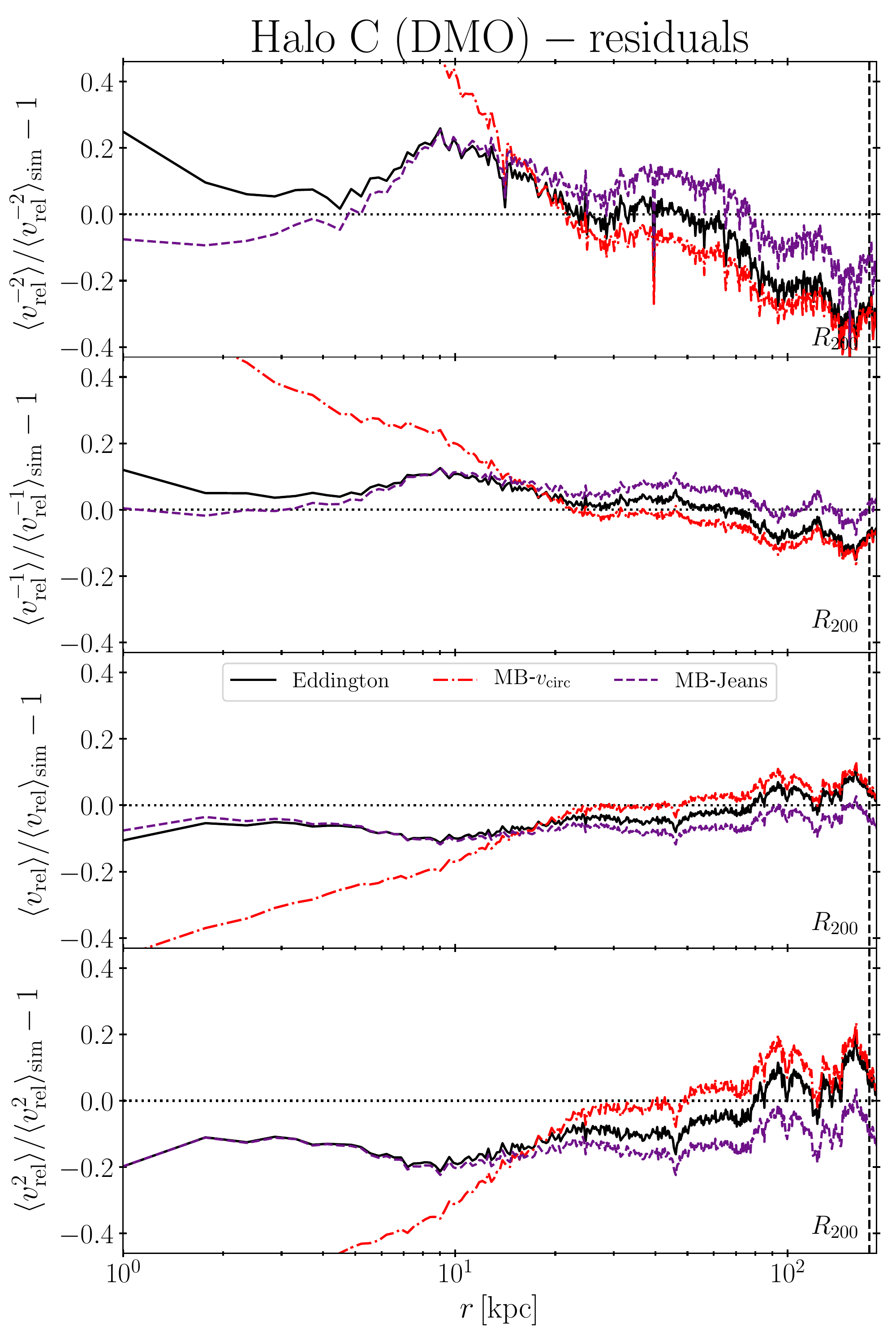}
\caption{\small Comparison of the moments of the \textit{relative} speed distribution from the Eddington method and from the simulation for Halo~C (left panel), and associated residuals with respect to the simulation (right panel), for the DMO run. Simulation outputs are shown as blue solid curves. Predictions obtained with the Eddington inversion method are displayed as black solid curves, while the two MB models ``$v_{\mathrm{circ}}$" and ``Jeans" are shown as red dot-dashed and purple dashed, respectively. These line styles match those in the speed distribution figures, \citefigs{fig:fv-comparison-HALOB}, \ref{fig:fv-comparison-HALOC}, and \ref{fig:fv-comparison-Mochima}.}
\label{fig:v-rel-moments-comparison-HALOC-DMO}
\end{center}
\end{figure}

\begin{figure}[t]
\begin{center}
\includegraphics[width=0.49\textwidth]{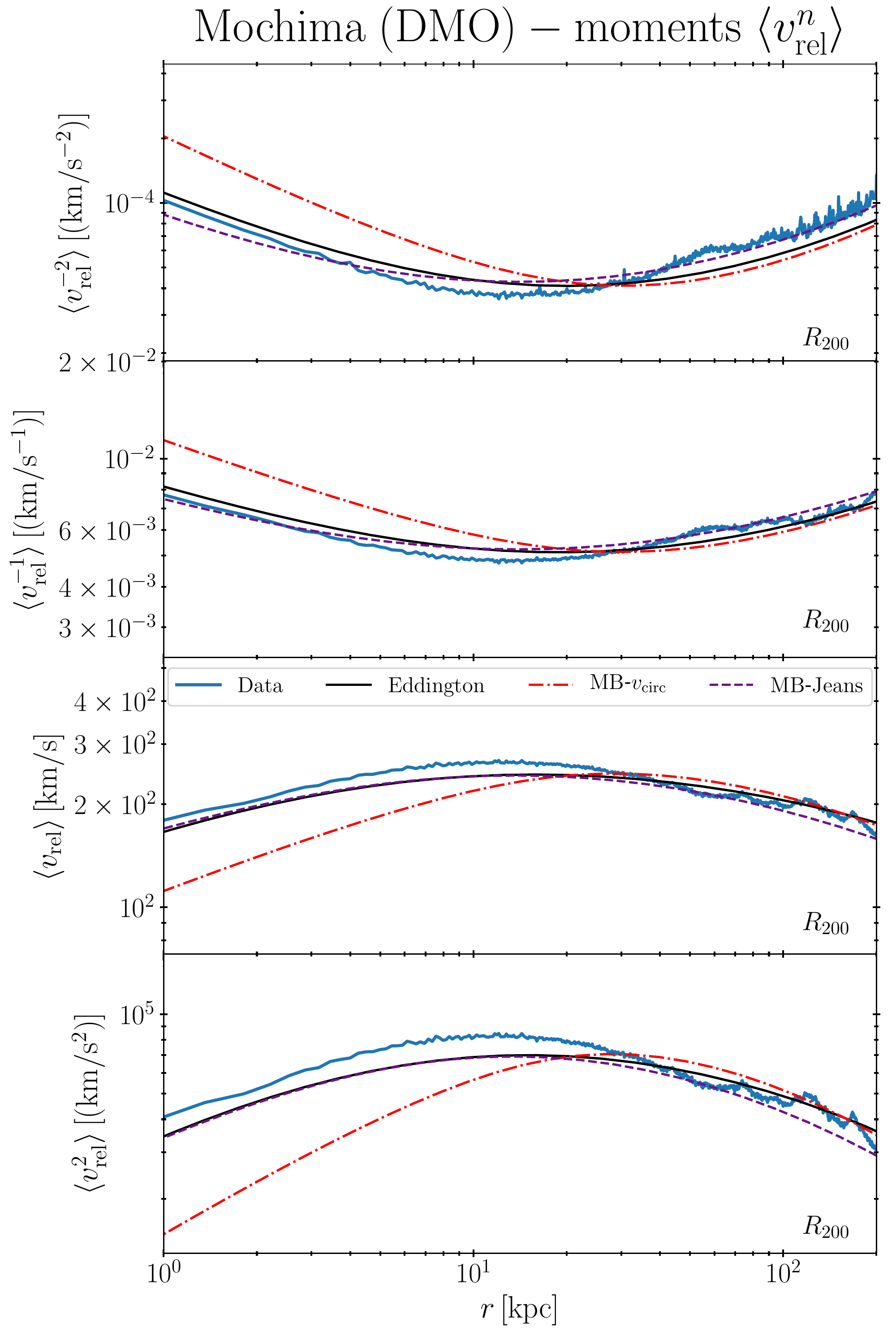} \hfill
\includegraphics[width=0.49\textwidth]{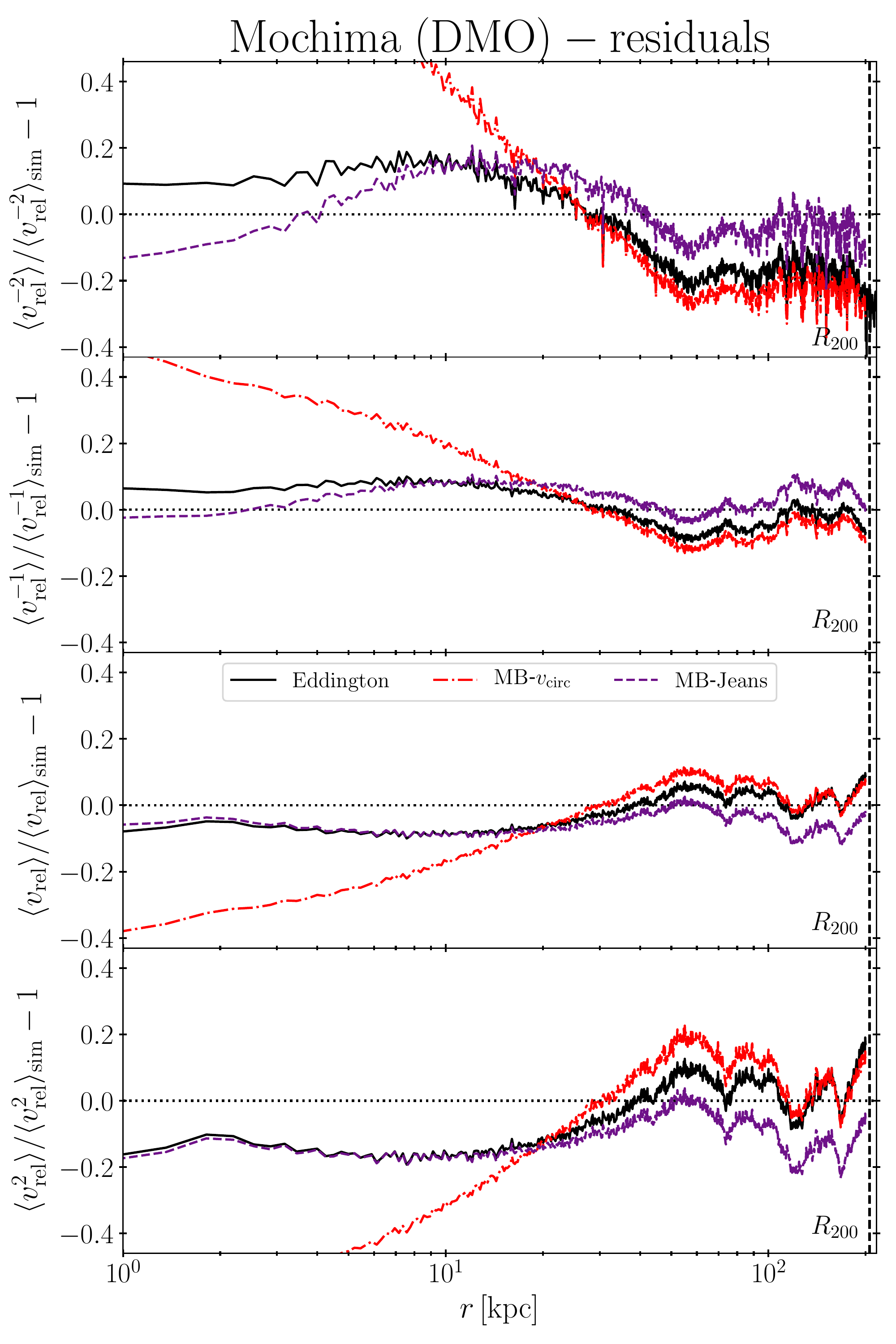}
\caption{\small Same as \citefig{fig:v-rel-moments-comparison-HALOC-DMO}, but for the Mochima simulation.}
\label{fig:v-rel-moments-comparison-Mochima-DMO}
\end{center}
\end{figure}

\begin{figure}[t]
\begin{center}
\includegraphics[width=0.49\textwidth]{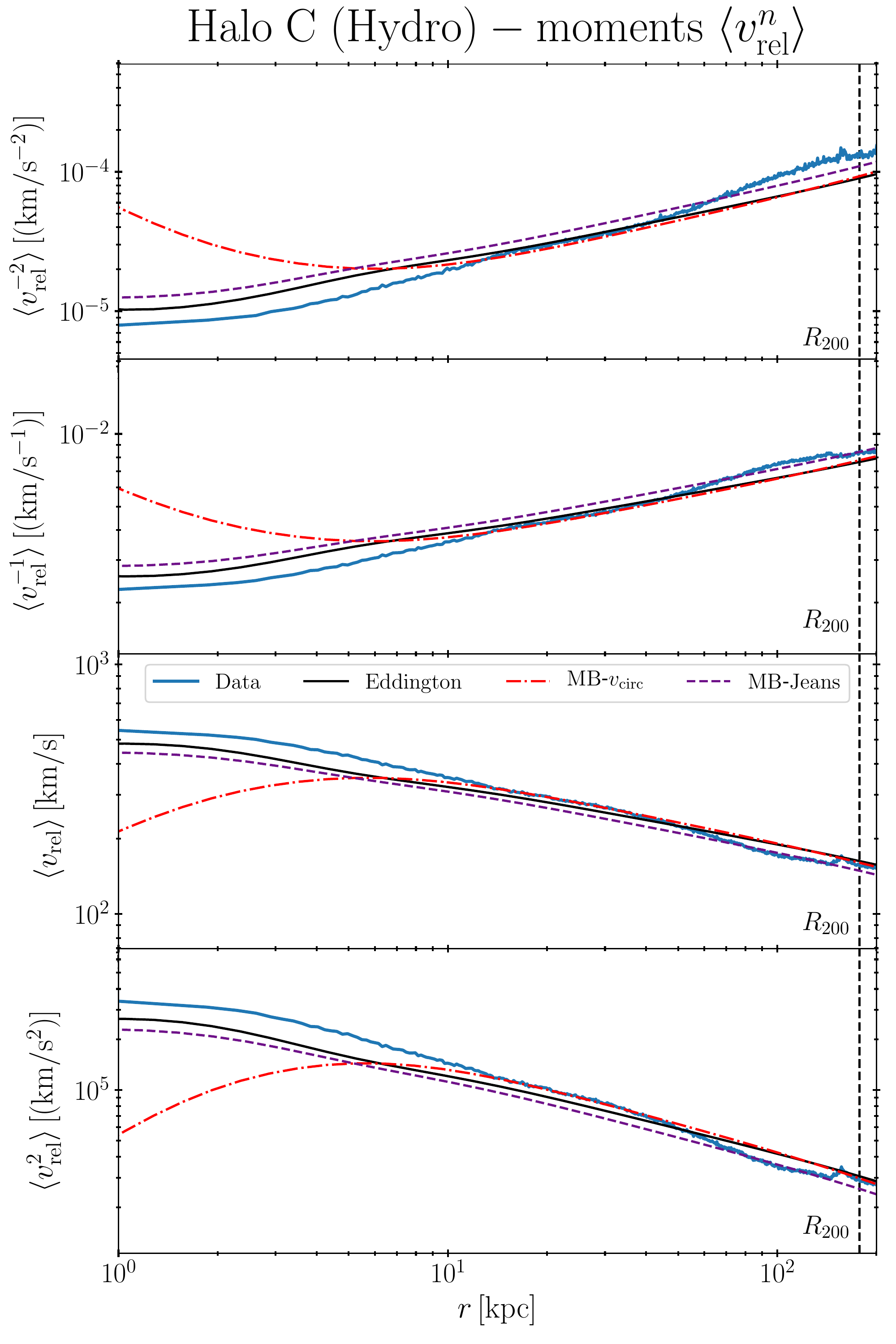} \hfill
\includegraphics[width=0.49\textwidth]{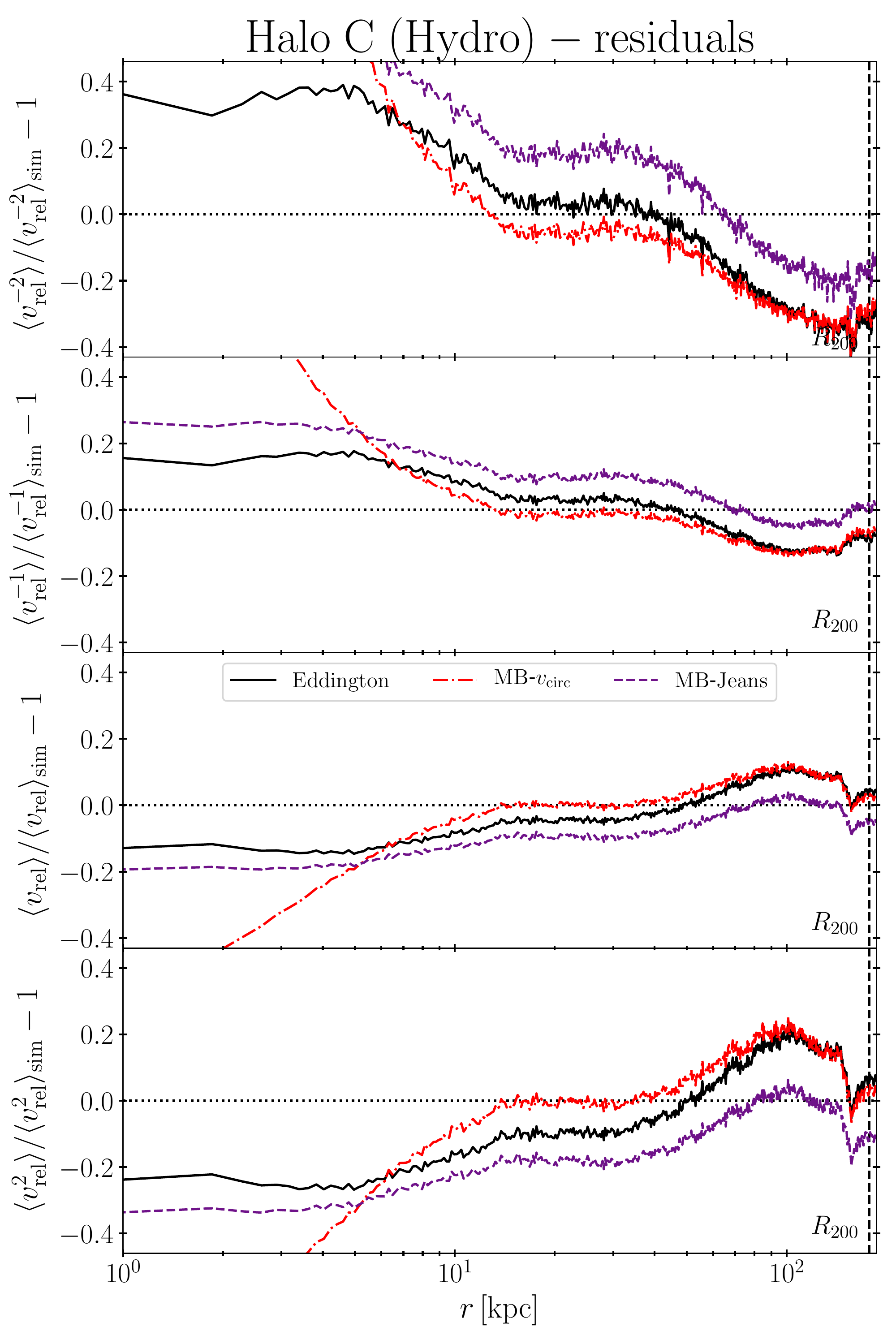}
\caption{\small Comparison of the moments of the \textit{relative} speed distribution from the Eddington method and from the simulation for Halo~C (left panel), and associated residuals with respect to the simulation (right panel), for the hydro run. Simulation outputs are shown as blue solid curves. Predictions obtained with the Eddington inversion method are displayed as black solid curves, while the two MB models ``$v_{\mathrm{circ}}$" and ``Jeans" are shown as red dot-dashed and purple dashed, respectively. These line styles match those in the speed distribution figures, \citefigs{fig:fv-comparison-HALOB}, \ref{fig:fv-comparison-HALOC}, and \ref{fig:fv-comparison-Mochima}.}
\label{fig:v-rel-moments-comparison-HALOC-hydro}
\end{center}
\end{figure}

\begin{figure}[t]
\begin{center}
\includegraphics[width=0.49\textwidth]{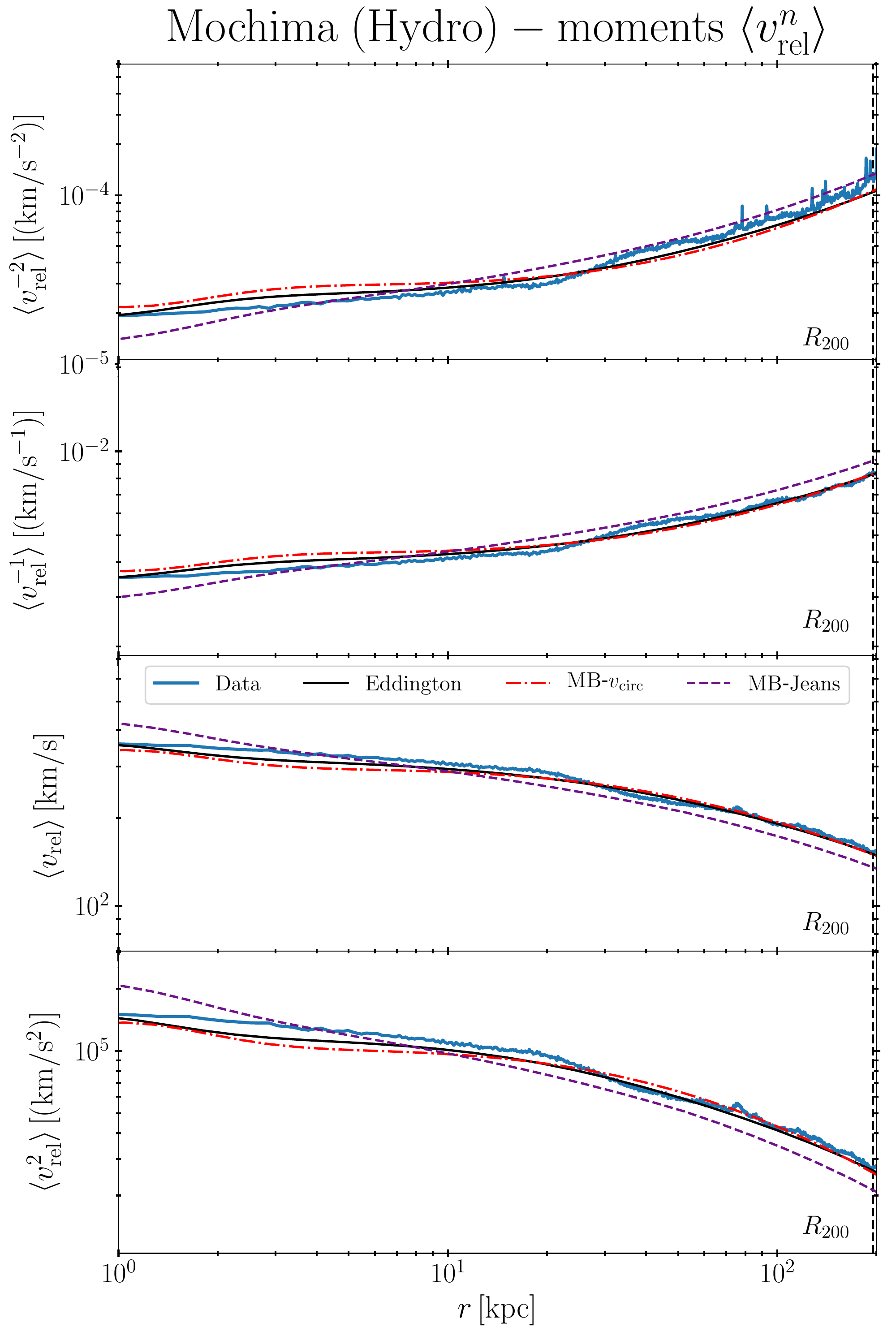} \hfill
\includegraphics[width=0.49\textwidth]{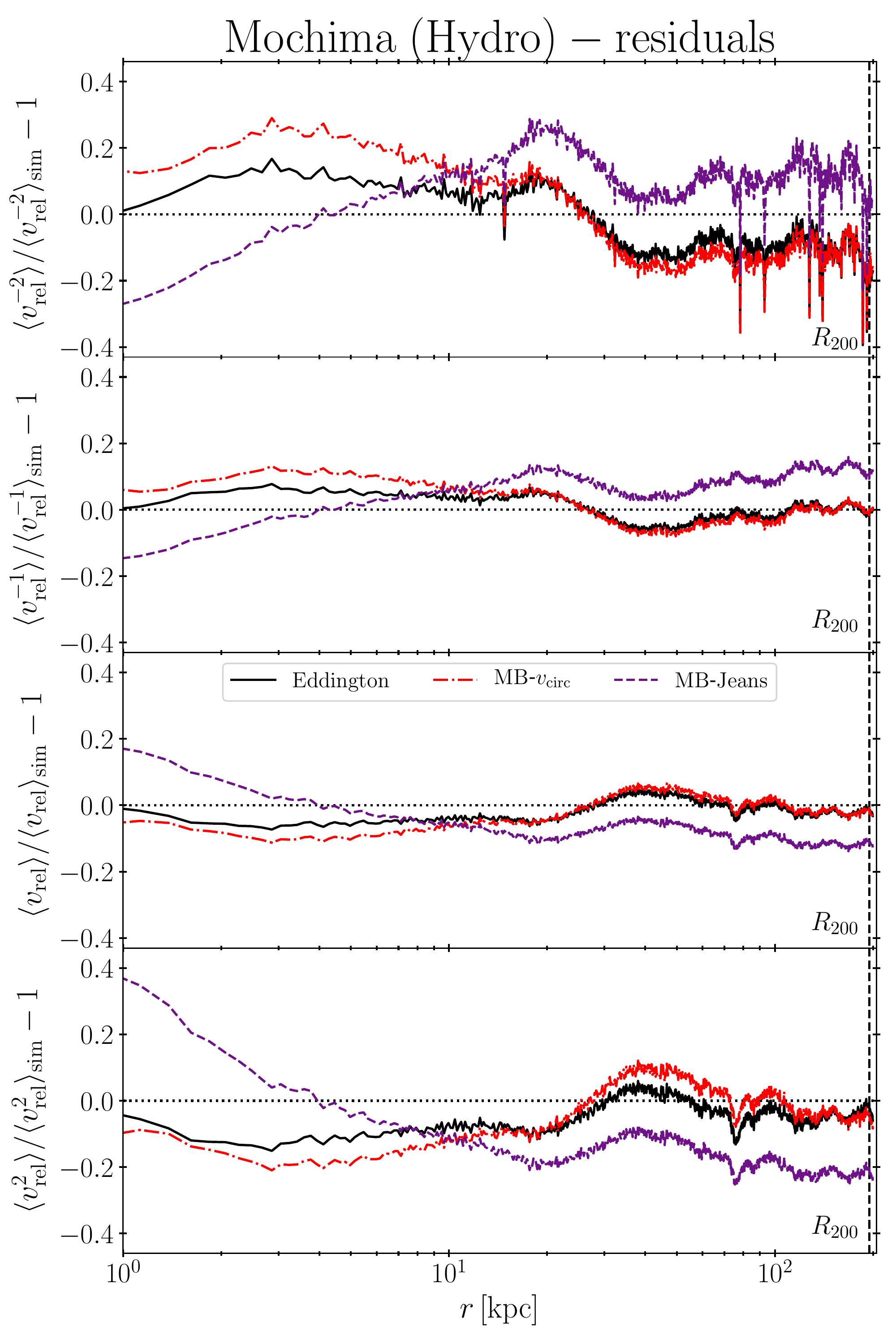}
\caption{\small Same as \citefig{fig:v-rel-moments-comparison-HALOC-hydro}, but for the Mochima simulation.}
\label{fig:v-rel-moments-comparison-Mochima-hydro}
\end{center}
\end{figure}

\clearpage

\bibliography{comp_theory_sims_published}
\bibliographystyle{JHEP}

\end{document}